\documentclass[nofootinbib,aps,prd,superscriptaddress,preprintnumbers,%
 showpacs,showkeys,floatfix,amssymb,amsfonts]{revtex4-1}  
\usepackage{graphicx}
\usepackage{bm}
\usepackage{array}
\usepackage{multirow}
\usepackage{amsmath}
\usepackage{amssymb}
\usepackage{amsfonts}
\usepackage{float}
\usepackage{dsfont}  
\usepackage{slashed}  
\usepackage{booktabs}
\usepackage{multirow} 
\usepackage{subfigure}
\usepackage[sort&compress]{natbib}
\usepackage{ulem}
\usepackage{empheq}
\usepackage{color}
\usepackage[svgnames,x11names,table]{xcolor}
\usepackage[colorlinks=true,linkcolor=MediumBlue,citecolor=MediumBlue,urlcolor=MediumBlue]{hyperref}
\usepackage[justification=raggedright,singlelinecheck=false]{caption}
\newcommand{\MSb}{{\overline{\rm MS}}}
\newcommand{\Efpm}{E_f^+}

\newcommand{\Eipm}{E_i^+}
\newcommand{\Eimm}{E_i^-}

\begin{document}
\title{Generalized Parton Distributions from Lattice QCD with Asymmetric \\[1ex] Momentum Transfer: Unpolarized Quarks at Nonzero Skewness}
\author{Min-Huan Chu}
\email{minhuan.chu@amu.edu.pl}
\affiliation{Faculty of Physics and Astronomy, Adam Mickiewicz University, ul.\ Uniwersytetu Pozna\'nskiego 2, 61-614 Pozna\'{n}, Poland}
\author{Manuel Cola\c{c}o}
\affiliation{Faculty of Physics and Astronomy, Adam Mickiewicz University, ul.\ Uniwersytetu Pozna\'nskiego 2, 61-614 Pozna\'{n}, Poland}
\author{Shohini Bhattacharya}
\affiliation{Department of Physics, University of Connecticut, Storrs, CT 06269, USA}
\author{Krzysztof Cichy}
\affiliation{Faculty of Physics and Astronomy, Adam Mickiewicz University, ul.\ Uniwersytetu Pozna\'nskiego 2, 61-614 Pozna\'{n}, Poland}

\author{Martha Constantinou}
\affiliation{Department of Physics,  Temple University,  Philadelphia,  PA 19122 - 1801,  USA}
\author{Andreas Metz}
\affiliation{Department of Physics,  Temple University,  Philadelphia,  PA 19122 - 1801,  USA}
\author{Fernanda Steffens}
\affiliation{Institut f\"ur Strahlen- und Kernphysik, Rheinische Friedrich-Wilhelms-Universit\"at Bonn,\\ Nussallee 14-16, 53115 Bonn}
%
%%%%%%%%%%%%%%%%%%%%%%%%%%%%%
\begin{abstract}
\,\\[3ex]
We extend the formalism of asymmetric frames of reference for generalized parton distributions (GPDs) to the case of nonzero skewness, i.e., including longitudinal momentum transfer.
The framework, based on Lorentz-invariant amplitudes and previously developed  and numerically implemented for unpolarized, helicity and transversity GPDs at zero skewness, gives efficient access to a broad range of kinematics, making full mapping of GPDs from the lattice realistic.
The general-skewness formalism is tested using lattice data with both transverse and longitudinal or only longitudinal momentum transfer, the latter being a special case with a reduced number of independent amplitudes.
We extract the amplitudes in coordinate space and express the GPDs $H$ and $E$ in terms of these amplitudes.
This is followed by reconstruction of quasi-distributions and their matching to the light cone. We further identify and discuss the principal challenges for nonzero skewness GPDs.
\end{abstract}

%%%%%%%%%%%%%%%%%%%%%%%%%%%%%
\maketitle

%%%%%%%%%%%%%%%%%%%%%%%%%%%%%%%%%%%%%%%%%%%%%%%%%%%%%%%%%%
\section{Introduction}
%%%%%%%%%%%%%%%%%%%%%%%%%%%%%%%%%%%%%%%%%%%%%%%%%%%%%%%%%%
The internal structure of hadrons has been the subject of intense study for over 50 years.
Several functions are used to quantify this structure.
The simplest are the parton distribution functions (PDFs), which describe the momentum fractions $x$ carried by hadron’s constituents, i.e., partons. 
Further elaboration of the hadronic structure can be obtained by various generalizations of PDFs, in particular generalized parton distributions (GPDs)~\cite{Mueller:1998fv, Ji:1996ek, Radyushkin:1996nd}.
Both PDFs and GPDs are defined via matrix elements of bi-local quark/gluon operators, with partonic fields separated along the light-front direction.
The difference between them lies in the initial and final states of the hadron, which coincide for PDFs but differ for GPDs, i.e., a nonzero momentum transfer is turned on in the latter.
This naturally establishes that the PDF limit of GPDs corresponds to the forward limit.
However, GPDs contain much more information than PDFs, and not all of them possess a PDF limit.
This wealth of information includes three-dimensional ``tomographic'' images of partons' position in the transverse plane~\cite{Burkardt:2000za, Ralston:2001xs, Diehl:2002he, Burkardt:2002hr}, decomposition of a hadron's angular momentum contributed by all partons~\cite{Ji:1996ek}, and mechanical properties of the system~\cite{Polyakov:2002wz, Polyakov:2002yz, Polyakov:2018zvc}.
For more information on GPDs, several reviews can be consulted~\cite{Goeke:2001tz, Diehl:2003ny, Ji:2004gf, Belitsky:2005qn, Boffi:2007yc, Guidal:2013rya, Mueller:2014hsa, Kumericki:2016ehc,Lorce:2025aqp}.

Importantly, it is not only the invariant momentum transfer, denoted by the Mandelstam variable $t$, that matters for the physics of GPDs.
Different physical properties are probed when this momentum transfer is purely in the transverse plane and when it includes a longitudinal component. The two cases are distinguished by the skewness variable $\xi$, expressing the difference between the momentum fractions $x$ carried by the initial- and the final-state parton.
The partonic interpretation of the process depends on both $x$ and $\xi$.
Assuming $\xi>0$ for definiteness, when $\xi < x < 1$ ($-1<x<-\xi$), GPDs describe a situation where the hadron emits a quark (antiquark) carrying a fraction $|x|+\xi$ of the hadron's momentum and the parton is reabsorbed by the hadron having changed its momentum fraction via interaction with a virtual photon to $|x|-\xi$. 
Meanwhile, if $-\xi<x<\xi$, the emission of a quark-antiquark pair with momentum fraction $2\xi$ is described.
The former case is a natural generalization of the PDF case and is governed by the standard Dokshitzer-Gribov-Lipatov-Altarelli-Parisi (DGLAP) evolution equations \cite{Dokshitzer:1977sg,Gribov:1972ri,Altarelli:1977zs}, while the latter one has no counterpart in the PDF setup and is, thus, genuinely new in GPDs.
Since the physical picture of the case $|x|<\xi$ is similar to the one encountered for mesonic distribution amplitudes (DAs), its evolution is governed by the evolution equations of the latter, i.e., the Efremov-Radyushkin-Brodsky-Lepage (ERBL) equations \cite{Efremov:1979qk,Lepage:1980fj}.
Thus, zero-skewness GPDs contain exclusively the DGLAP region, while nonzero-skewness GPDs have both the DGLAP and the ERBL region (as long as $\xi<1$).
In this paper, we concentrate on the $\xi\neq0$ case.

Experimental information on GPDs can be obtained from hard exclusive scattering processes such as deep virtual Compton scattering (DVCS)~\cite{Mueller:1998fv, Ji:1996ek, Radyushkin:1996nd, Ji:1996nm, Collins:1998be} and hard exclusive meson production~\cite{Radyushkin:1996ru, Collins:1996fb, Mankiewicz:1997uy}. However, extracting GPDs from such reactions in a model-independent manner is very complicated, mainly because in the observable quantities, such as Compton form factors, the GPDs enter in a convolution over the partonic momentum fraction $x$. This integration smears out the $x$-dependence and entangles it with the skewness $\xi$, making it difficult to isolate the functional form of the GPDs themselves~\cite{Bertone:2021yyz, Moffat:2023svr}. 
This difficulty becomes especially severe when $\xi \neq 0$, which sets GPDs apart from ordinary PDFs as discussed above.
Theoretically, the $\xi$-dependence of GPDs encodes important aspects of their analytic structure. 
The QCD evolution of GPDs with the renormalization scale is naturally expressed in terms of conformal moments—Mellin moments weighted by Gegenbauer polynomials—which diagonalize the leading-order evolution equations in the ERBL region~\cite{Ji:1996nm,Radyushkin:1997ki}. Each conformal moment evolves independently with scale according to its own anomalous dimension. This structure reflects the approximate conformal symmetry of QCD at short distances and provides a systematic framework for describing the scale dependence of GPDs at $\xi \neq 0$, where conventional DGLAP evolution in $x$ alone is insufficient. In addition, in the small-$\xi$ limit, the behavior of the GPDs is governed by Regge trajectories, leading to characteristic power-law scaling that manifests in observables such as DVCS amplitudes~\cite{Radyushkin:1998es,Diehl:2004cx}. This scaling behavior plays a central role in guiding phenomenological models of GPDs and in constraining their behavior across kinematic regimes.
Furthermore, GPDs at $\xi \neq 0$ provide access to form factors of the QCD energy-momentum tensor, which encode fundamental mechanical properties of hadrons such as internal pressure and shear forces. These form factors are directly linked to the Mellin moments of GPDs and cannot be fully reconstructed from the forward limit alone. Moreover, the $\xi$-dependence is also essential for isolating the so-called D-term contribution, which plays a key role in ensuring the polynomiality of Mellin moments of GPDs~\cite{Ji:1996ek,Radyushkin:1997ki,Schweitzer:2002yu}. Studying GPDs across a range of skewness values thus offers a powerful consistency check on theoretical frameworks, as polynomiality provides a model‑independent constraint derived from Lorentz invariance.
A related line of work also realizes these analytic features of GPDs at finite skewness, employing a conformal-moment representation generated by $t$-channel string exchanges in Anti-de Sitter (AdS) space. This approach naturally incorporates Regge behavior at small $\xi$, conformal-moment evolution, and polynomiality, while providing quantitative comparisons to lattice moments and $x$-space GPDs. For details, see Refs.~\cite{Mamo:2024jwp, Mamo:2024vjh, Hechenberger:2025rye, Hechenberger:2025wnz}. Lattice QCD can provide a valuable and complementary approach to this problem. While experimental data constrain GPDs primarily at small and positive skewness, lattice computations can access a broader range, including both large and negative values of $\xi$, which are difficult to reach experimentally. As we will show, this opens up new opportunities to explore the full kinematic structure of GPDs and to test theoretical expectations beyond the experimentally accessible regime. 

Thus, it is a natural question whether GPDs can be accessed in the non-perturbative setup of lattice QCD.
Similar to PDFs or any other partonic distributions, direct access in standard Euclidean lattice QCD is not possible due to the real-time component of light-front correlations.
Such direct access is only possible for the lowest Mellin moments of such distributions, accessed through the operator product expansion of local operators, see, e.g., Refs.~\cite{Hagler:2003jd, QCDSF-UKQCD:2007gdl, Alexandrou:2011nr, Alexandrou:2013joa,Constantinou:2014tga,Green:2014xba,Alexandrou:2017ypw,Alexandrou:2017hac,Hasan:2017wwt,Gupta:2017dwj,Capitani:2017qpc,Alexandrou:2018sjm,Shintani:2018ozy,Jang:2018djx,Bali:2018qus,Bali:2018zgl,Alexandrou:2019ali,Jang:2019jkn,Constantinou:2020hdm,Alexandrou:2022dtc,Tsuji:2023llh,Djukanovic:2023beb,Djukanovic:2023jag,Hackett:2023rif,Alexandrou:2025vto}.

However, recent years have brought about a new class of methods in which access to the $x$-dependence is possible, in principle.
Even though proposals along these lines were put up already long ago~\cite{Liu:1993cv,Braun:1994jq,Aglietti:1998ur,Detmold:2005gg,Braun:2007wv}, this very prolific thread gained momentum only after the seminal proposal of Ji to calculate quasi-distributions in 2013~\cite{Ji:2013dva,Ji:2014gla,Ji:2020ect}.
The essence of the idea is to replace the calculation of the lattice-inaccessible light-front correlations with one of equal-time spatial correlations, readily computable on a lattice with a Euclidean metric.
Quasi-distributions reduce to their light-cone physical counterparts in the infinite-momentum limit.
Obviously, this limit is in itself inaccessible on the lattice, but quasi-distributions at a finite momentum $P^3$ can be ``translated'' to the infinite-momentum frame using the so-called matching procedure.
The latter exploits the fact that quasi- and light-cone distributions have the same infrared structure and differ only in the ultraviolet regime.
Consequently, their difference can be calculated in perturbation theory~\cite{Xiong:2013bka,Ma:2014jla, Izubuchi:2018srq}.
Ji's proposal led to a revival of some of the older proposals~\cite{Liu:1993cv,Braun:1994jq,Aglietti:1998ur,Detmold:2005gg,Braun:2007wv} and also to new or related ones~\cite{Chambers:2017dov,Radyushkin:2017cyf,Radyushkin:2019mye,Ma:2014jla,Ma:2017pxb,Detmold:2021uru}.
To illustrate the dynamical progress, we note example applications of these new frameworks for quark and gluon PDFs, e.g.\ Refs.~\cite{Lin:2014zya,Alexandrou:2015rja,Chen:2016utp,Alexandrou:2016jqi,Alexandrou:2017huk,Chen:2017mzz,Orginos:2017kos,Lin:2017ani,Alexandrou:2018pbm,Lin:2018pvv,Alexandrou:2018eet,Liu:2018uuj,Zhang:2018nsy,Sufian:2019bol,Alexandrou:2019lfo,Izubuchi:2019lyk,Joo:2019jct,Joo:2019bzr,Cichy:2019ebf,Bringewatt:2020ixn,DelDebbio:2020rgv,Chai:2020nxw,Joo:2020spy,Bhat:2020ktg,Alexandrou:2020uyt,Alexandrou:2020qtt,Lin:2020ssv,Fan:2020nzz,Gao:2020ito,Lin:2020fsj,Karpie:2021pap,Alexandrou:2021oih,Egerer:2021ymv,HadStruc:2021qdf,Gao:2021dbh,Gao:2022iex,LatticeParton:2022xsd,HadStruc:2022yaw,HadStruc:2022nay,Gao:2022uhg,Delmar:2023agv,Holligan:2024umc,Holligan:2024wpv}, notably including also higher twist \cite{Bhattacharya:2020cen,Bhattacharya:2020xlt,Bhattacharya:2020jfj,Bhattacharya:2021boh,Bhattacharya:2021moj}, mesonic and baryonic DAs~\cite{Zhang:2017zfe,Zhang:2017bzy,Bali:2018spj,Xu:2018mpf,Zhang:2020gaj,Hua:2020gnw,Detmold:2021qln,LatticeParton:2022zqc,Gao:2022vyh,Holligan:2023rex,LatticeParton:2024vck,LatticeParton:2024zko} and transverse-momentum-dependent PDFs and wave functions~\cite{LatticeParton:2020uhz,Shanahan:2020zxr,Zhang:2020dbb,Li:2021wvl,Shanahan:2021tst,Schlemmer:2021aij,LatticePartonLPC:2022eev,LatticePartonCollaborationLPC:2022myp,Zhang:2022xuw,Alexandrou:2023ucc,LatticeParton:2023xdl,LatticePartonLPC:2023pdv,LatticeParton:2024mxp}. More recent approaches include a procedure that utilizes gradient flow to determine moments of parton distribution functions of any order~\cite{Shindler:2023xpd}, as well as using boosted fields in the Coulomb gauge without a Wilson line~\cite{Gao:2023lny}.
For more thorough accounts of this progress, see, e.g.\ the reviews~\cite{Cichy:2018mum,Ji:2020ect,Constantinou:2020pek,Cichy:2021lih,Cichy:2021ewm}.

Separately, we discuss the progress in lattice extraction of GPDs using the novel approaches.
The matching for general skewness in a simple transverse momentum cutoff scheme was worked out already shortly after the proposal to calculate quasi-distributions~\cite{Ji:2015qla, Xiong:2015nua}.
Later, it was adapted to more practical renormalization schemes~\cite{Liu:2019urm, Radyushkin:2019owq, Ma:2022ggj,Yao:2022vtp}.
Early work included also model studies~\cite{Bhattacharya:2018zxi, Bhattacharya:2019cme, Ma:2019agv, Luo:2020yqj, Shastry:2022obb}.
The first lattice QCD calculations were reported for the pion~\cite{Chen:2019lcm} and the nucleon~\cite{Alexandrou:2020zbe,Lin:2020rxa,Alexandrou:2021bbo,CSSMQCDSFUKQCD:2021lkf,Lin:2021brq,Bhattacharya:2023nmv}, all of them in symmetric frames of reference, where the momentum transfer is evenly split between the source and the sink state.
In such a setup, accessing each momentum transfer value necessitates a new calculation, making the full mapping of GPDs' dependence on $t$ and $\xi$ prohibitively expensive.
To avoid this caveat, asymmetric frames of reference were introduced in Ref.~\cite{Bhattacharya:2022aob}.
The asymmetric frame is chosen so that all momentum transfer is ascribed to the source, while the sink momentum is fixed to $P = (P^0, 0, 0, P^3)$, where $P$ denotes the average hadron four‑momentum. For a general discussion of GPDs in different frames, we also refer to~\cite{Braun:2023alc}. This choice allows us to access several combinations of $(t,\,\xi)$ in a single lattice calculation.
With this development, mapping out GPDs in a wide kinematic range becomes a realistic task.
The asymmetric-frame approach was initially applied to the unpolarized case at zero skewness, resulting in the reconstruction of the nucleon's $x$-dependent distributions~\cite{Bhattacharya:2022aob,Cichy:2023dgk}, the extraction of several Mellin moments~\cite{Bhattacharya:2023ays} and the first application of the pseudo-GPD approach~\cite{Bhattacharya:2024qpp}.
Results for the pion were also reported \cite{Ding:2024saz}.
Recently, the lattice unpolarized GPDs data were used to explore the possible synergies with experimental data to obtain reliable tomographic pictures and information on the quark angular momentum~\cite{Cichy:2024afd}.
Further, the formalism for the axial vector case was developed, along with lattice results for the $x$-dependent distributions of the nucleon~\cite{Bhattacharya:2023jsc} and Mellin moments of helicity GPDs~\cite{Bhattacharya:2024wtg}.
The framework has recently become complete for twist-2 GPDs with the tensor case~\cite{Bhattacharya:2025yba}.
Other recent work on GPDs from the lattice involved an investigation of systematic effects from renormalization and matching \cite{Holligan:2023jqh} in the hybrid scheme \cite{Ji:2020brr}.
Finally, the short-distance factorization approach (pseudo-GPDs) was used to extract several Mellin moments of unpolarized $H$ and $E$ GPDs, using data at both zero and nonzero skewness \cite{HadStruc:2024rix}.

In this paper, we extend the methodology of Ref.~\cite{Bhattacharya:2022aob} to incorporate the case of unpolarized GPDs at nonzero skewness, i.e.\ including momentum transfer also in the direction of nucleon's motion. In particular, we consider the special case of only longitudinal momentum transfer, where the number of independent Dirac structures contributing to the underlying matrix elements is reduced. 
Further, we test the methodology with lattice data corresponding to the same ensemble as used in our previous zero skewness work. We extract Lorentz-invariant amplitudes and use them to access matrix elements of GPDs, paying attention to their combined $(t,\xi)$-dependence.
Then, we reconstruct the $x$-dependence and perform matching to light-cone GPDs, revealing and discussing difficulties of these stages.

We organize the remainder of the paper as follows. Section II reviews the principles of asymmetric frames and presents the general methodology for nonzero skewness and details of our lattice setup.
In Section III, we discuss our numerical results, first in coordinate space and then, upon matching, in momentum space.
Section IV concludes the paper and offers an outlook on future work.

%%%%%%%%%%%%%%%%%%%%%%%%%%%%%%%%%%%%%%%%%%%%%%%%%%%%%%%%%%
\section{GPDs from asymmetric frames}
\label{sec:gpd_asy}
%%%%%%%%%%%%%%%%%%%%%%%%%%%%%%%%%%%%%%%%%%%%%%%%%%%%%%%%%%
\subsection{Generalities}
%%%%%%%%%%%%%%%%%%%%%%%%%%%%%%%%%%%%%%%%%%%%%%%%%%%%%%%%%%
GPDs describe aspects of a hadron's internal structure through their dependence on three variables:
\begin{itemize}
\item $x$ -- parton's momentum fraction, 
\item $t\equiv \Delta^2$ -- the invariant momentum transfer squared ($\Delta^\mu$ being the momentum transfer 4-vector),
\item $\xi_{\rm LC} \equiv -\Delta^+/2P^+$ -- skewness, representing the longitudinal part of momentum transfer ($P^\mu$ -- average 4-momentum; hadron moving along the light-cone plus direction; light-cone coordinates defined as $v^\pm=(v^0\pm v^3)/\sqrt{2}$ for a generic 4-vector $v^\mu$); we use the subscript LC to differentiate from skewness in the quasi-GPD case.
\end{itemize}

In Refs.~\cite{Bhattacharya:2022aob,Bhattacharya:2023jsc,Bhattacharya:2025yba}, we developed the formalism to extract GPDs in asymmetric frames of reference, respectively for unpolarized, axial vector and tensor cases.
All three cases were illustrated with lattice calculations of amplitudes and GPDs at zero skewness, i.e., with purely transverse momentum transfer.
Here, we recall the most important aspects of the unpolarized case and generalize the relevant expressions to nonzero skewness: we consider the general case with both transverse and longitudinal momentum transfers, as well as the special case with only longitudinal transfer.

We start by recalling the underlying motivation for employing asymmetric frames of reference wherein the final momentum of a hadron is fixed and all momentum transfer is ascribed to the source momentum.
The traditionally used symmetric frames of reference split this momentum transfer equally between the source and sink momenta, implying that both of these momenta are $\Delta$-dependent.
From the lattice-QCD perspective, the symmetric frame necessitates a separate calculation for every case of $\Delta$, making the full mapping of $(x,t,\xi)$-dependent GPDs prohibitively expensive.
Meanwhile, the asymmetric frame that we use is defined by:
\begin{equation}
\label{eq:pfa}
\vec{P}_f =  \left(0,0,P_f^3\right),
\end{equation}
\begin{equation}
\label{eq:pia}
\vec{P}_i=\vec{P}_f - \vec{\Delta} =  \left(-\Delta^1,-\Delta^2,P_f^3-\Delta^3\right),
\end{equation}
where the subscript $i/f$ denotes the initial/final state, $\vec{\Delta}=(\Delta^1,\Delta^2,\Delta^3)$ represents the spatial part of $\Delta^\mu$ and the average momentum is defined as $P=(P_i+P_f)/2$.
The key aspect is that the final momentum is fixed and thus, different cases of $\Delta$ can be obtained in a single calculation, i.e., from a computed fixed-sink sequential propagator.

In Ref.~\cite{Bhattacharya:2022aob}, we considered the general vector matrix element,
\begin{equation}
\label{eq:ME}
F^{\mu}(z, P, \Delta) = \langle P_f | \bar{\psi} (-\tfrac{z}{2}) \gamma^\mu \, {\cal W}(-\tfrac{z}{2}, \tfrac{z}{2})  \psi (\tfrac{z}{2}) |P_i\rangle ,
\end{equation}
%%%%%%%%%%%%%%%%%%%
where $z^\mu \equiv (z^0 = 0, z^\perp = 0^\perp, z^3 \neq 0)$, $\psi/\bar{\psi}$ are quark/antiquark fields, ${\cal W}$ is a straight Wilson line that ensures gauge invariance and $|P_{i/f}\rangle$ are initial/final nucleon states with momenta $P_{i/f}$.
The matrix element $F^{\mu} (z,P,\Delta)$, underlying unpolarized GPDs, can be parametrized in terms of eight linearly-independent Dirac structures and eight frame-independent amplitudes, denoted $A_i \equiv A_i (z\cdot P, z \cdot \Delta, \Delta^2, z^2)$, which depend on the Lorentz invariants $z\cdot P$, $z\cdot\Delta$, $\Delta^2$ and $z^2$, 
%%%%%%%%%%%%%%%%%%%%
\begin{align}
\label{eq:MEparametrization}
F^{\mu} (z,P,\Delta) & = \bar{u}(P_f,\lambda') \bigg [ \dfrac{P^{\mu}}{m} A_1 + m z^{\mu} A_2 + \dfrac{\Delta^{\mu}}{m} A_3 + i m \sigma^{\mu z} A_4 + \dfrac{i\sigma^{\mu \Delta}}{m} A_5 \nonumber \\[1ex]
& \hspace{5cm} + \dfrac{P^{\mu} i\sigma^{z \Delta}}{m} A_6 + m z^{\mu} i\sigma^{z \Delta} A_7 + \dfrac{\Delta^{\mu} i\sigma^{z \Delta}}{m} A_8  \bigg ] u(P_i, \lambda) \, ,
\end{align}
%%%%%%%%%%%%%%%%%%%%
with summation over repeated indices and the following compact notation:
$\sigma^{\mu \nu} \equiv \tfrac{i}{2} (\gamma^\mu \gamma^\nu - \gamma^\nu \gamma^\mu)$,  
$\sigma^{\mu z} \equiv \sigma^{\mu \rho} z_\rho$, 
$\sigma^{\mu \Delta} \equiv \sigma^{\mu \rho} \Delta_\rho$, $\sigma^{z \Delta} \equiv \sigma^{\rho \tau} z_\rho \Delta_\tau$. 

We decompose the momentum transfer as $\Delta = \Delta_T + \Delta_L$, with $\Delta_T = (0, \Delta^1, \Delta^2, 0)$ and $\Delta_L = (\Delta^0, 0, 0, \Delta^3)$. In the special case where the transfer is purely longitudinal, i.e. $\Delta^1 = \Delta^2 = 0$ and $\Delta^0 \neq 0,\, \Delta^3 \neq 0$, the above parametrization simplifies considerably and the number of independent Dirac structures is reduced to three, i.e.,
%%%%%%%%%%%%%%%%%%%%
\begin{align}
\label{eq:MEparametrization2}
F^{\mu,L} (z,P,\Delta_L) & = \bar{u}(P_f,\lambda') \bigg [ \dfrac{P^{\mu}}{m} A_1^L + m z^{\mu} A_2^L +  i m \sigma^{\mu z} A_4^L   \bigg ] u(P_i, \lambda) \,.
\end{align}
%%%%%%%%%%%%%%%%%%%%
The Lorentz-invariant amplitudes of this case, denoted with the superscript $L$, are linear combinations of the eight amplitudes of the general case, see Appendix \ref{sec:app_longitudinaldecomp} for more details. However, the general-case amplitudes cannot be disentangled from one another and the data provides access only to the redefined amplitudes $A_i^L$ ($i=1\,,2\,,4$).

We also note that our numerical analysis employs various combinations of signs for the components of 4‑vectors like $P^\mu$, $\Delta^\mu$ and $z^\mu$. Hence, it is important to know the symmetry properties of the amplitudes under various transformations. This was extensively discussed in Appendix B of Ref.~\cite{Bhattacharya:2022aob} and here, we recall the properties that are crucial in the nonzero-skewness case, i.e., under time reversal. For the amplitude $A_n$ ($n=1,\ldots,8$), these read
\begin{equation}
A^*_n (- \bar{z} \cdot \bar{P}, -\bar{z} \cdot \bar{\Delta}, \bar{\Delta}^2, \bar{z}^2)  = s_n A_n (z \cdot P, z \cdot \Delta, \Delta^2, z^2) \,, \label{eq:symmetries_A}   
\end{equation}
where for a generic 4-vector $v^\mu=(v^0,\vec{v})$, $\bar{v}^\mu=(v^0,-\vec{v})$, and the symmetry properties are expressed by the sign coefficient $s_n$, where $s_n=1$ (-1) for the amplitudes with odd (even) $n$. 
We note in passing that, for $\xi$-symmetry, both Hermiticity and time-reversal constraints must be imposed~\cite{Bhattacharya:2022aob}. 
Applying these constraints leads to three out of the eight amplitudes vanishing for $\xi = 0$, namely 
$A_3$, $A_4$, and $A_8$, since these amplitudes are skewness-odd and do not have poles at zero skewness. 
This was numerically confirmed in Ref.~\cite{Bhattacharya:2022aob}. 
In the present work with $\xi \neq 0$, however, we find these amplitudes are potentially nonzero.
The other amplitudes, $A_1$, $A_2$, $A_5$, $A_6$, and $A_7$ are skewness-even.

%%%%%%%%%%%%%%%%%%%%%%%%%%%%%%%%%%%%%%%%%%%%%%%%%%%%%%%%%%
\subsection{Decomposition of matrix elements in terms of amplitudes}
\label{sec:decomp}
%%%%%%%%%%%%%%%%%%%%%%%%%%%%%%%%%%%%%%%%%%%%%%%%%%%%%%%%%%
Matrix elements (MEs) are frame-dependent and below, we show the decomposition in the asymmetric frame for the vector case, i.e., operator insertions including the $\gamma_\mu$ ($\mu=0,\,1,\,2,\,3$) Dirac structure.
From now on, we specify to Euclidean spacetime metric and use lower Dirac indices to avoid confusion with general Minkowski-spacetime expressions.
To disentangle the amplitudes $A_1$ - $A_8$, eight independent MEs are needed and can be obtained by applying four different parity projectors:
\begin{eqnarray}
\mathrm{unpolarized:\quad}\Gamma_0 &=& \frac{1}{4} \left(1 + \gamma_0\right)\,, \\
\mathrm{polarized\ in\ the\ }k\rm{-direction:\quad}\Gamma_k &=& \frac{1}{4} \left(1 + \gamma_0\right) i \gamma_5 \gamma_k\,, \quad k=1,2,3\,.
\end{eqnarray}

For the combination of $\mu$ and $\nu$ insertion and projection Dirac indices, we denote the extracted MEs by
\begin{equation}
\label{eq:Pi}
\Pi_\mu(\Gamma_\nu) = K\, {\rm Tr}\left[\Gamma_\nu \, \left(\frac{-iP_f \hspace*{-0.33cm}\slash \,\,+m}{2m} \right) \, \tilde{F}^\mu \left(\frac{-iP_i \hspace*{-0.27cm}\slash \,\,+m}{2m} \right) \right],
\end{equation}
where $\tilde{F}^\mu$ is the parametrization \eqref{eq:MEparametrization} or \eqref{eq:MEparametrization2} without the spinors $\bar{u}$, $u$, and
$K$ is a kinematic factor that reads 
\begin{equation}
K = \frac{2 m^2}{\sqrt{E_f E_i (E_f+m) (E_i+m)}},
\end{equation}
with $m$ being the nucleon mass and $E_{i/f}$ the energies of the initial/final state, $E_{i/f}=\sqrt{m^2+\vec{P}_{i/f}^2}$.

Below, we recall the expressions from Appendix C of Ref.~\cite{Bhattacharya:2022aob} for general skewness and we write them in a convenient form employing
the energies $E_{i/f}$, $E^\pm=E_f\pm E_i$, $E_{f/i}^\pm\equiv E_{f/i}\pm m$, the nucleon mass, components of the momentum transfer vector $\vec{\Delta}$ and the third component of the final momentum $P_{f3}$.\\

\newpage
\begin{center}
\textit{Momentum transfer in both transverse and longitudinal directions ($\Delta_T\neq0$, $\Delta_L\neq0$)}
\end{center}
\begin{eqnarray}
\Pi_0(\Gamma_0) &=& K\,  \Bigg( A_1 \frac {E^+(\Efpm\Eipm - P_{f3}
              (P_{f3}-\Delta_3))} {8 m^3} 
              + A_3 \frac {E^-(\Efpm\Eipm - P_{f3} (P_{f3}-\Delta_3))} {4 m^3}  
    + zA_4 \frac {\Efpm \Delta_3 - E^- P_{f3}} {4 m} \nonumber \\[1ex]   
    && \qquad + A_5 \frac {(E^- E^+
              \Efpm - (\Efpm+\Eipm) P_{f3} \Delta_3)} {4 m^3}  
    + zA_6 \frac {E^+ \left (-3 \Delta_3 P_{f3}^2 + \Delta_3^2 P_{f3} + 2
           E_f E^- P_{f3} + \Eimm \Efpm \Delta_3 \right) } {8 m^3}  \nonumber \\[1ex]
   && \qquad  + zA_8 \frac {
           E^- \left (-3 \Delta_3 P_{f3}^2 + \Delta_3^2 P_{f3} + 
          2 E_f E^- P_{f3} + \Eimm
              \Efpm \Delta_3 \right)} {4 m^3}   \Bigg) \hspace*{2cm} \label{eq:Pi0G0_ns} \\[3ex]
\Pi_0(\Gamma_1) &=& i\, K\, \Delta_2\, \Bigg( 
A_1 \frac {E^+ P_{f3} } {8 m^3} + A_3 \frac {E^- P_{f3} } {4m^3}  + zA_4 \frac {  \Efpm } {4 m} 
- A_5 \frac { 
           (\Efpm+\Eipm) P_{f3} } {4 m^3}    \nonumber \\[1ex]
   && \qquad\qquad  - zA_6 \frac {E^+   \left ((\Efpm)^2 + 
          P_{f3} (P_{f3} - \Delta_3) \right)} {8 m^3}   
          - zA_8 \frac {E^-  \left ((\Efpm)^2 + P_{f3} (P_{f3} - \Delta_3) \right)} {4 m^3}
  \Bigg) \hspace*{1cm} \label{eq:Pi0G1_ns}\\[3ex]
\Pi_0(\Gamma_2) &=& -i\, K\, \Delta_1\, \Bigg(
 A_1 \frac { E^+ P_{f3}} {8m^3}  
+ A_3\frac { E^-P_{f3}} {4 m^3} 
           + zA_4  \frac { \Efpm } {4 m}
           -A_5 \frac {  (\Efpm+\Eipm) P_{f3} } {4 m^3} \nonumber \\[1ex]
   && \qquad\qquad  - zA_6\frac { E^+ \left ((\Efpm)^2 + P_{f3} (P_{f3} - \Delta_3) \right) } {8 m^3} 
   - zA_8 \frac {E^- \left ((\Efpm)^2 + P_{f3} (P_{f3} - \Delta_3) \right)} {4 m^3} 
  \Bigg) \label{eq:Pi0G2_ns}\\[1ex]
 \Pi_0(\Gamma_3) &=& 0 \label{eq:Pi0G3_ns}\\[2ex]
\Pi_1(\Gamma_0) &=& i\, K\,\Delta_1\,  \Bigg(
 A_1 \frac { \Efpm \Eipm - P_{f3} (P_{f3}-\Delta_3)} {8 m^3} 
 - A_3 \frac { \Efpm \Eipm - P_{f3} (P_{f3}-\Delta_3)} {4 m^3} 
    + zA_4 \frac {P_{f3}} {4 m}   \nonumber \\[1ex]
   && \qquad\qquad + A_5 \frac {
        (P_{f3} (P_{f3} - \Delta_3) - \Eimm \Efpm)} {4 m^3}
   +zA_6 \frac {  -3 \Delta_3 P_{f3}^2 + \Delta_3^2 P_{f3} + 2 E_f
              E^- P_{f3} + \Eimm \Efpm \Delta_3 } {8 m^3} \nonumber \\[1ex]
   && \qquad\qquad  +zA_8 \frac { P_{f3} \left (\Delta_1^2+\Delta_2^2-(E^-)^2 \right) + E^- \Efpm \Delta_3 } {4 m^3} \Bigg) \label{eq:Pi1G0_ns}\\[3ex]
\Pi_1(\Gamma_1) &=&  K\, \Delta_1 \Delta_2\, \Bigg(
-A_1 \frac { P_{f3} } {8 m^3} + A_3 \frac { P_{f3}} {4 m^3} + A_5 \frac { P_{f3}} {4 m^3}  \nonumber \\[1ex]
   && \qquad\qquad\quad  + zA_6 \frac{\left ((\Efpm)^2 + 
          P_{f3} (P_{f3} - \Delta_3) \right) } {8 m^3} 
          - zA_8 \frac { \left ((\Efpm)^2 + 
          P_{f3} (P_{f3} - \Delta_3) \right) } {4 m^3}
  \Bigg) \label{eq:Pi1G1_ns}\\[3ex]
\Pi_1(\Gamma_2) &=&  K\,  \Bigg(
A_1 \frac { P_{f3} \Delta_1^2} {8
        m^3} - A_3 \frac { P_{f3} \Delta_1^2} {4 m^3} + zA_4 \frac { (\Efpm \Eipm + P_{f3} (P_{f3} - \Delta_3)) } {4
           m} 
 -A_5 \frac { P_{f3} \left (\Delta_1^2 - 2mE^- \right) + ((\Efpm)^2- P_{f3}^2)
               \Delta_3 } {4 m^3} 
 \nonumber \\[1ex]
   && \qquad  - zA_6 \frac { \left ((\Efpm)^2 + P_{f3} (P_{f3} - \Delta_3) \right) 
        \Delta_1^2} {8 m^3} 
        + zA_8 \frac { \left ((\Efpm)^2 + 
          P_{f3} (P_{f3} - \Delta_3) \right)  \Delta_1^2} {4
        m^3}  
  \Bigg) \label{eq:Pi1G2_ns}\\[3ex]  %
\Pi_1(\Gamma_3) &=& K\,\Delta_2\,  \Bigg( 
 zA_4 \frac { P_{f3}} {4 m} +A_5 \frac { (\Efpm)^2 - P_{f3}^2} {4 m^3} 
  \Bigg) \label{eq:Pi1G3_ns}
\end{eqnarray}
\begin{eqnarray}
\Pi_2(\Gamma_0) &=& i\, K\,\Delta_2\,  \Bigg(
 A_1 \frac { \Efpm \Eipm - P_{f3} (P_{f3}-\Delta_3)} {8 m^3} 
 - A_3 \frac { \Efpm \Eipm - P_{f3} (P_{f3}-\Delta_3)} {4 m^3} 
    + zA_4 \frac {P_{f3}} {4 m}   \nonumber \\[1ex]
   && \qquad\qquad + A_5 \frac {
        (P_{f3} (P_{f3} - \Delta_3) - \Eimm \Efpm)} {4 m^3}
   +zA_6 \frac {  -3 \Delta_3 P_{f3}^2 + \Delta_3^2 P_{f3} + 2 E_f
              E^- P_{f3} + \Eimm \Efpm \Delta_3 } {8 m^3} \nonumber \\[1ex]
   && \qquad\qquad  +zA_8 \frac { P_{f3} \left (\Delta_1^2+\Delta_2^2-(E^-)^2 \right) + E^- \Efpm \Delta_3 } {4 m^3} \Bigg) \label{eq:Pi2G0_ns}\\[1ex]
  \Pi_2(\Gamma_1) &=&  -K\,  \Bigg(
A_1 \frac { P_{f3} \Delta_2^2} {8
        m^3} - A_3 \frac { P_{f3} \Delta_2^2} {4 m^3} + zA_4 \frac { (\Efpm \Eipm + P_{f3} (P_{f3} - \Delta_3)) } {4
           m} 
 -A_5 \frac { P_{f3} \left (\Delta_2^2 - 2mE^- \right) + ((\Efpm)^2- P_{f3}^2)
               \Delta_3 } {4 m^3} 
 \nonumber \\[1ex]
   && \qquad  - zA_6 \frac { \left ((\Efpm)^2 + P_{f3} (P_{f3} - \Delta_3) \right) 
        \Delta_2^2} {8 m^3} 
        + zA_8 \frac { \left ((\Efpm)^2 + 
          P_{f3} (P_{f3} - \Delta_3) \right)  \Delta_2^2} {4
        m^3}  
  \Bigg) \label{eq:Pi2G1_ns}\\[1ex]
\Pi_2(\Gamma_2) &=&  -K\, \Delta_1 \Delta_2\, \Bigg(
-A_1 \frac { P_{f3} } {8 m^3} + A_3 \frac { P_{f3}} {4 m^3} + A_5 \frac { P_{f3}} {4 m^3}  \nonumber \\[1ex]
   && \qquad\qquad\quad  + zA_6 \frac{\left ((\Efpm)^2 + 
          P_{f3} (P_{f3} - \Delta_3) \right) } {8 m^3} 
          - zA_8 \frac { \left ((\Efpm)^2 + 
          P_{f3} (P_{f3} - \Delta_3) \right) } {4 m^3}
  \Bigg) \label{eq:Pi2G2_ns}\\[1ex]
\Pi_2(\Gamma_3) &=& -K\,\Delta_1\,  \Bigg( 
 zA_4 \frac { P_{f3}} {4 m} +A_5 \frac { (\Efpm)^2 - P_{f3}^2} {4 m^3} 
  \Bigg) \label{eq:Pi2G3_ns}\\[3ex]
  \Pi_3(\Gamma_0) &=& i\, K\,  \Bigg(
- A_1 \frac {   (2 P_{f3} - \Delta_3) (\Efpm \Eipm - 
          P_{f3} (P_{f3}-\Delta_3))} {8
        m^3} + zA_2 \frac {   \Efpm \Eipm -
          P_{f3} (P_{f3}-\Delta_3) } {4 m} 
 \nonumber \\[1ex]
   && \qquad  - A_3 \frac {   \Delta_3 (\Efpm \Eipm - 
          P_{f3} (P_{f3}-\Delta_3))} {4 m^3}   + A_5 \frac {  
            P_{f3} \left (\Delta_1^2+\Delta_2^2-(E^-)^2 \right) + E^- \Efpm \Delta_3 } {4
        m^3}  
 \nonumber \\[1ex]
   && \qquad  
+ zA_6 \frac {   (2 P_{f3} - \Delta_3) \left (3 \Delta_3 P_{f3}^2 - 
          \Delta_3^2 P_{f3} - 2 E_fE^- P_{f3} - \Eimm \Efpm \Delta_3 \right) } {8 m^3}  \nonumber \\[1ex]
   && \qquad 
 + z^2A_7 \frac {   -3 \Delta_3
              P_{f3}^2 + \Delta_3^2 P_{f3} + 
          2 E_f E^- P_{f3} + \Eimm \Efpm \Delta_3  } {4
        m^3}   \nonumber \\[1ex]
   && \qquad + zA_8 \frac {   \Delta_3 \left (3
              \Delta_3 P_{f3}^2 - \Delta_3^2 P_{f3} -
          2 E_f E^- P_{f3} - \Eimm \Efpm \Delta_3 \right)
        } {4 m^3} 
  \Bigg) \label{eq:Pi3G0_ns}\\[3ex]
\Pi_3(\Gamma_1) &=&  K\,\Delta_2\,  \Bigg(
 A_1 \frac { P_{f3} (2 P_{f3} - 
          \Delta_3)} {8
           m^3} 
- zA_2 \frac { P_{f3} } {4 m} + A_3 \frac { P_{f3} \Delta_3} {4 m^3}  - A_5 \frac { (\Efpm)^2 + P_{f3} (P_{f3} - \Delta_3)} {4 m^3} 
 \nonumber \\[1ex]
   && \qquad   - zA_6 \frac { (2 P_{f3} - \Delta_3)
         \left ((\Efpm)^2 + P_{f3} (P_{f3} - \Delta_3) \right) } {8 m^3} 
+ z^2A_7 \frac { (\Efpm)^2 +  P_{f3} (P_{f3} - \Delta_3) } {4 m^3}  \nonumber \\[1ex]
   && \qquad - zA_8 \frac { \Delta_3 \left ((\Efpm)^2 + 
          P_{f3} (P_{f3} - \Delta_3) \right) } {4 m^3} 
  \Bigg) \label{eq:Pi3G1_ns}%\\[3ex]
\end{eqnarray}

\begin{eqnarray}
\Pi_3(\Gamma_2) &=&  -K\,\Delta_1\,  \Bigg(
 A_1 \frac { P_{f3} (2 P_{f3} - 
          \Delta_3)} {8
           m^3} 
- zA_2 \frac { P_{f3} } {4 m} + A_3 \frac { P_{f3} \Delta_3} {4 m^3}  - A_5 \frac { (\Efpm)^2 + P_{f3} (P_{f3} - \Delta_3)} {4 m^3} 
 \nonumber \\[1ex]
   && \qquad   - zA_6 \frac { (2 P_{f3} - \Delta_3)
         \left ((\Efpm)^2 + P_{f3} (P_{f3} - \Delta_3) \right) } {8 m^3} 
+ z^2A_7 \frac { (\Efpm)^2 +  P_{f3} (P_{f3} - \Delta_3) } {4 m^3}  \nonumber \\[1ex]
   && \qquad - zA_8 \frac { \Delta_3 \left ((\Efpm)^2 + 
          P_{f3} (P_{f3} - \Delta_3) \right) } {4 m^3} 
  \Bigg) \label{eq:Pi3G2_ns}\\[3ex]
\Pi_3(\Gamma_3) &=& 0 \label{eq:Pi3G3_ns}
\end{eqnarray}
\vspace*{3mm}
\begin{center}
\textit{Momentum transfer in only the longitudinal direction ($\Delta_T=0$, $\Delta_L\neq0$)}
\end{center}
\begin{eqnarray}
\label{eq:Pi0G0L}
\Pi_0(\Gamma_0) &=&  K \Bigg(A_1^L \frac{ E^+  (E_f^+  E_i^+ +P_{f3} 
   (\Delta_3 -P_{f3} ))}{8 m^3}+ z A_4^L \frac{ (E_f^+  \Delta_3 -E^-  P_{f3} )}{4 m} \Bigg)\\[3ex]
\Pi_1(\Gamma_2) &=& 
 K \Bigg(z A_4^L \frac{ (E_f^+  E_i^+ +P_{f3}  (P_{f3} -\Delta_3 ))}{4 m} \Bigg)\\[3ex]
\Pi_2(\Gamma_1) &=& - K \Bigg(z A_4^L \frac{ (E_f^+  E_i^+ +P_{f3}  (P_{f3} -\Delta_3 ))}{4 m}\Bigg) \\[3ex]
\label{eq:Pi3G0L}
\Pi_3(\Gamma_0) &=& i\, K \Bigg( A_1^L \frac{  (\Delta_3 -2 P_{f3} ) (E_f^+  E_i^+ +P_{f3}  (\Delta_3 -P_{f3} ))}{8 m^3}
+ z A_2^L \frac{ (E_f^+  E_i^+ +P_{f3}  (\Delta_3 -P_{f3} ))}{4 m} \Bigg)\\\nonumber
\end{eqnarray}
All other matrix elements vanish in the limit of zero transverse momentum transfer ($\Delta_T=0$).
We note that equations (\ref{eq:Pi0G0L})-(\ref{eq:Pi3G0L}) are the same as the four non-vanishing equations among (\ref{eq:Pi0G0_ns})-(\ref{eq:Pi3G3_ns}), upon the replacement $A_i\rightarrow A_i^L$ ($i=1,2,4$) and $A_i\rightarrow0$ ($i=3,5,6,7,8$).

%%%%%%%%%%%%%%%%%%%%%%%%%%%%%%%%%%%%%%%%%%%%%%%%%%%%%%%%%%
\subsection{Quasi-GPDs in terms of amplitudes}
\label{sec:GPD_amp}
%%%%%%%%%%%%%%%%%%%%%%%%%%%%%%%%%%%%%%%%%%%%%%%%%%%%%%%%%%
Similar to Ref.~\cite{Bhattacharya:2022aob}, we consider two definitions of quasi‑GPDs: the ``standard'' one (denoted by subscript 0), employing only the $\gamma_0$ Dirac insertion, and a Lorentz-invariant (LI subscript) one, which utilizes in addition the $\gamma_1$ and $\gamma_2$ Dirac structures.
Equations (41)-(42) and (46)-(47) of Ref.~\cite{Bhattacharya:2022aob} present Minkowski-spacetime expressions for the LI and asymmetric frame standard definitions, respectively.
Here, we write their Euclidean versions and we explicitly use the skewness parameter for quasi-distributions, i.e.\ for hadron moving in the spatial third direction, 
\begin{equation}
\xi=-\Delta_3/2P_3.    
\end{equation}
The quasi-skewness $\xi$ is equal to $\xi_{\rm LC}$ up to power corrections of $\mathcal{O}(m^2/P_3^2)$ \cite{Liu:2019urm, Bhattacharya:2019cme} and below, we will refer to it simply as skewness.

The standard definition of GPDs in the asymmetric frame reads
\begin{align}
H_0\!=\!A_1+\frac{2E^-}{E^+}A_3+\frac{1}{1-\xi}\frac{E^- m^2}{E^+ P_3}zA_4 + \left(-P_3(1+\xi)+\frac{E_f E_i - m^2}{(1-\xi)P_3}\right)zA_6 
+ \frac{2E^-}{E^+}\left(-P_3(1+\xi)+\frac{E_f E_i - m^2}{(1-\xi)P_3}\right)zA_8,
\label{eq:stan_H}
\end{align}
\begin{align}
E_0=-A_1-\frac{2E^-}{E^+}A_3-\frac{1}{1-\xi}\frac{2E_f m^2}{E^+ P_3}zA_4 + 2A_5 + \left(2P_3+\frac{2m^2}{(1-\xi)P_3}\right)zA_6 
+ \frac{2E^-}{E^+}\left(2P_3+\frac{2m^2}{(1-\xi)P_3}\right)zA_8.
\label{eq:stan_E}
\end{align}
We note that the coefficients of $A_1$ and $A_5$ remain skewness-independent, while the coefficient of $A_6$ becomes dependent on $\xi$.
Additionally, terms with $A_3$, $A_4$ and $A_8$ appear, with $\xi$-dependent coefficients.\footnote{Note that in Ref.~\cite{Bhattacharya:2022aob}, terms with $A_3$, $A_4$ and $A_8$ were formally included, but in the $\xi=0$ case, these amplitudes vanish by symmetry.}

The Lorentz-invariant definition has the form
\begin{align}
H_{\rm LI}=A_1-2\xi A_3,
\label{eq:LI_H}
\end{align}
\begin{align}
E_{\rm LI}=-A_1+2\xi A_3+2A_5+2P_3zA_6-4\xi P_3zA_8.
\label{eq:LI_E}
\end{align}
Here, the difference with respect to the $\xi=0$ case is, solely, the appearance of $A_3$ and $A_8$ terms with $\xi$-dependent coefficients.
With respect to the standard definition, the expressions are much simpler.
In this context, we mention that in the limit $z^2 \to 0$ ($P_3 \to \infty$), one finds full equivalence between the standard and LI definitions of the quasi GPDs.
We also note that only the LI definitions have definite skewness symmetry, being skewness-even (as combinations of $\xi$-even amplitudes and $\xi$-odd ones multiplied by $\xi$).
In the numerical part, we compare both definitions.
In principle, both should lead to the same light-cone GPDs upon matching, possibly at a different range of convergence towards the physical distributions.
In Ref.~\cite{Radyushkin:2023ref}, another parametrization of the vector matrix element was proposed, with a clearer separation of twist-2 and higher-twist contributions.
It implies that the standard definition contains contaminations by such higher-twist terms, possibly suggesting slower convergence and favoring the LI definition. However, the sign of these contributions is not known a priori, so potentially, the formally higher-twist amplitudes can also decrease the overall size of the contamination. Thus, the convergence to the light cone can only be resolved empirically in a non-perturbative calculation.
Indeed, in the zero skewness case it was found \cite{Cichy:2023dgk} that the $E_{\rm LI}$ GPD converges much faster than its standard counterpart, while the rate of convergence of both definitions of the $H$ GPD was similar.
Interestingly, also the statistical signal for $E_{\rm LI}$ was found to be better than for the standard definition, implying that the additional MEs of $\gamma_{1/2}$ insertions subtract correlated noise.

We now turn to the specific case where all the momentum transfer is in the longitudinal direction (i.e., \( \Delta^\mu = \Delta_L^\mu \)); see Appendix~\ref{sec:app_longitudinaldecomp} for details. For this particular setup, we focus on the LI definition—namely, the definition of the GPD that reproduces the same combination of amplitudes as in the light-cone case. (The non-LI definition differs from the LI form by the amplitude \(A^L_4\).)
In this case, mapping the decomposition in Eq.~(\ref{eq:MEparametrization2}) to the parametrization of the GPDs, we find that the amplitude \( A^L_1 \) is related to both the GPDs \( H \) and \( E \) via:
\begin{align}
    H^L_{\rm LI} \equiv \big ( H + \dfrac{1}{2}\alpha \xi (H + E) \big )_{\rm LI} = A^L_1,
    \label{e:GPD_forward}
\end{align}
where $\alpha = \tfrac{-(z \cdot \Delta)( z \cdot P)}{z^2 P^2 - (z \cdot P)^2}$. 
Several comments are in order.
First, the relation in Eq.~(\ref{e:GPD_forward}) highlights that, in a kinematic setup where the momentum transfer is purely longitudinal, the structures associated with \( H \) and \( E \) mix. As a result, it becomes difficult to disentangle the two GPDs.
Second, in the light-cone limit \( z^2 = 0 \), one finds 
\(\alpha\big|_{z^2=0} = -2\xi\), from which the known relation 
\(H^L \propto H - \frac{\xi^2}{1 - \xi^2} \, E\)~\cite{Diehl:2002he} is recovered. 
(Moreover, one can check that with \(\alpha\big|_{z^2=0} = -2\xi\), Eqs.~\eqref{eq:LI_H}, \eqref{eq:LI_E}, and \eqref{e:GPD_forward} are consistent.)
This result independently confirms that, in the \(\Delta_T = 0\) case, 
the two GPDs cannot be disentangled. 
Third, if one argues that $\alpha \propto \xi$ is numerically small for all relevant values of $z^2$, then the admixture term $\tfrac{1}{2}\alpha \xi (H+E)$ is suppressed, so that the dominant contribution to $A^L_1$ comes from the GPD $H$ alone.
In other words, in this regime, the extraction of \( H \) from the matrix element is robust and practically uncontaminated by \( E \).
As a consistency check, we note that in the forward limit, both \( \alpha \) and \( \xi \) vanish, leading to the clean identification: $H = A^L_1$.
In this limit, the amplitude \( A^L_1 \) also reduces appropriately to its forward counterpart, \( A^L_1 \rightarrow A_1 \) (see Appendix~\ref{sec:app_longitudinaldecomp}), confirming the consistency of the decomposition with the known structure of the correlator in the forward limit.

\subsection{Bare matrix elements, renormalization and matching to physical GPDs}
\label{sec:lattice}
In this subsection, we discuss the lattice extraction of bare matrix elements, their renormalization and the steps needed to go to final momentum-space GPDs -- reconstruction of the $x$-dependence and matching.

The lattice data used in this work was obtained from an ensemble of gauge field configurations generated using lattice QCD simulations with $ N_f = 2 + 1 + 1 $ twisted-mass fermions including clover improvement and an Iwasaki-improved gauge action \cite{Alexandrou:2018egz}. The quark masses were tuned to yield a pion mass of approximately 260\,MeV. The lattice has a spacing of $a \approx 0.093$~fm and a volume of $L^3 Ta^4$, with $L = 32$ denoting the spatial extent and $T = 64$ the temporal extent.

The lattice calculation involves determination of nucleon MEs of the form of Eq.~(\ref{eq:ME}), containing nonlocal operator insertions with straight Wilson lines along the $z$-direction, connecting quark fields separated by a distance $z$. The sink momentum $\vec{P}_f$ is aligned along the $z$-axis and the momentum transfer contains longitudinal or both longitudinal and transverse components. 
We always consider the flavor non-singlet $u-d$ combination.
The computation employs randomly distributed source positions on each gauge configuration, with each source serving as the origin for a Dirac operator inversion to enhance statistical precision. The matrix elements are evaluated on $\{96,\,95,\,120\}$ gauge configurations, with $\{8,\,8,\,32\}$ source positions per configuration, corresponding to final-state momenta $P_{f3} = \{\pm1,\,\pm2,\,\pm3\}(2\pi/L)$ in lattice units, or equivalently $|P_{f3}| = \{0.42,\,0.83,\,1.25\}$ GeV in physical units. Of particular relevance to this study is the largest momentum, $|P_{f3}| = 1.25$ GeV, as our primary interest lies in accessing quasi-GPDs at the highest attainable nucleon boost. 

The matrix element corresponding to the $\gamma_\mu$ insertion and the $\Gamma_\kappa$ parity projector is extracted from the ratio of three-point and two-point correlation functions,
\begin{align}
\label{eq:ratio}
R_{\mu}(\Gamma_{\kappa},P_f,P_i; t_s, \tau;z) = \frac{C_{\mu}^{3\text{pt}}(\Gamma_{\kappa}, P_f, P_i; t_s, \tau;z)}{C^{2\text{pt}}(\Gamma_0, P_f; t_s)}
\sqrt{\frac{C^{2\text{pt}}(\Gamma_0, P_i; t_s - \tau) C^{2\text{pt}}(\Gamma_0, P_f; \tau) C^{2\text{pt}}(\Gamma_0, P_f; t_s)}{C^{2\text{pt}}(\Gamma_0, P_f; t_s - \tau) C^{2\text{pt}}(\Gamma_0, P_i; \tau) C^{2\text{pt}}(\Gamma_0, P_i; t_s)}}.
\end{align}
Here, $t_s$ denotes the source-sink separation ($t_s=10a\approx0.93$ fm;  the study of excited states contamination is beyond the scope of this work, but will be addressed in the future), and $\tau$ is the operator insertion time. We extract the ground-state matrix elements $\Pi_{\mu}(\Gamma_{\kappa})$ by identifying plateau regions in $\tau$ where the ratio $R_{\mu}$ becomes insertion-time-independent. 

In lattice QCD, smearing of the quark fields is essential for improving the overlap with the nucleon ground state \( N_0(\vec{P}) \), as it enhances the ground-state projection \( |\langle N_0(\vec{P}) | \chi | 0 \rangle|^2 \) by shaping the interpolating operator \( \chi \) -- which creates a nucleon from the vacuum -- to better resemble the spatial profile of the ground-state nucleon. This improves the suppression of excited-state contributions in correlation functions. 
However, conventional smearing methods -- such as Gaussian smearing, which is also employed in this work -- are optimized for nucleons at rest $N_0(\vec{p} = 0)$ and thus suppress the overlap with states of large momentum. This poses a significant challenge for accessing high-momentum nucleon states. To address this limitation, momentum smearing techniques have been developed -- see Ref.~\cite{Bali:2016lva} for a detailed formulation. In this approach, a momentum vector \( \vec{k} \) is incorporated directly into the quark smearing function, modifying its momentum profile to favor quark fields carrying momentum components compatible with the target nucleon state. This enhances the overlap with boosted nucleons, improving the signal-to-noise ratio.
The momentum smearing was shown to be very efficient in the context of non-local operators, both for PDFs~\cite{Alexandrou:2016jqi} and GPDs~\cite{Alexandrou:2020zbe}.
Thus, we adopt it also in the present work. 

To improve the signal, we average over two-point correlation functions that are equivalent in the lattice computations, namely those related by independent sign flips of the components of $\vec{P}_f$ and $\vec{P}_i$, and permutation of the transverse components of $\vec{P}_i$, taking into account our momentum smearing setup.
The latter always involves eight cases for $\vec{P}_i$, corresponding to momentum smearing vectors $\vec{k}=(2,0,\pm P_{f3}),\,(-2,0,\pm P_{f3}),\,(0,2,\pm P_{f3}),\,(0,-2,\pm P_{f3})$.
The sink two-point functions are obtained in two momentum smearing setups, with $\vec{k}=(0,0,\pm P_{f3})$, and hence, are averaged over two possibilities.
Meanwhile, the average over source two-point correlators involves eight cases when the transverse momentum transfer is in one direction, 16 cases for $\Delta_T=(\Delta,\Delta)$ and two groups of 16 cases for $\Delta_T=(\Delta_1,\Delta_2\neq\Delta_1)$.
The appearance of two inequivalent groups of source correlators when $\Delta_1\neq\Delta_2$ is an important and subtle consideration that is related to the fact that, e.g.\ for $\Delta_T=(1,2)$, the influence of momentum smearing for $\vec{k}=(2,0,P_{f3})$ and $\vec{k}=(0,2,P_{f3})$ affects differently the amplitudes of correlation functions in their spectral decompositions -- consequently, the resulting two-point correlators are inequivalent due to an overall factor.
Note also that we do not average $C_{\mu}^{3\text{pt}}(\Gamma_{\kappa}, P_f, P_i; t_s, \tau;z)$ over momentum smearing setups, because in the asymmetric frame, sign flips of components and permutations do not lead to equivalent MEs.

Now, we discuss the details of our kinematic setup. Table~\ref{tab:kinematic}(a) presents the values of $t$ and $\xi$ and the associated representative kinematic configurations for all our sink momenta with $P_{f3} = \{\pm1,\,\pm2,\,\pm3\}(2\pi/L)$. Only a single representative $\vec{\Delta}$ is listed for each class of $(\Delta_1,\Delta_2)$ related by sign changes and permutations of these components, as such transformations leave $t$ invariant. 
We also show the number of measurements performed for each kinematic case, defined as the product of the number of gauge configurations, source positions, and equivalent momentum-smearing setups.
Table~\ref{tab:kinematic}(a) further includes the corresponding initial-state momenta $P_{i3} = \{\pm1,\,\pm2,\,\pm4\,,\pm3\}(2\pi/L)$, illustrating that the GPDs symmetry under $\xi \leftrightarrow -\xi$, which corresponds to the replacement $P_{i3} \leftrightarrow P_{f3}$, can only be strictly verified when $\Delta_T = 0$ -- see the first rows of $\xi=-1/5$ and $\xi=1/5$. For $\Delta_T \neq 0$, the values of $t$ differ, see e.g.\ $(\Delta_1,\Delta_2)=(1,1)$ of $\xi=-1/5$ and $\xi=1/5$, thereby obstructing a direct verification of the symmetry.

\renewcommand{\arraystretch}{1.3}
\begin{table}[t!]
\centering
\noindent
\begin{minipage}[t]{0.58\textwidth}
\resizebox{\linewidth}{!}{
\scalebox{0.15}{
\begin{tabular}{|c|c|c|c|c|c|}
\hline
$\xi$ & $t$ (GeV$^2$) & $(\Delta_1,\Delta_2,\Delta_3)$ & $N_{\rm meas}$ & $P_{i3}$ & $P_{f3}$ \\
\hline
\multirow{6}{*}{$-1/2$} 
      & $-0.445$ & $(0,0,\pm2)$ & 30720 & & \\
      \cline{2-4}
      & $-0.686$ & $(1,0,\pm2)$ & 30720 & & \\
      & $-0.914$ & $(1,1,\pm2)$ & 61440 & & \\
      & $-1.337$ & $(2,0,\pm2)$ & 30720 & $\pm1$ & $\pm3$ \\
      & $\bold{-1.534}$ & $\bold{(1,2,\pm2)}$ & \textbf{122880} & & \\
      & $-2.253$ & $(3,0,\pm2)$ & 30720 & & \\
\hline
\multirow{6}{*}{$-1/5$}
      & $-0.089$ & $(0,0,\pm1)$ & 30720 & & \\
      \cline{2-4}
      & $\bold{-0.295}$ & $\bold{(1,0,\pm1)}$ & \textbf{30720} & & \\
      & $-0.493$ & $(1,1,\pm1)$ & 61440 & & \\
      & $-0.865$ & $(2,0,\pm1)$ & 30720 & $\pm2$ & $\pm3$ \\
      & $-1.041$ & $(1,2,\pm1)$ & 122880 & & \\
      & $-1.696$ & $(3,0,\pm1)$ & 30720 & & \\
\hline
\multirow{6}{*}{$1/7$}
      & $-0.061$ & $(0,0,\mp1)$ & 30720 & & \\
      \cline{2-4}
      & $-0.203$ & $(1,0,\mp1)$ & 30720 & & \\
      & $-0.343$ & $(1,1,\mp1)$ & 61440 & & \\
      & $-0.613$ & $(2,0,\mp1)$ & 30720 & $\pm4$ & $\pm3$ \\
      & $-0.745$ & $(1,2,\mp1)$ & 122880 & & \\
      & $-1.248$ & $(3,0,\mp1)$ & 30720 & & \\
\hline
\multirow{3}{*}{$1/5$}
      & $-0.089$ & $(0,0,\mp1)$ & 6080 & & \\
      \cline{2-4}
      & $-0.366$ & $(1,1,\mp1)$ & 12160 & $\pm3$ & $\pm2$ \\
      & $-0.628$ & $(2,0,\mp1)$ & 6080 & & \\
\hline
$1/2$ & $-0.679$ & $(1,1,\mp2)$ & 12288 & $\pm3$ & $\pm1$ \\
\hline
\end{tabular}
}
}
\vspace*{2mm}
%\captionsetup{justification=justified}
\caption{\textbf{(a)} Kinematic setups considered in this work. We list the momentum transfer squared $t$, the skewness $\xi$, the momentum-transfer vector $\vec{\Delta} = (\Delta_1, \Delta_2, \Delta_3)$, the number of measurements $N_{\rm meas}$, and the initial and final nucleon momenta $P_{i3}$ and $P_{f3}$. The cases with vanishing transverse momentum transfer ($\Delta_T = 0$) are explicitly demarcated in the first row for each of the first four values of $\xi$.
Only a single representative $\vec{\Delta}$ is shown for each equivalence class under sign flips and permutations of $(\Delta_1,\Delta_2)$, which leave $t$ invariant. The cases written in bold font, $\vec{\Delta} = (1,0,\pm1)$ and  $\vec{\Delta} = (1,2,\pm2)$, are further elaborated in Tables~I (b) and (c), respectively. Note that all momentum or momentum transfer components are given in units of $2\pi/L$.}
\label{tab:kinematic}
\end{minipage}%
\hfill
\raisebox{4.5cm}[0pt][0pt]{% %%%%%%%%%%% RAISE AND LOWER RIGHT-SIDE TABLES (x cm)
  \begin{minipage}[t]{0.39\textwidth}

\resizebox{0.92\linewidth}{!}{
\scalebox{0.1}{
\begin{tabular}{|c|c|c|}
\hline 
$P_{f3}$  & $\vec{P_i}$ & $\vec{k}$   \\
\hline
        &    $ (1, 0, 2)$  &  $ (2, 0, 3)$  \\  
\cline{2-2} \cline{3-3}
        &    $ (-1, 0, 2)$   &   $ (-2, 0, 3)$ \\  
\cline{2-2} \cline{3-3}
   \smash{\raisebox{0.9ex}{$3$}}     &  $(0, 1, 2) $    &  $ (0, 2, 3)$  \\  
\cline{2-2} \cline{3-3}
        &   $(0, -1, 2) $   &   $ (0, -2, 3)$  \\  
\hline
        &   $(1, 0, -2) $   &   $ (2, 0, -3)$  \\  
\cline{2-2} \cline{3-3}
    \smash{\raisebox{-1.5ex}{$-3$}}     &    $ (-1, 0, -2)$   &$ (-2, 0, -3)$     \\  
\cline{2-2} \cline{3-3}
        &    $ (0, 1, -2)$   &  $ (0, 2, -3)$  \\  
\cline{2-2} \cline{3-3}
        &  $(0, -1, -2) $    & $ (0, -2, -3)$   \\  
\hline
\end{tabular}
}
}
\vspace*{-1mm}
\caption*{\textbf{(b)} Momentum-smearing vectors $\vec{k}$ employed for different permutations/signs of $\vec{P}_i$ and a given $P_{f3}$, for an example where $\Delta_T$ has one nonvanishing component.}

\vspace{2mm}

\resizebox{0.92\linewidth}{!}{

\begin{tabular}{|c|c|c|c|}
\hline 
$P_{f3}$  & $\vec{P_i}$ & $\vec{k}^{(A)}$ & $\vec{k}^{(B)}$ \\
\hline
        &    $ (1, 2, 1)$  &  $ (2, 0, 3)$ & $ (0, 2, 3)$\\  
\cline{2-2} \cline{3-3} \cline{4-4}
        &    $ (-1, 2, 1)$   &   $ (-2, 0, 3)$&$ (0, 2, 3)$ \\  
\cline{2-2} \cline{3-3} \cline{4-4}
   \smash{\raisebox{0.9ex}{$3$}}     &  $(1, -2, 1) $    &  $ (2, 0, 3)$& $ (0, -2, 3)$ \\  
\cline{2-2} \cline{3-3} \cline{4-4}
        &   $(-1, -2, 1) $   &   $ (-2, 0, 3)$&$ (0, -2, 3)$  \\  
\hline
        &   $(1, 2, -1) $   &   $ (2, 0, -3)$& $ (0, 2, -3)$ \\  
\cline{2-2} \cline{3-3} \cline{4-4}
    \smash{\raisebox{-1.5ex}{$-3$}}     &    $ (-1, 2, -1)$   &$ (-2, 0, -3)$&  $ (0, 2, -3)$   \\  
\cline{2-2} \cline{3-3} \cline{4-4}
        &    $ (1, -2, -1)$   &  $ (2, 0, -3)$& $ (0, -2, -3)$ \\  
\cline{2-2} \cline{3-3} \cline{4-4}
        &  $ (-1, -2, -1)$    & $ (-2, 0, -3)$&  $ (0, -2, -3)$ \\  
\hline
        &    $ (2, 1, 1)$  & $ (0, 2, 3)$ &  $ (2, 0, 3)$\\  
\cline{2-2} \cline{3-3} \cline{4-4}
        &    $ (-2, 1, 1)$   & $ (0, 2, 3)$&   $ (-2, 0, 3)$\\  
\cline{2-2} \cline{3-3} \cline{4-4}
   \smash{\raisebox{0.9ex}{$3$}}     &  $(2, -1, 1) $    &  $ (0, -2, 3)$&  $ (2, 0, 3)$\\  
\cline{2-2} \cline{3-3} \cline{4-4}
        &   $(-2, -1, 1) $   &  $ (0, -2, 3)$&   $ (-2, 0, 3)$\\  
\hline
        &   $(2, 1, -1) $   & $ (0, 2, -3)$ &   $ (2, 0, -3)$\\  
\cline{2-2} \cline{3-3} \cline{4-4}
    \smash{\raisebox{-1.5ex}{$-3$}}     &    $ (-2, 1, -1)$   &  $ (0, 2, -3)$  &$ (-2, 0, -3)$ \\  
\cline{2-2} \cline{3-3} \cline{4-4}
        &    $ (2, -1, -1)$   & $ (0, -2, -3)$ &  $ (2, 0, -3)$\\  
\cline{2-2} \cline{3-3} \cline{4-4}
        &  $ (-2, -1, -1)$    &  $ (0, -2, -3)$ & $ (-2, 0, -3)$\\ 
\hline
\end{tabular}
}
\vspace*{-1mm}
\caption*{\textbf{(c)} Momentum-smearing vectors $\vec{k}$ employed for different permutations/signs of $\vec{P}_i$ and a given $P_{f3}$, for an example with $\Delta_1\neq\Delta_2$. 
The upper index of $\vec{k}$ indicates separation into two inequivalent groups of source two-point functions, e.g., the setup with $\vec{P}_i=(1,2,1)$ and $\vec{k}^{(A)}=(2,0,3)$ is equivalent to $\vec{P}_i=(2,1,1)$ and $\vec{k}^{(A)}=(0,2,3)$, but not to $\vec{P}_i=(1,2,1)$ and $\vec{k}^{(B)}=(0,2,3)$.}
\end{minipage}
}
\end{table}

Tables~\ref{tab:kinematic}(b) and~(c) provide representative examples of the bold-highlighted kinematic configurations in Table~\ref{tab:kinematic}(a) -- $\vec{\Delta}=(1,0,\pm1)(2\pi/L)$ and $\vec{\Delta}=(1,2,\pm2)(2\pi/L)$.
As shown in these two tables, the smearing vector \( \vec{k} = (2, 0, P_{f3})(2\pi/L) \), up to permutations and sign flips of its components, was used. Although, in principle, it is possible to fully optimize \( \vec{k} \) individually for each value of $t$, the associated computational cost would be prohibitive. Therefore, fixed values of \( \vec{k} \) were used across all momentum-transfer configurations.
In Table~\ref{tab:kinematic}(b), each case of $\vec{P}_i$ is taken from a momentum-smearing setup along the same transverse direction and according to the sign of the nonzero $\vec{k}$-component. 
In turn, Table~\ref{tab:kinematic}(c) highlights the aforementioned subtlety of the setup with different magnitudes of $\Delta_1$ and $\Delta_2$.
Namely, for a given $\vec{P}_i$, there are two momentum-smearing vectors $\vec{k}$ that can be employed in the analysis.
We denote these by a superscript $(A)$ or $(B)$.
Since momentum smearing affects the amplitudes of correlation functions in their spectral decompositions, results for a given $\vec{P}_i$ are inequivalent when using 
$\vec{k}^{(A)}$ and $\vec{k}^{(B)}$.
This aspect needs to be taken into account when averaging over source two-point correlators, i.e., sixteen possibilities for $\vec{P}_i$ are equivalent when the corresponding $\vec{k}^{(A)}$ is used and the same sixteen cases of $\vec{P}_i$ give another group of equivalent correlators with the corresponding $\vec{k}^{(B)}$.
 
This study also explored additional kinematic configurations that are not shown in Table~\ref{tab:kinematic}(a). This is because the signal of the correlators decays exponentially with increasing nucleon momentum, while the noise remains approximately independent of the momentum.
Consequently, the signal-to-noise ratio deteriorates exponentially in cases with large $|\vec{P}_i|$. The expected poor statistical quality was observed for the case with \( |P_{f3}| = 3(2\pi/L) \) and skewness \( \xi = 1/4 \), corresponding to \( |P_{i3}| = 5(2\pi/L) \), as well as for momentum transfers of the form \( \vec{\Delta} = (1,3,1)(2\pi/L) \) and \( (\Delta_1,\Delta_2) = (4,0)(2\pi/L) \), all of which involve large values of \( |\vec{P}_i| \). A substantially larger number of measurements would be necessary to obtain statistically reliable signals in such high-momentum states.

Before proceeding to extraction of light-cone GPDs, quasi-GPDs obtained in coordinate space need to be subjected to three further stages.
Firstly, bare MEs need to be renormalized.
Secondly, coordinate-space renormalized quasi-GPDs have to be translated into momentum space.
Finally, momentum-space quasi-GPDs can be matched into their light-cone counterparts, which proceeds by subtracting perturbatively calculated differences in their ultraviolet regimes, profiting from their identical infrared (non-perturbative) structure.
At this stage, conversion to the $\MSb$ scheme also takes place, together with evolution to the standard scale of 2 GeV.

In this work, we follow the renormalization procedure established in our earlier papers, see Refs.~\cite{Alexandrou:2017huk, Alexandrou:2019lfo, Alexandrou:2020zbe}, using a non-local variant of the regularization-independent momentum subtraction (RI/MOM) scheme \cite{Martinelli:1994ty}.
The renormalization conditions are imposed on the amputated vertex functions of the non-local operator, separately at each value of $z$, and on the quark propagator.
The renormalization scale, $\mu^R$, is chosen to minimize discretization effects (hypercubic artifacts) and the renormalization functions are chirally extrapolated.
Having tested several choices for $\mu^R$ and having established small effects of them, we take a single scale of around 4.1 GeV as our choice. This scale, together with the third component of the 4-momentum appearing in the renormalization conditions, $p_3^R\approx2.2$ GeV, is also input to the matching procedure.
We note that in this paper, we concentrate on the extraction of lattice data for nonzero skewness from asymmetric frames of reference and we do not adopt latest developments in the renormalization of quasi-distributions, such as the hybrid scheme~\cite{Ji:2020brr}. Improvements in this aspect are left for further work.

The translation of renormalized GPDs from coordinate to momentum space is a Fourier transform over Wilson line lengths.
However, formally, it is an integral from minus to plus infinity, while the lattice data are necessarily discrete and truncated to a finite length.
This is a manifestation of the well-known inverse problem in the context of reconstruction of $x$-dependent partonic distributions~\cite{Karpie:2019eiq}.
Similarly to our zero-skewness work in asymmetric frames, we employ the Backus-Gilbert (BG) approach~\cite{BackusGilbert} to reconstruct momentum-space distributions.
The BG method is based on a model-independent criterion to select a unique distribution from the formally infinitely many distributions that are consistent with discrete and truncated lattice data.
The criterion is to minimize the variance of the solution with respect to the statistical variation of the lattice data.
We note that although it is a formal solution to the inverse problem (i.e., only one distribution satisfies the criterion), it does not circumvent the need for better lattice data.
That is, although the inverse problem will always be present in any lattice calculation, its numerical impact may be significantly alleviated with more continuous data spanning larger intervals of $z$.

The final ingredient of our procedure is the matching from quasi distributions, i.e., the BG-reconstructed $x$-dependent quasi-GPDs, to physical distributions defined on the light front (in Minkowski metric).
Again, we proceed similarly to Ref.~\cite{Bhattacharya:2022aob}, employing the matching derived in Ref.~\cite{Liu:2019urm}.
In this approach, the RI-renormalized quasi-GPD is perturbatively factorized into the light-front GPD and also scheme-converted to the $\MSb$ scheme and evolved to a scale conventionally chosen to be 2 GeV. 

With respect to zero skewness, the matching expressions are more complicated at $\xi\neq0$, due to contributions not only from the standard DGLAP region, but also from the ERBL region specific to $\xi\neq0$ GPDs. We refer to Ref.~\cite{Liu:2019urm} for the actual equations, involving also RI/MOM counterterms in the $\slashed{p}$ variant employed in our renormalization.
We also mention that upon improving renormalization by adopting the hybrid scheme, matching equations also change and we plan to adopt this strategy in our future work.

%%%%%%%%%%%%%%%%%%%%%%%%%%%%%%%%%%%%%%%%%%%%%%%%%%%%%%%%%%%%%%%%%%

\section{Numerical results}
\subsection{Matrix elements and amplitudes}
In this section, we present the numerical results for the proton matrix elements $\Pi_i(\Gamma_j)$ and the corresponding amplitudes extracted from the decomposition outlined in Eqs.~(\ref{eq:Pi0G0_ns}-\ref{eq:Pi3G3_ns}).

\begin{figure}[h!]
\centering
\includegraphics[scale=0.5]{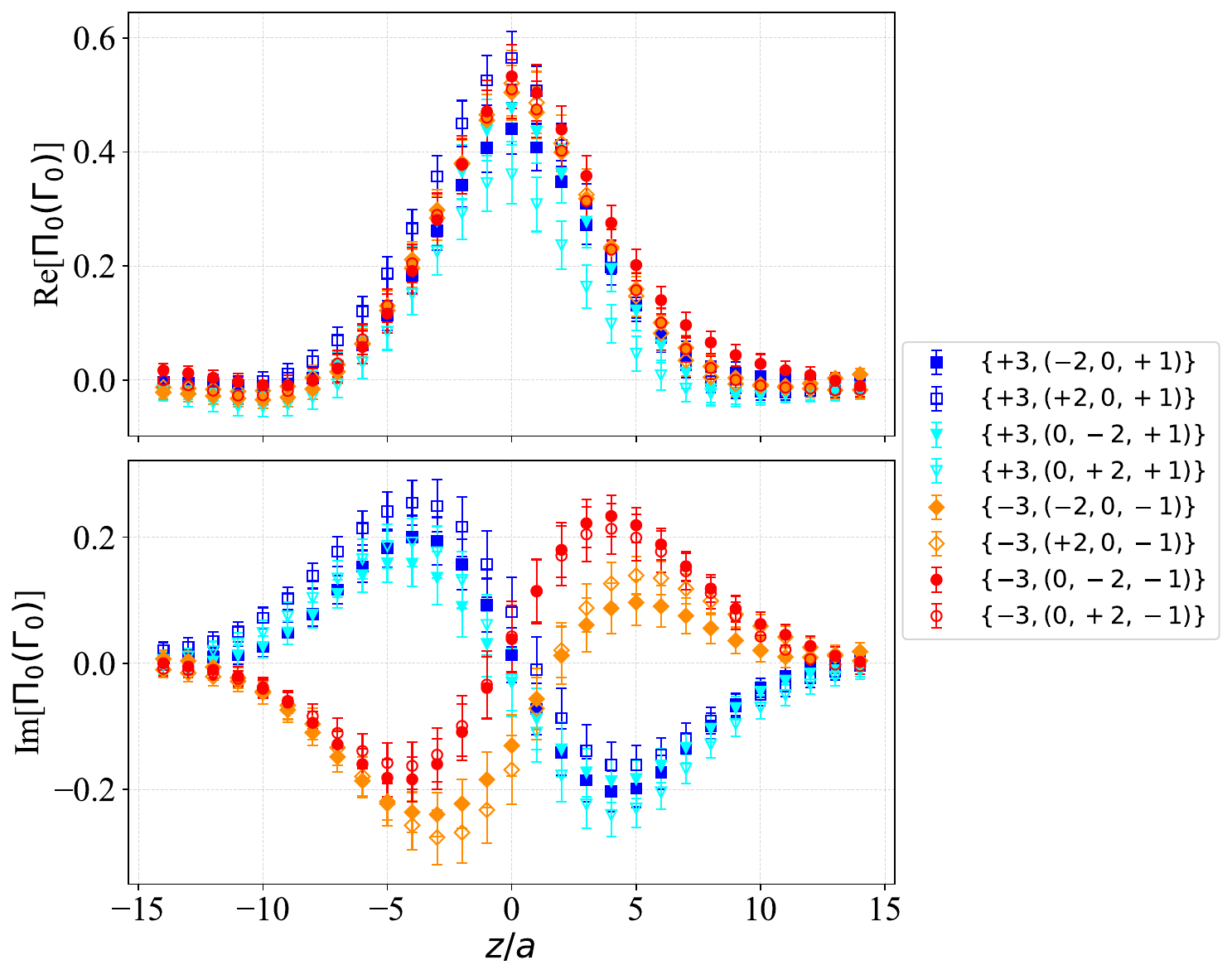}
\caption{Bare matrix elements $\Pi_0(\Gamma_0)$ at momentum combinations $\{P_{f3}, \Delta_1, \Delta_2, \Delta_3\}$ in units of $2\pi/L$. Upper and lower panels correspond to the real and imaginary parts, respectively.}
\label{fig:ME_g0_proj13}
\end{figure}

\begin{figure}[h!]
\centering
\includegraphics[scale=0.5]{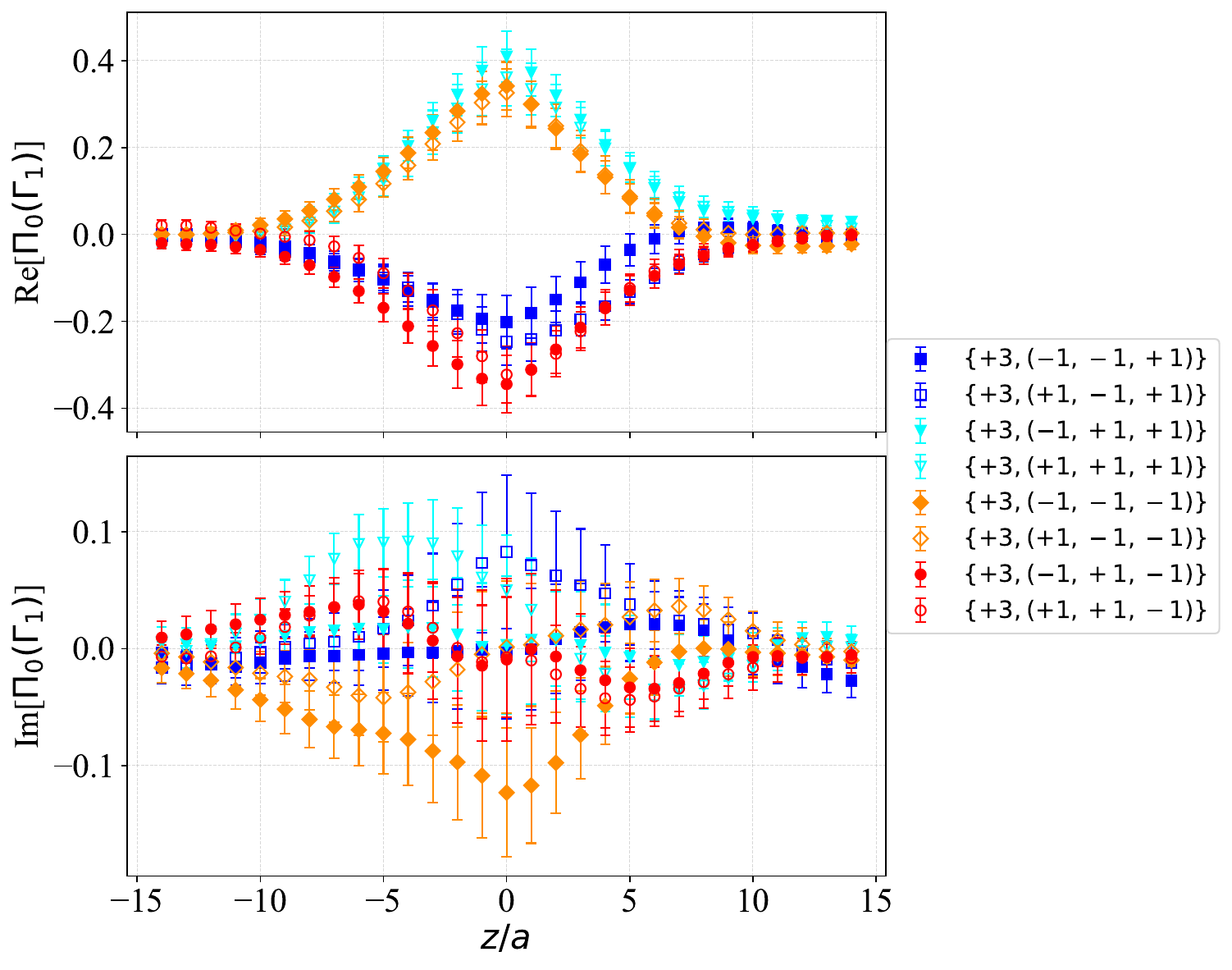}
\caption{Bare matrix elements $\Pi_0(\Gamma_1)$ at momentum combinations $\{P_{f3}, \Delta_1, \Delta_2, \Delta_3\}$ in units of $2\pi/L$. Upper and lower panels correspond to the real and imaginary parts, respectively.}
\label{fig:ME_g0_proj4}
\end{figure}

We begin with the results of bare matrix elements $\Pi_{\mu}(\Gamma_{\nu})$. Representative ones for $\Pi_0(\Gamma_0)$ and $\Pi_0(\Gamma_1)$ are shown in Fig.~\ref{fig:ME_g0_proj13} and Fig.~\ref{fig:ME_g0_proj4}, respectively. While the bare matrix elements may superficially exhibit a symmetry with respect to $\pm z$ and $\pm P_3$, Eq.~(\ref{eq:symmetries_A}) establishes symmetry properties solely for the amplitudes.

%%%%%%%%%%%%%%%%%%%%%%%%%%%%%%%%%%%%%%%%%%%%%%%%%%%%%%%%%%%%%%%%%%
%%%%%%%%%%%%%%%%%%%%%%%%%%%%%%%%%%%%%%%%%%%%%%%%%%%%%%%%%%%%%%%%%%
%%%%%%%%%%%%%%%%%%%%%%%%%%%%%%%%%%%%%%%%%%%%%%%%%%%%%%%%%%%%%%%%%%
\begin{figure}[h!]
\centering
\includegraphics[scale=0.6]{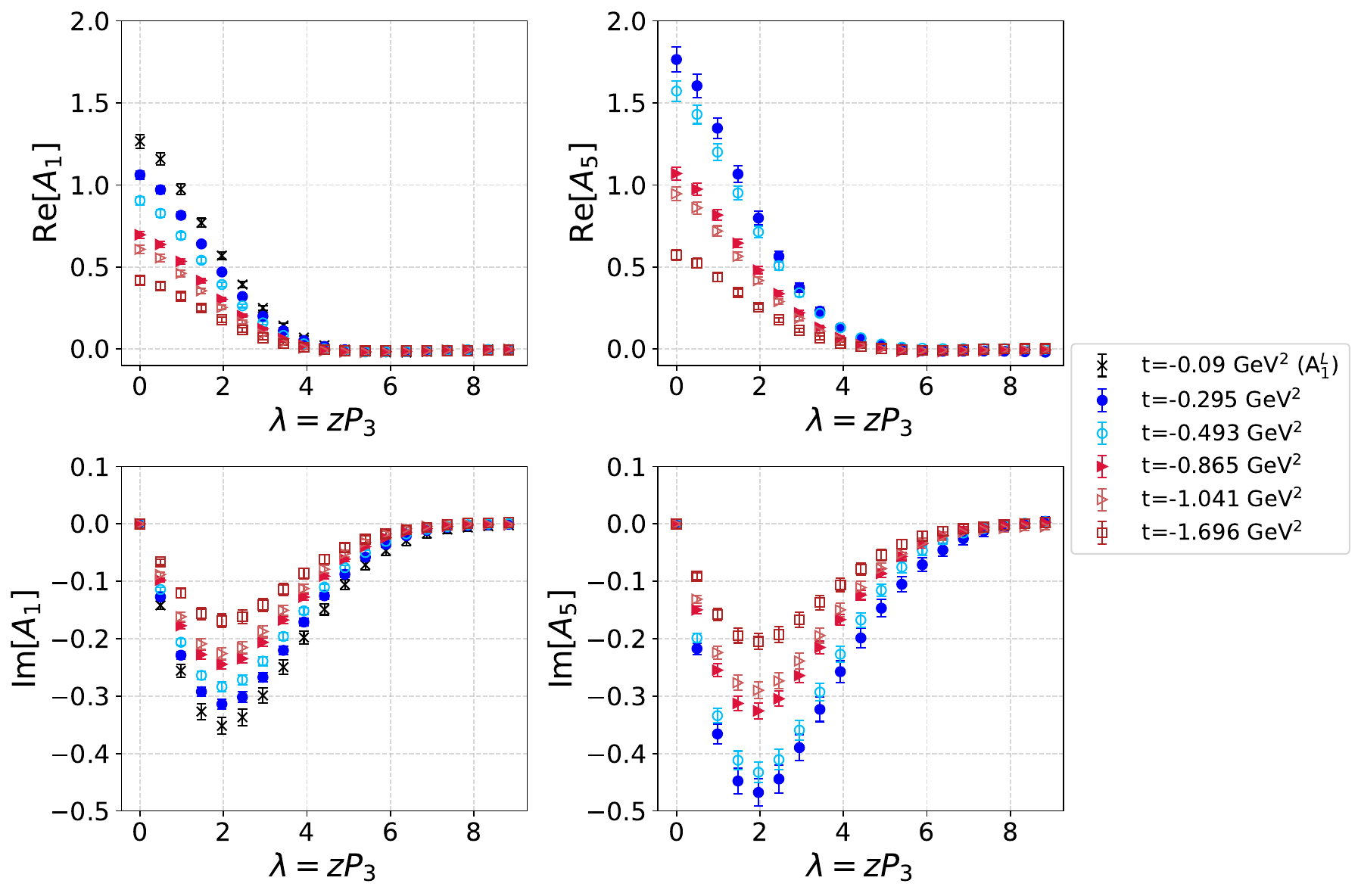}
\caption{Momentum-transfer dependence of amplitudes $A_1$ and $A_5$ at fixed skewness $\xi = -1/5$. $A_1^L$ from $\Delta_T=0$ case is included and noted in the legend (numerically, $A_1^L\approx A_1$, see the main text for details). The horizontal axis is $\lambda=zP_3$. Upper and lower panels correspond to the real and imaginary parts, respectively.}
\label{fig:amp_15_t_dep}
\end{figure}
%%%%%%%%%%%%%%%%%%%%%%%%%%%%%%%%%%%%%%%%%%%%%%%%%%%%%%%%%%%%%%%%%%
%%%%%%%%%%%%%%%%%%%%%%%%%%%%%%%%%%%%%%%%%%%%%%%%%%%%%%%%%%%%%%%%%%
\begin{figure}
\centering
\includegraphics[scale=0.6]{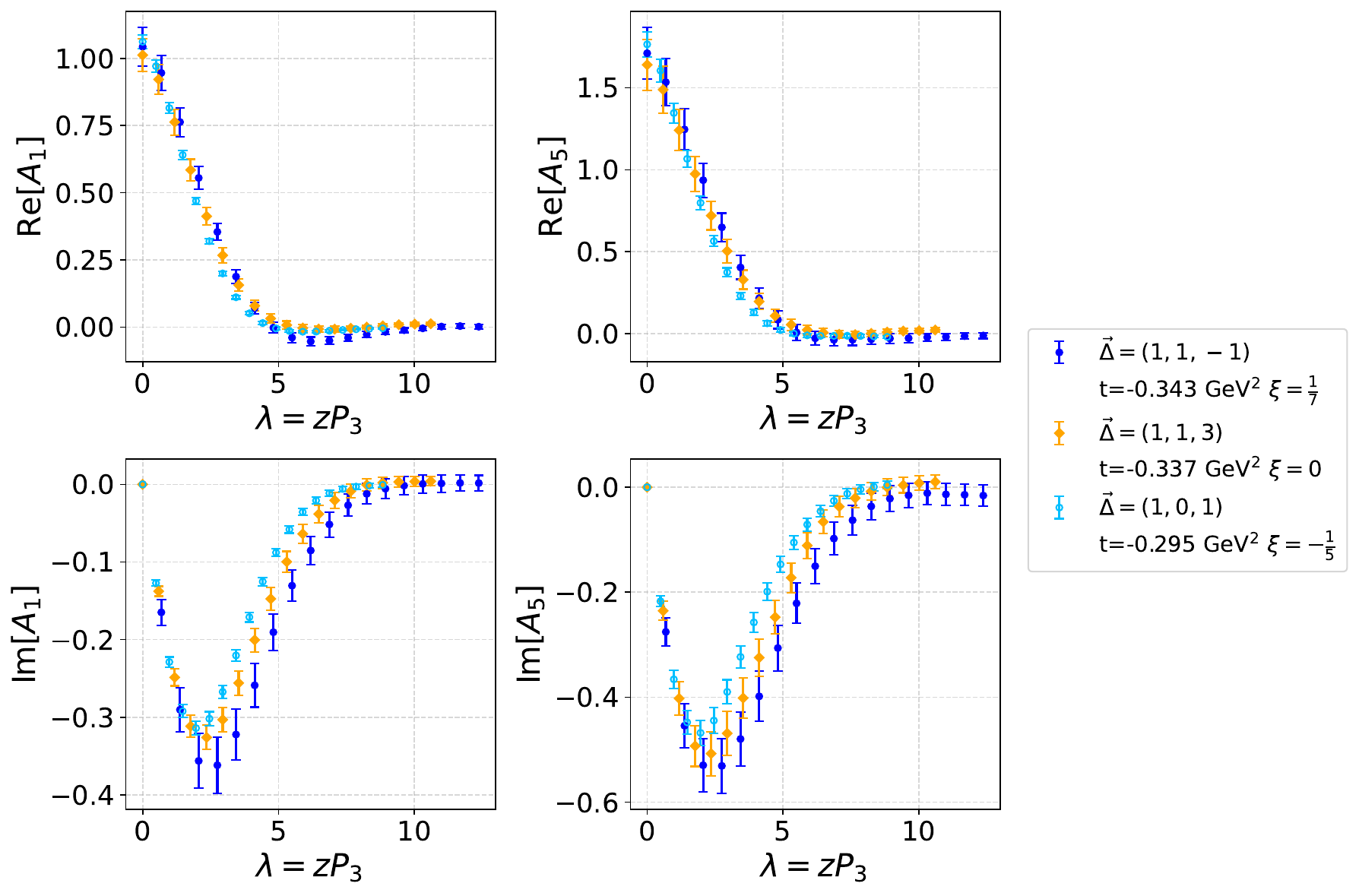}
\includegraphics[scale=0.6]{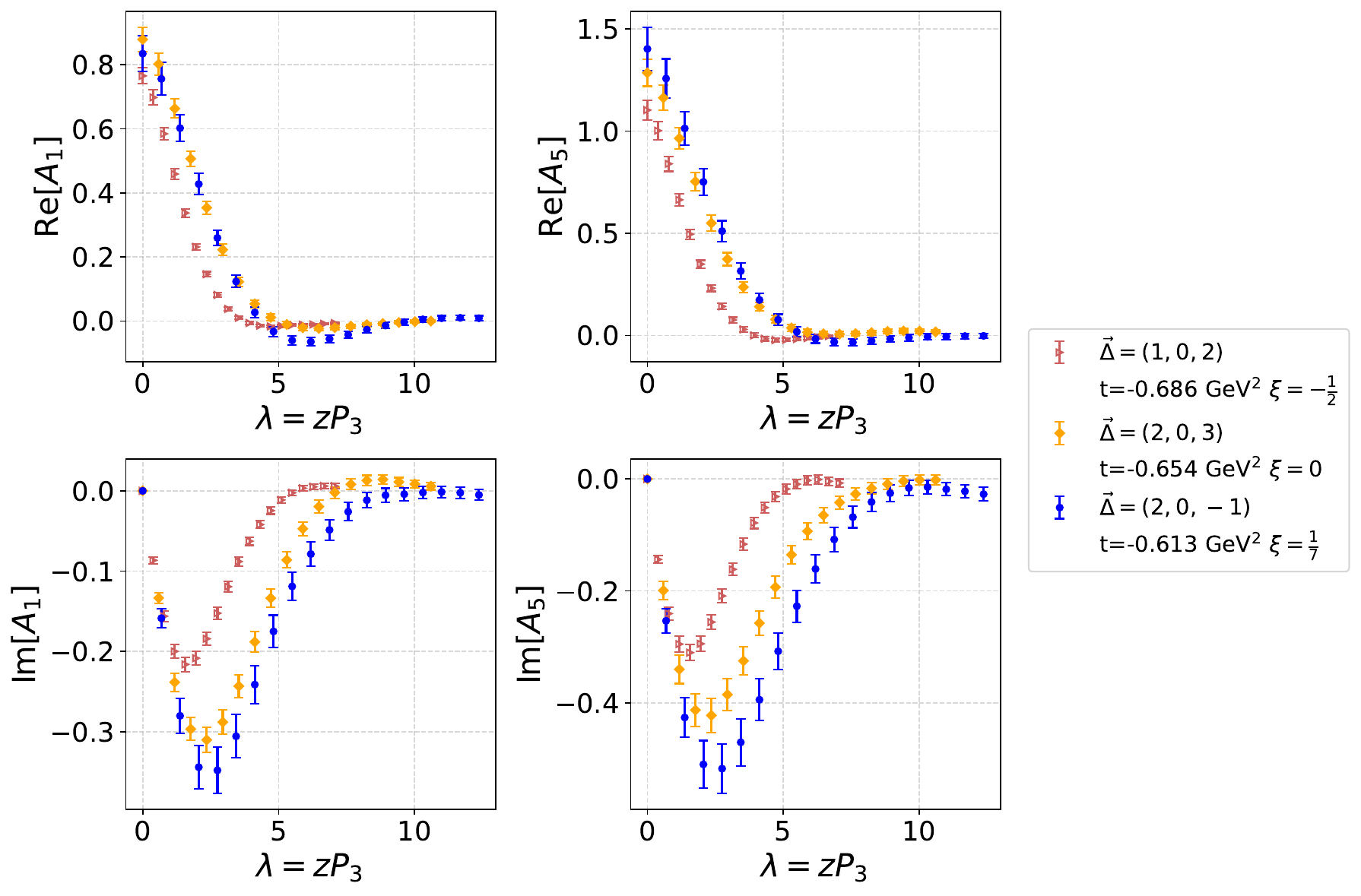}
\caption{Skewness dependence of amplitudes $A_1$ and $A_5$ at approximately fixed $t$ corresponding to different $\vec{\Delta}$ vectors for different $\xi$. Upper and lower panels correspond to the real and imaginary parts, respectively.}
\label{fig:amp_15_xi_dep}
\end{figure}
%%%%%%%%%%%%%%%%%%%%%%%%%%%%%%%%%%%%%%%%%%%%%%%%%%%%%%%%%%%%%%%%%%
%%%%%%%%%%%%%%%%%%%%%%%%%%%%%%%%%%%%%%%%%%%%%%%%%%%%%%%%%%%%%%%%%%
\begin{figure}
\centering
\includegraphics[scale=0.6]{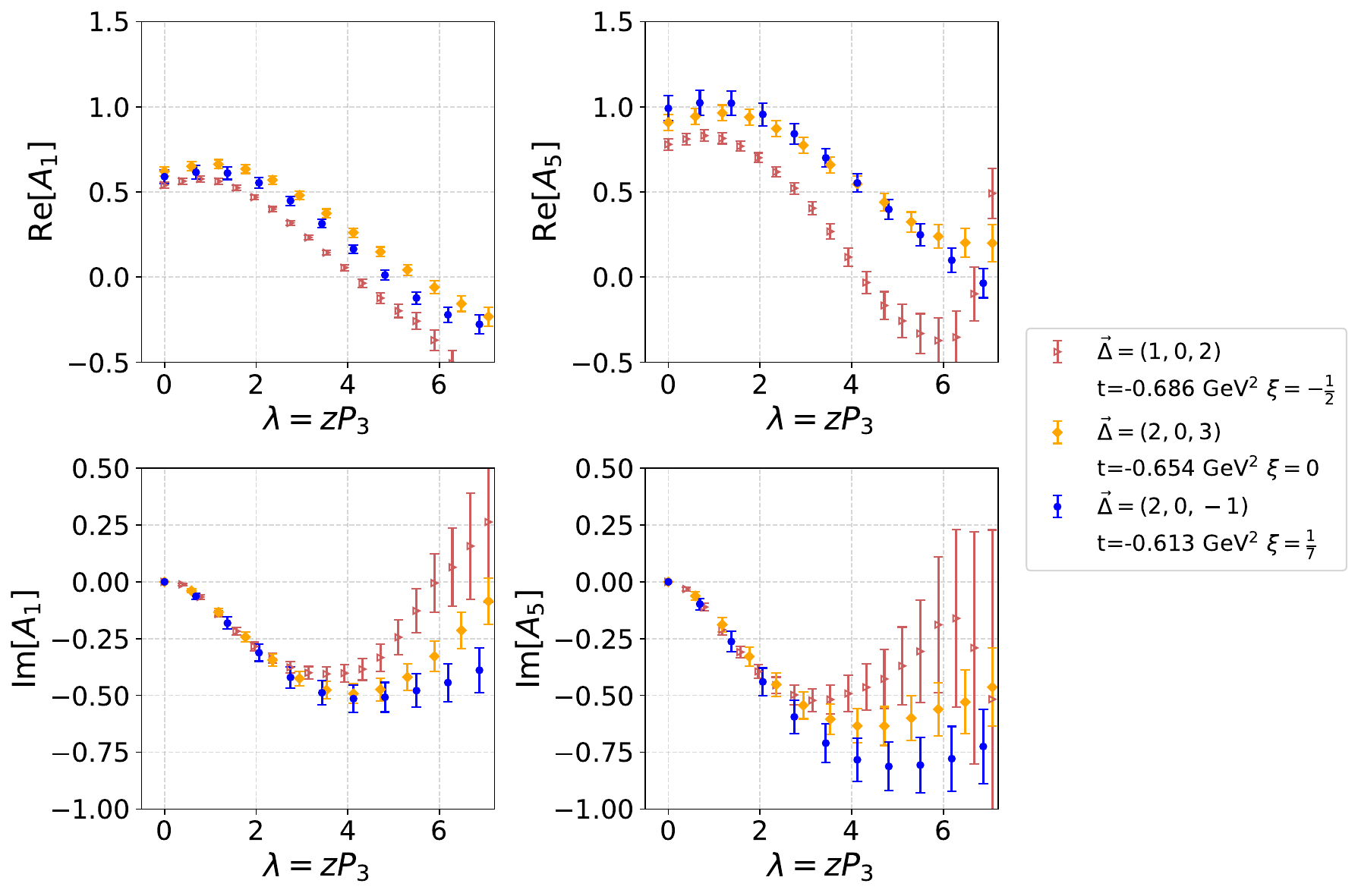}
\caption{Skewness dependence of renormalized amplitudes $A_1$ and $A_5$ at approximately fixed $t$ corresponding to different $\vec{\Delta}$ vectors for different $\xi$. Upper and lower panels correspond to the real and imaginary parts, respectively.}
\label{fig:amp_15_xi_dep_rn}
\end{figure}
%%%%%%%%%%%%%%%%%%%%%%%%%%%%%%%%%%%%%%%%%%%%%%%%%%%%%%%%%%%%%%%%%%

%%%%%%%%%%%%%%%%%%%%%%%%%%%%%%%%%%%%%%%%%%%%%%%%%%%%%%%%%%%%%%%%%%
\begin{figure}
\centering
\includegraphics[scale=0.6]{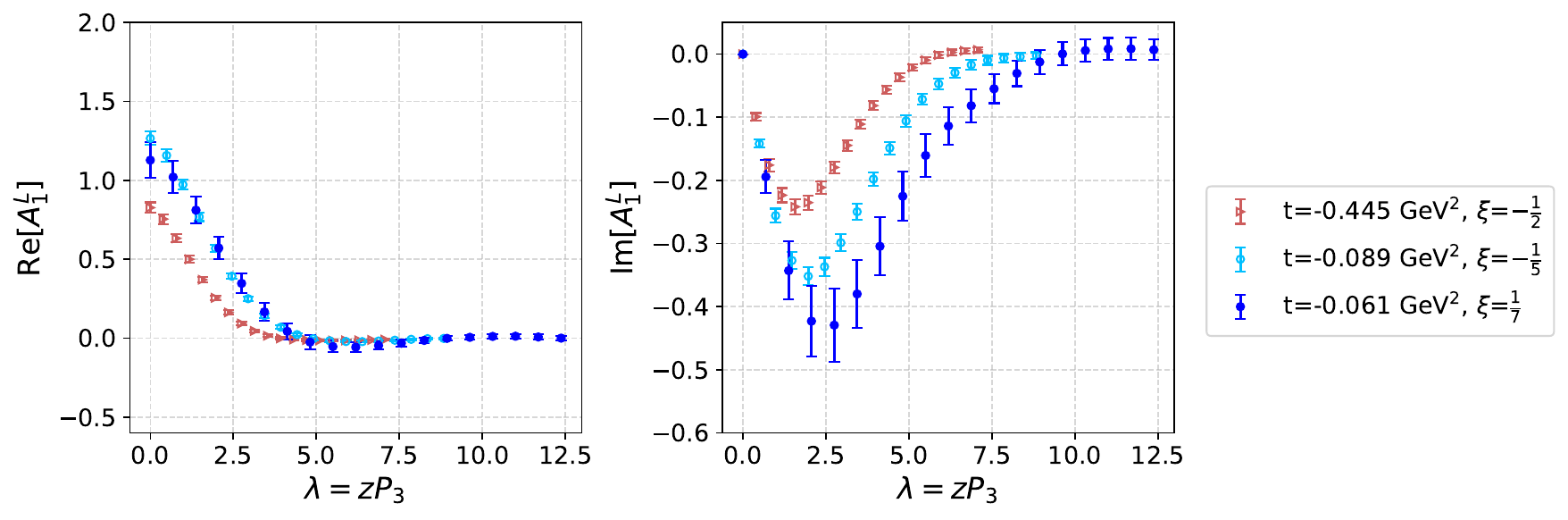}
\caption{Skewness dependence of amplitudes $A^L_1$ from the longitudinal setup $\vec{\Delta}=(0,0,\Delta_3)$. Left and right panels correspond to the real and imaginary parts, respectively.}
\label{fig:amp_A1_Q00_xi_dep}
\end{figure}
%%%%%%%%%%%%%%%%%%%%%%%%%%%%%%%%%%%%%%%%%%%%%%%%%%%%%%%%%%%%%%%%%%

We extract the eight amplitudes $A_1$, $zA_2$, $A_3$, $zA_4$, $A_5$, $zA_6$, $z^2A_7$ and $zA_8$ (i.e., together with their corresponding powers of $z$ to make decompositions $z$-independent) from the 16 matrix elements using Eqs.~(\ref{eq:Pi0G0_ns}–\ref{eq:Pi3G3_ns}). These equations constitute an exactly determined linear system. There are 16 equations, but only 8 of them are linearly independent. To extract the amplitudes, we employ two approaches, which are outlined below, with further technical details provided in Appendix~\ref{sec:app_LSM}. The first approach is the least-squares method, in which the amplitudes are obtained by minimizing the sum of squared differences between the measured matrix elements and the values predicted by the linear model. A similar procedure has been adopted in~\cite{HadStruc:2024rix} for amplitude extraction. Furthermore, we also consider an alternative method to reduce redundancy in the system. Specifically, we omit any equations whose coefficients are all zero, as they provide no constraint on the amplitudes. Moreover, if two or more equations share identical coefficients (or differ only by an overall sign), we treat them as a single effective equation by averaging their corresponding matrix elements, applying sign corrections where appropriate. This reduces the original system to a minimal set of linearly independent equations without loss of information.

We have verified that, in practice, the least-squares method yields results that are numerically consistent with the aforementioned reduction approach. This agreement is not coincidental: the least-squares fit minimizes the residual globally and assigns weights to all available equations based on their coefficients. As a result, equations with vanishing coefficients do not contribute to the minimization and are effectively ignored. Likewise, equations with identical or sign-flipped coefficients are automatically combined through their cumulative contribution to the objective function, which is equivalent to averaging their associated matrix elements. Therefore, the least-squares method inherently incorporates the effects of redundancy, sign symmetries, and zero constraints. Specifically in symmetric kinematic setups, such as $\Delta_1 = \Delta_2$ or $\Delta_1 \Delta_2 = 0$, these simplifications occur frequently, and both methods produce equivalent results. Even in asymmetric cases where $\Delta_1 \ne \Delta_2$, the two approaches remain consistent within statistical uncertainties. Owing to its generality and automation, we adopt the least-squares method for amplitude extraction throughout the remainder of this work.

\begin{figure}[h!]
\centering
\includegraphics[scale=0.55]{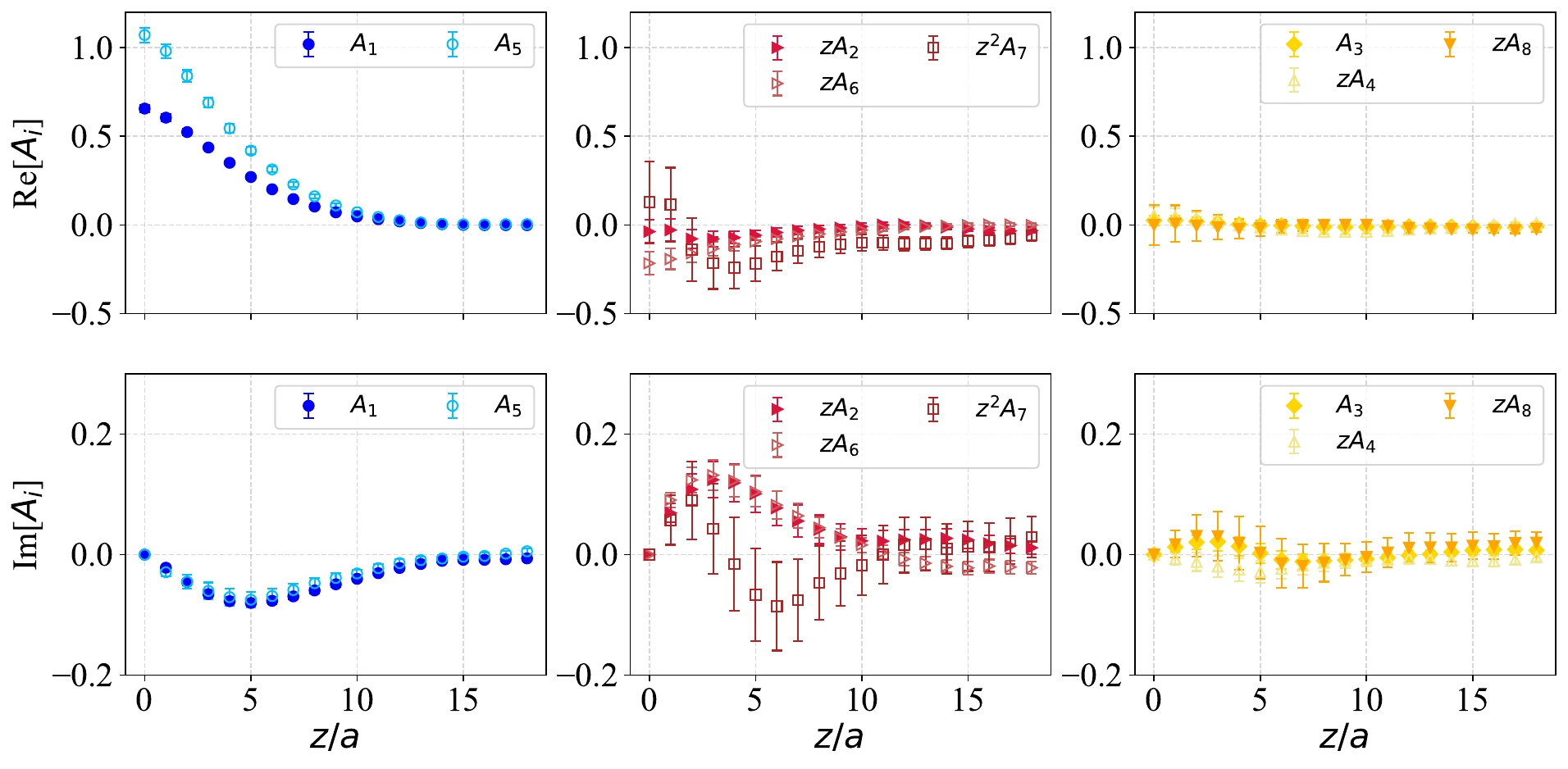}
\caption{Amplitudes $A_1$--$A_8$ extracted from bare matrix elements at $t = -0.493\;\text{GeV}^2$, corresponding to $\vec{\Delta}$ averaged from $(\pm1,\pm1,\pm1)$ (in units  of $2\pi/L$), and skewness $\xi = -1/5$.}
\label{fig:amp_1-8}
\end{figure}

\begin{figure}[h!]
\centering
\includegraphics[scale=0.47]{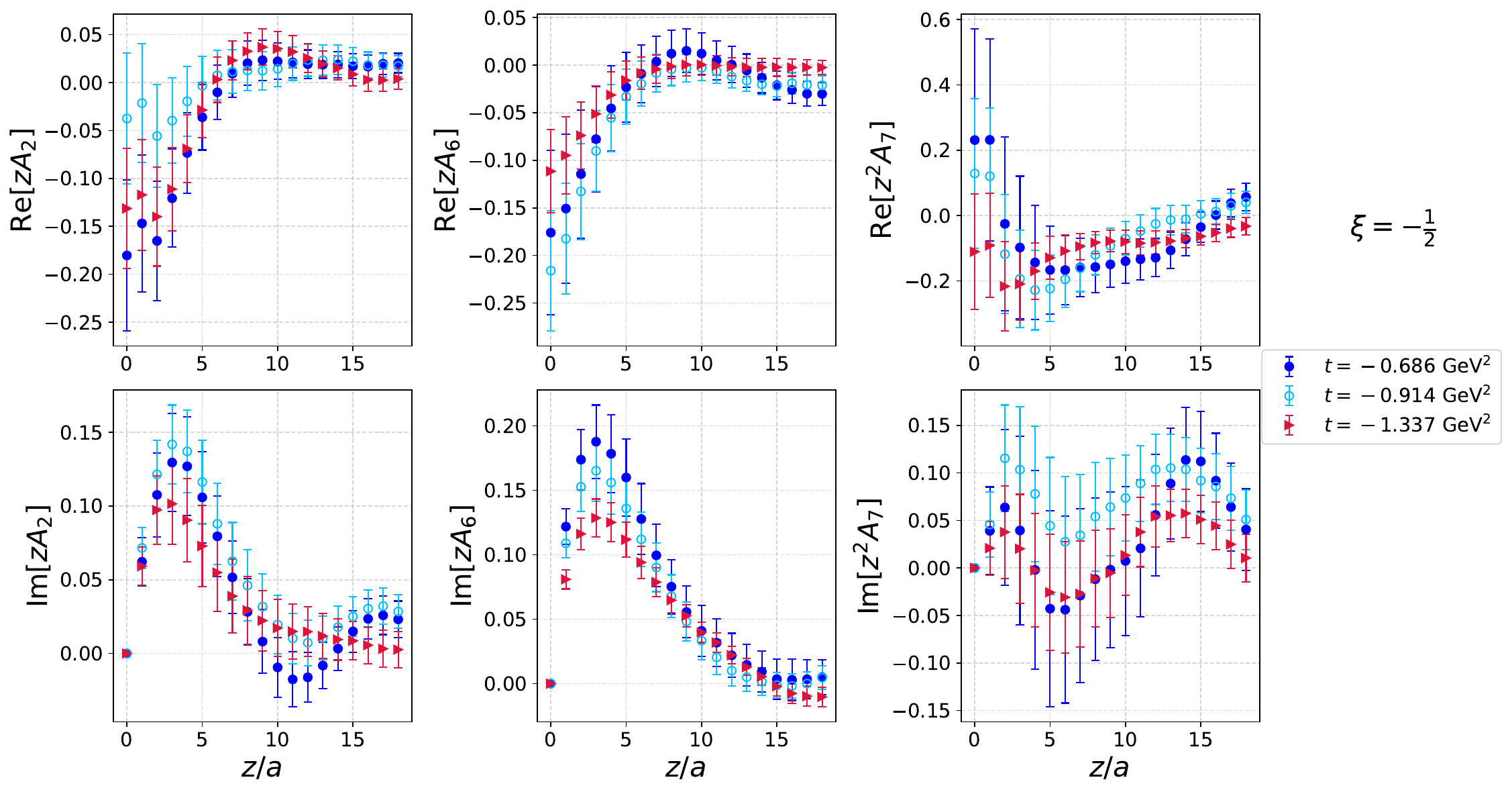}
\caption{Amplitudes $A_2$, $A_6$ and $A_7$ extracted from bare matrix elements at several values of $t$ and skewness $\xi = -1/2$.}
\label{fig:amp_267}
\end{figure}

As discussed in Sec.~\ref{sec:gpd_asy}, the extracted amplitudes satisfy the symmetry relations under sign reversals of $z$, $P$, and $\Delta$, as specified in Eq.~(\ref{eq:symmetries_A}), and we therefore average the results over all combinations of $\pm z$, $\pm P$, and $\pm \Delta$ to enhance statistical precision. A representative example of the eight amplitudes is presented in Fig.~\ref{fig:amp_1-8}, where $A_1$ and $A_5$ are found to dominate, while $A_3$, $A_4$, and $A_8$ are consistent with zero within the quoted uncertainties. In the $\xi=0$ case, a previous study~\cite{Bhattacharya:2022aob} reported that, among all real and imaginary components, only $A_1$, $A_5$, and $\mathrm{Im}[A_6]$ were significantly nonzero, with $A_3$, $A_4$, and $A_8$ vanishing exactly; in the present analysis, we similarly find that $A_3$, $A_4$, and $A_8$ remain compatible with zero, while $\mathrm{Im}[A_2]$ and $\mathrm{Im}[A_6]$ exhibit clear nonzero values, whereas the status of $\mathrm{Re}[A_2]$, $\mathrm{Re}[A_6]$, and $A_7$ is less conclusive based on a single kinematic setup. To investigate further, we examine their dependence on $t$ as shown in Fig.~\ref{fig:amp_267}, which indicates that 
$A_2$ and $A_6$ are nonzero (both the real and imaginary parts, with $\mathrm{Re}[A_2]$ showing only a very weak signal), while $A_7$ remains compatible with zero within statistical uncertainties. Additional details are provided in Appendix~\ref{sec:app_amp}, using the setup with $\xi=-1/5$. The above discussion reflects the current statistical precision of the lattice data, under which some amplitudes remain consistent with zero within uncertainties. The study in Ref.~\cite{HadStruc:2024rix} presents results for all amplitudes in both cases, suggesting that these small amplitudes could be resolved.

%%%%%%%%%%%%%%%%%%%%%%%%%%%%%%%%%%%%%%%%%%%%%%%%%%%%%%%%%%%%%%%%%%
%%%%%%%%%%%%%%%%%%%%%%%%%%%%%%%%%%%%%%%%%%%%%%%%%%%%%%%%%%%%%%%%%
%%%%%%%%%%%%%%%%%%%%%%%%%%%%%%%%%%%%%%%%%%%%%%%%%%%%%%%%%%%%%%%%%

After obtaining the amplitudes, we investigate the $t$-dependence of the dominant ones, $A_1$ and $A_5$, at fixed skewness $\xi=-1/5$, as shown in Fig.~\ref{fig:amp_15_t_dep}. We use $\lambda = z(P_{f3} + P_{i3})/2 = zP_3$ to represent the coordinate-space variable, which is also used in the subsequent figures. As discussed in Sec.~\ref{sec:decomp} and noted in Appendix~\ref{sec:app_longitudinaldecomp}, the definition of $A_1^L$ in terms of the standard eight amplitudes is given by
\begin{align}
A_1^L \equiv A_1 + \alpha A_3 - \frac{\alpha^2}{2} A_5 - \frac{\alpha}{2} (z \cdot \Delta_L) A_6 - \frac{\alpha^2}{2} (z \cdot \Delta_L) A_8.
\end{align}
In this expression, $A_3$, $A_6$, and $A_8$ are typically suppressed, leaving $A_5$ as the dominant correction. In our kinematic setups, $\alpha \approx 0.15$, implying $A_1^L \approx A_1$. We therefore present $A_1^L$ together with $A_1$ in Fig.~\ref{fig:amp_15_t_dep}. Both amplitudes show a monotonic decrease with increasing $-t$, consistent with the behavior observed at vanishing skewness~\cite{Bhattacharya:2022aob}.

Next, we investigate the $\xi$-dependence of the amplitudes. In the asymmetric frame adopted here, the value of $t$ is correlated with the longitudinal momentum transfer $\Delta_3$ and therefore varies with $\xi$ even at fixed $\Delta_T$. This feature does not exist in the symmetric frame, where one can have different $\xi$ at the same value of $t$ from just changing $\Delta_3$ at a fixed $\Delta_T$. Consequently, it is not possible in the asymmetric frame to construct multiple kinematic configurations with identical $t$ but different $\xi$. To address this, we select several momentum setups with approximately equal $t$ values while varying $\xi$. The corresponding results for $A_1$ and $A_5$ are shown in Fig.~\ref{fig:amp_15_xi_dep}, while Fig.~\ref{fig:amp_A1_Q00_xi_dep} illustrates the behavior of $A_1^L$ in the $\Delta_T=0$ case at three values of $\xi$. Additional examples are provided in Appendix~\ref{sec:app_amp}.

The results indicate that the amplitudes $A_1$ and $A_5$ primarily depend on the invariant momentum transfer $-t$. However, a more careful inspection reveals an important trend: increasing the skewness parameter $\xi$ tends to accelerate the decay of matrix elements. This behavior is particularly pronounced in the lower four plots in Fig.~\ref{fig:amp_15_xi_dep}, where the faster decay at $\xi=-1/2$ cannot be attributed solely to the largest $-t$. A similar trend is also observed in the upper four plots, where the configuration with the largest $\xi$, despite having the smallest $-t$, exhibits the most rapid decrease. These observations suggest that skewness plays a non-negligible role.
In addition, to elucidate the impact of $\xi$ on the decay behavior, the renormalized amplitudes $A_1$ and $A_5$ are shown in the same set of lower four plots in Fig.~\ref{fig:amp_15_xi_dep} as Fig.~\ref{fig:amp_15_xi_dep_rn}. It is evident that larger values of $\xi$ lead to a more rapid decrease of $zP_3$ in the renormalized amplitudes.

%%%%%%%%%%%%%%%%%%%%%%%%%%%%%%%%%%%%%%%%%%%%%%%%%%%%%%%%%%%%%%%%%%
\begin{figure}
\centering
\includegraphics[scale=0.6]{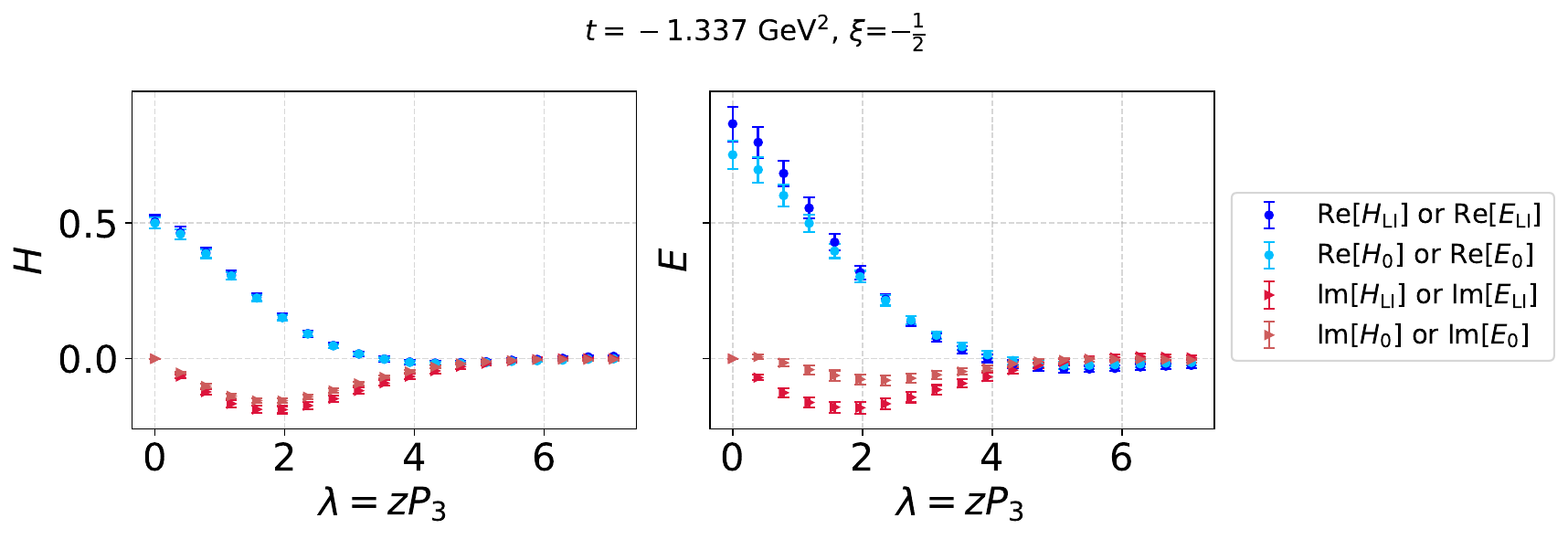}
\includegraphics[scale=0.58]{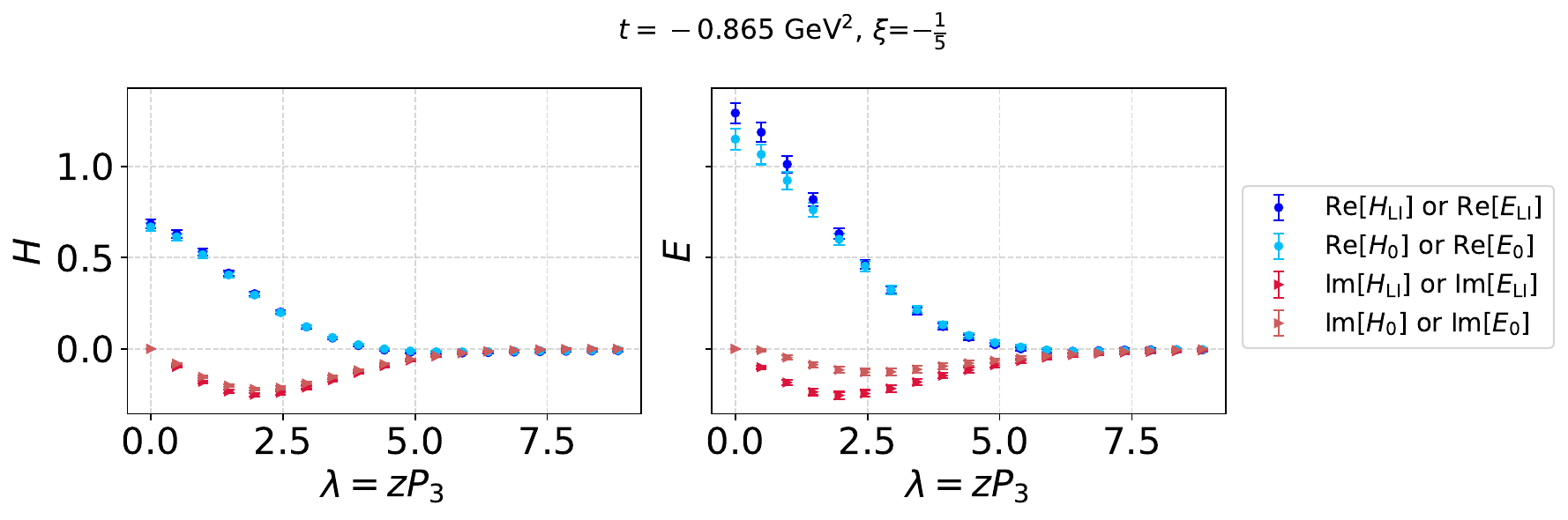}
\includegraphics[scale=0.6]{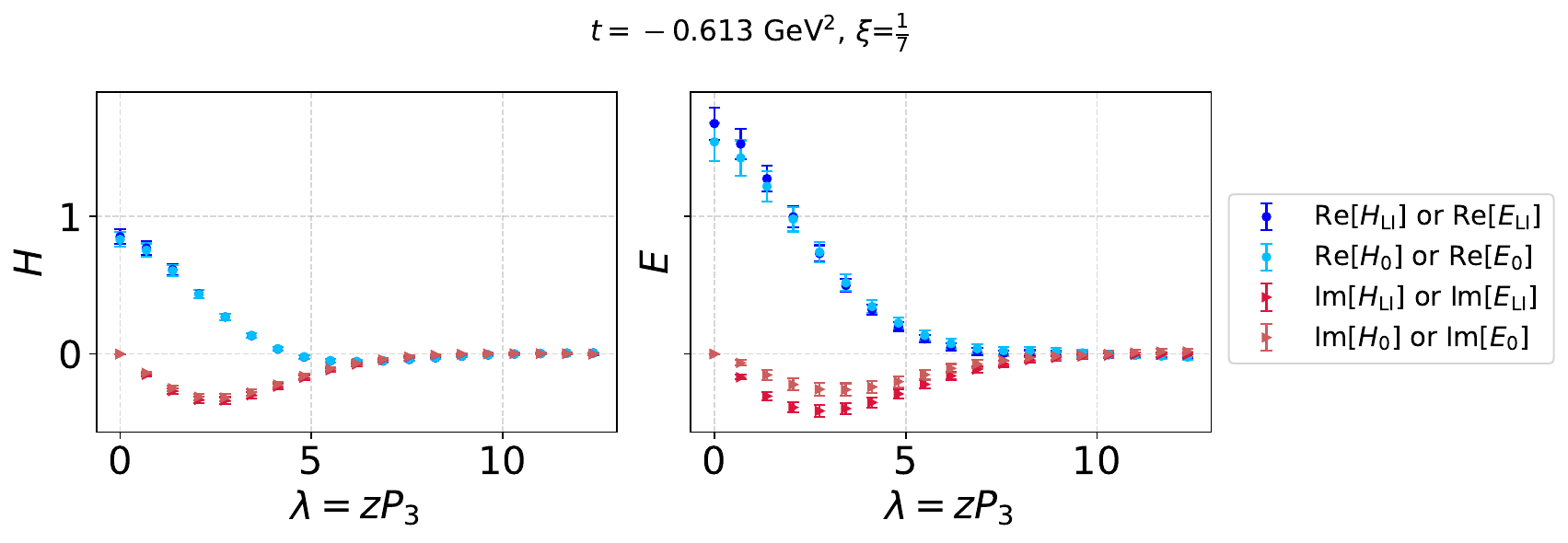}
\caption{Comparison between bare LI and standard GPDs $H$ and $E$ at selected values of $\xi$ and $t$, where $\vec{\Delta}=(2,0,\Delta_3)$ (in units of $2\pi/L$) is fixed in all plots. The upper and lower panels represent the real and imaginary parts, respectively.}
\label{fig:HE_listan_dep}
\end{figure}
%%%%%%%%%%%%%%%%%%%%%%%%%%%%%%%%%%%%%%%%%%%%%%%%%%%%%%%%%%%%%%%%%%

\subsection{Bare H and E GPDs in coordinate space}
After extracting the amplitudes, we obtain the bare standard $H$ and $E$ GPDs using Eqs.~(\ref{eq:stan_H}) and (\ref{eq:stan_E}), and the Lorentz-invariant (LI) GPDs using Eqs.~(\ref{eq:LI_H}) and (\ref{eq:LI_E}). As discussed in Sec.~\ref{sec:GPD_amp}, the LI $E$ GPD exhibits better convergence compared to its standard counterpart, while the $H$ GPD shows similar convergence behavior in both the standard and LI formulations. Therefore, the subsequent analysis primarily focuses on the LI GPDs.

First, in principle, the Lorentz-invariant and standard definitions of the $H$ and $E$ GPDs should coincide in the light-cone limit, i.e., at infinite $P_3$. We perform a detailed comparison of the LI and standard bare $H$ and $E$ GPDs in Fig.~\ref{fig:HE_listan_dep}. The results indicate excellent agreement for Re[$H$], Re[$E$], and Im[$H$], whereas only approximate agreement is observed for Im[$E$]. It also indicates that LI Im[$E$] has smaller relative uncertainties, as also observed in the $\xi=0$ case.

\begin{figure}
\centering
\includegraphics[scale=0.6]{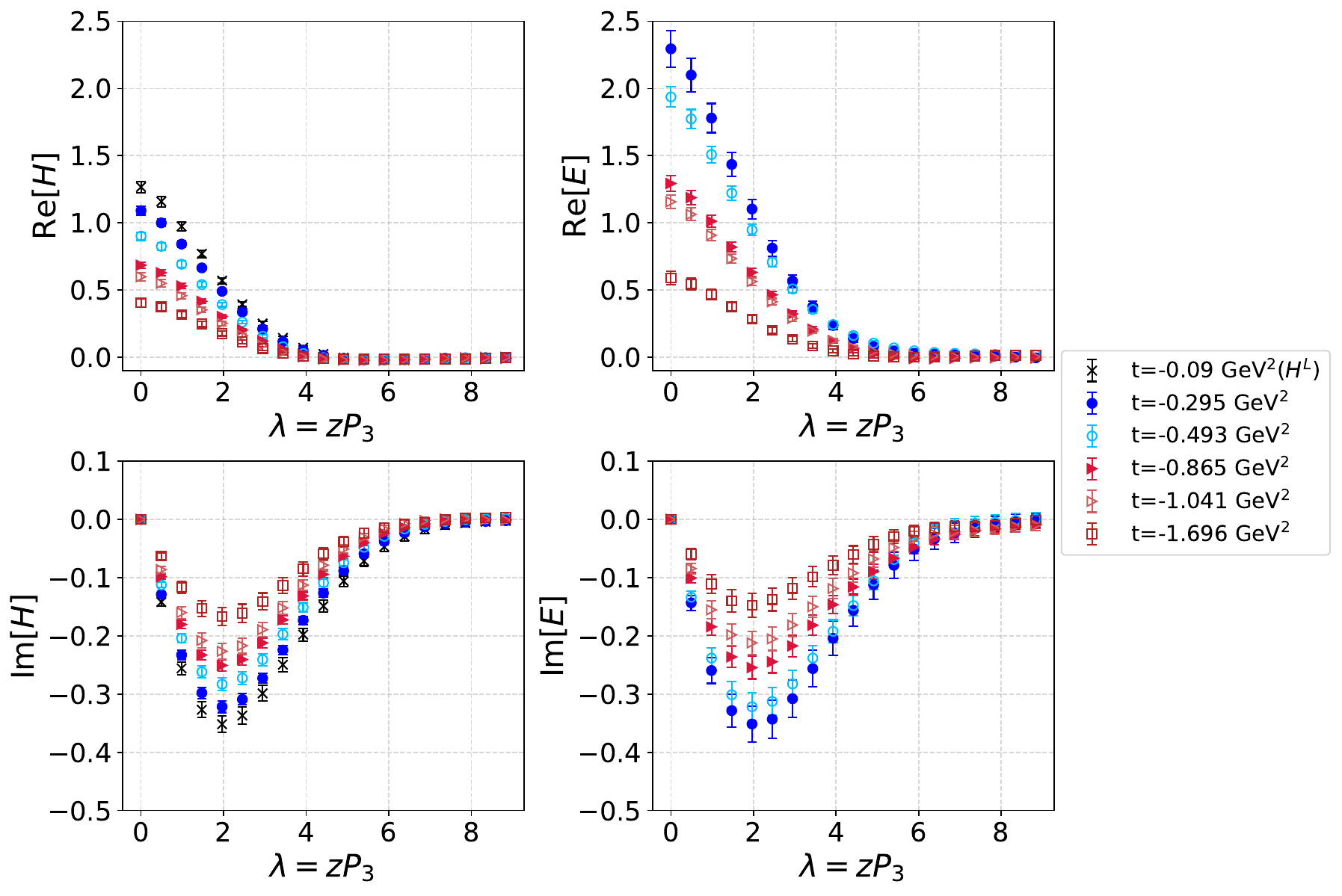}
\caption{Momentum transfer dependence of the bare LI GPDs $H$ and $E$ at $\xi=-1/5$. $H^L$ from the longitudinal setup $\Delta_T=0$ is included and noted in the legend. The upper and lower panels show the real and imaginary parts, respectively.}
\label{fig:HE_t_dep}
\end{figure}

Second, we examine the $t$ dependence of the bare LI GPDs $H$ and $E$ at a fixed skewness $\xi$, as illustrated in Fig.~\ref{fig:HE_t_dep}. It also includes $H^L_1$, which is numerically close to $H$. The figure clearly indicates that both $H$ and $E$ GPDs decrease in magnitude with increasing momentum transfer $-t$.

%%%%%%%%%%%%%%%%%%%%%%%%%%%%%%%%%%%%%%%%%%%%%%%%%%%%%%%%%%%%%%%%%%
%%%%%%%%%%%%%%%%%%%%%%%%%%%%%%%%%%%%%%%%%%%%%%%%%%%%%%%%%%%%%%%%%%

\begin{figure}
\centering
\includegraphics[scale=0.6]{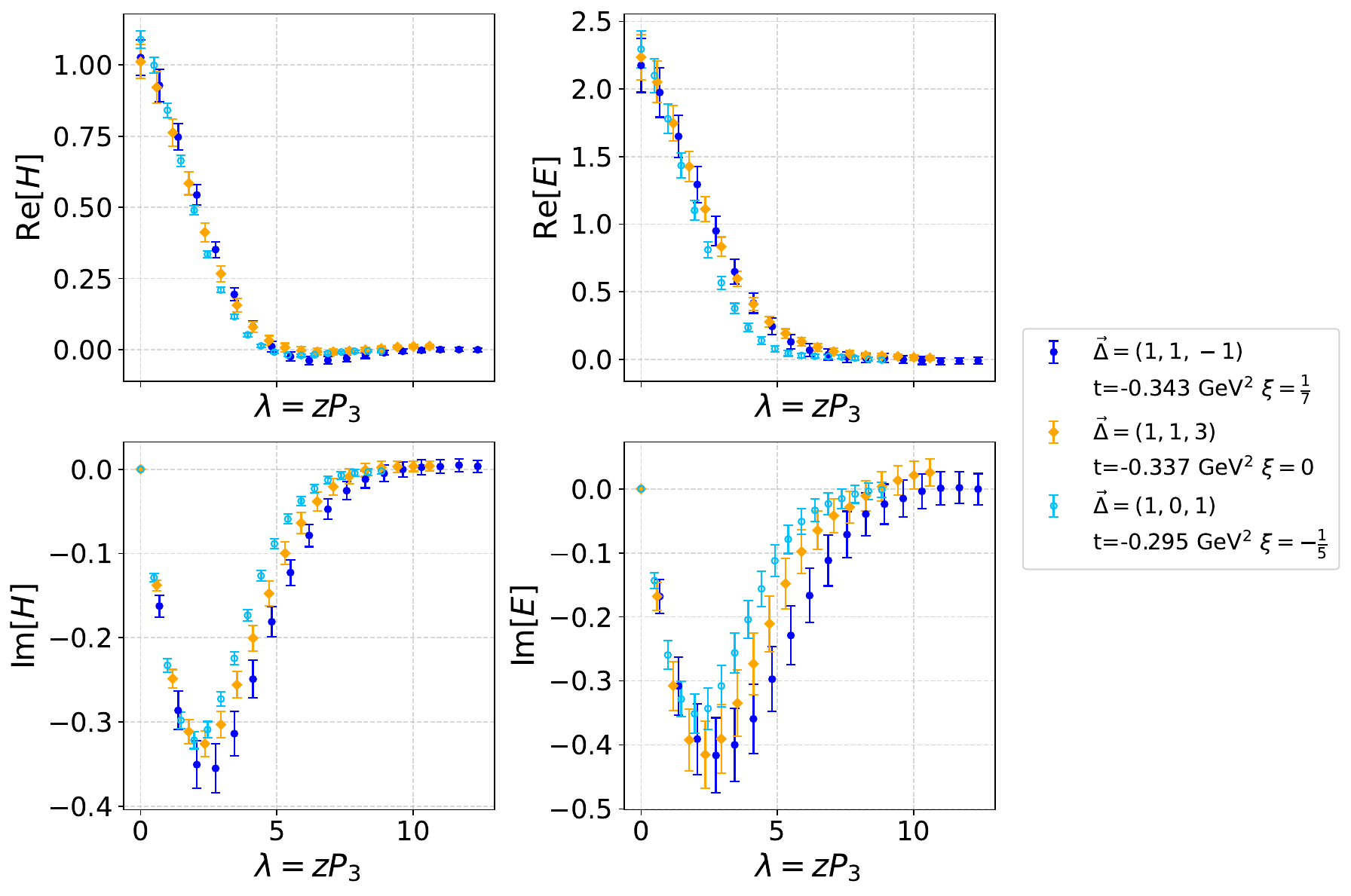}
\includegraphics[scale=0.6]{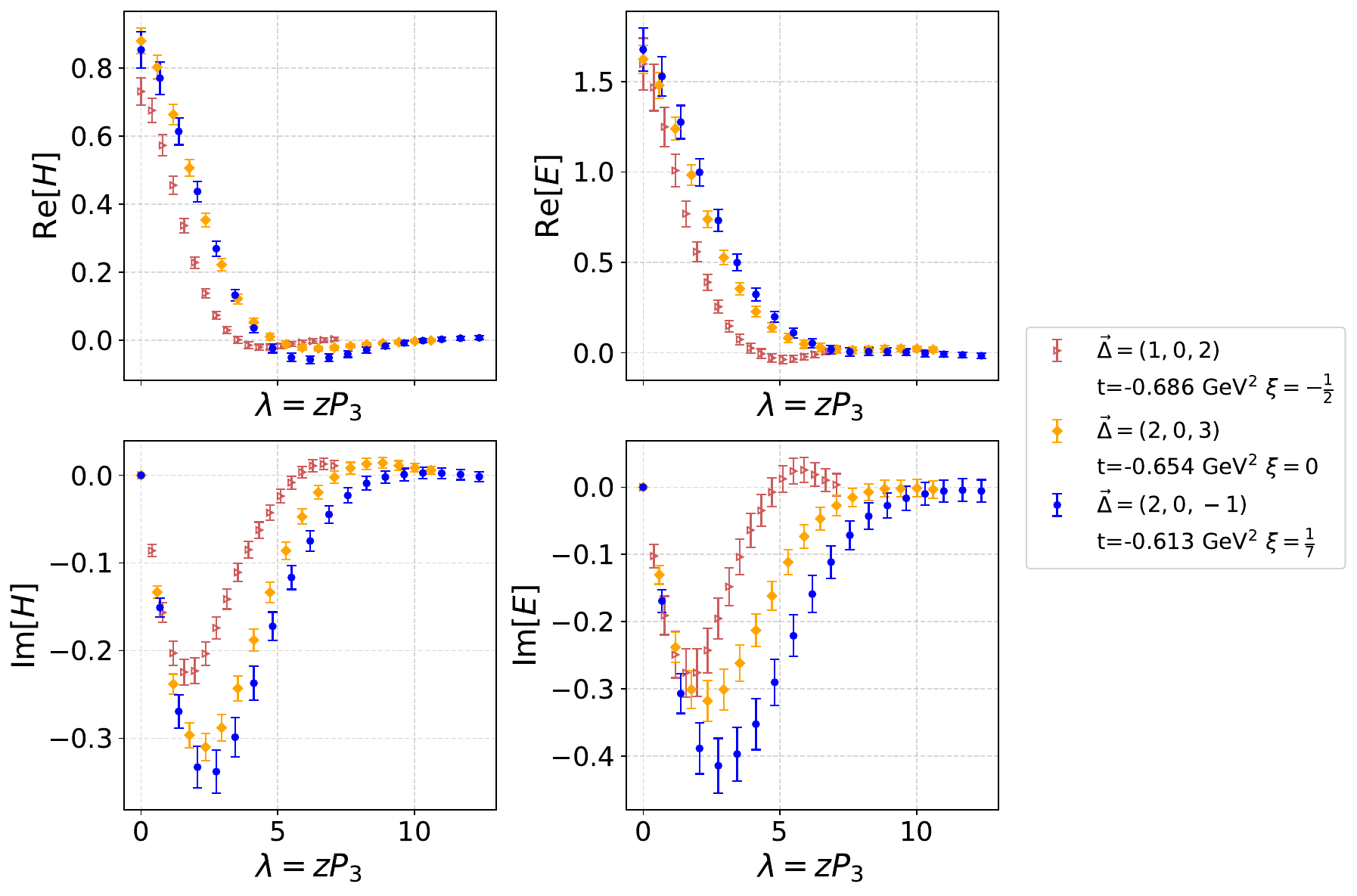}
\caption{Skewness dependence of the bare LI GPDs $H$ and $E$ at similar values of $t$ corresponding to $\vec{\Delta}$. The upper and lower panels show the real and imaginary parts, respectively.}
\label{fig:HE_xi_dep}
\end{figure}

Third, due to the asymmetric frame employed, identical values of $t$ for different $\xi$ cannot be achieved exactly. Therefore, analogous to the previous amplitude analysis, Fig.~\ref{fig:HE_xi_dep} depicts the $\xi$ dependence of the bare LI GPDs $H$ and $E$ evaluated at possibly close values of $t$. Additional examples are provided in Appendix~\ref{sec:app_HE}.

%%%%%%%%%%%%%%%%%%%%%%%%%%%%%%%%%%%%%%%%%%%%%%%%%%%%%%%%%%%%%%%%%%
%%%%%%%%%%%%%%%%%%%%%%%%%%%%%%%%%%%%%%%%%%%%%%%%%%%%%%%%%%%%%%%%%%

\begin{figure}
\centering
\includegraphics[scale=0.55]{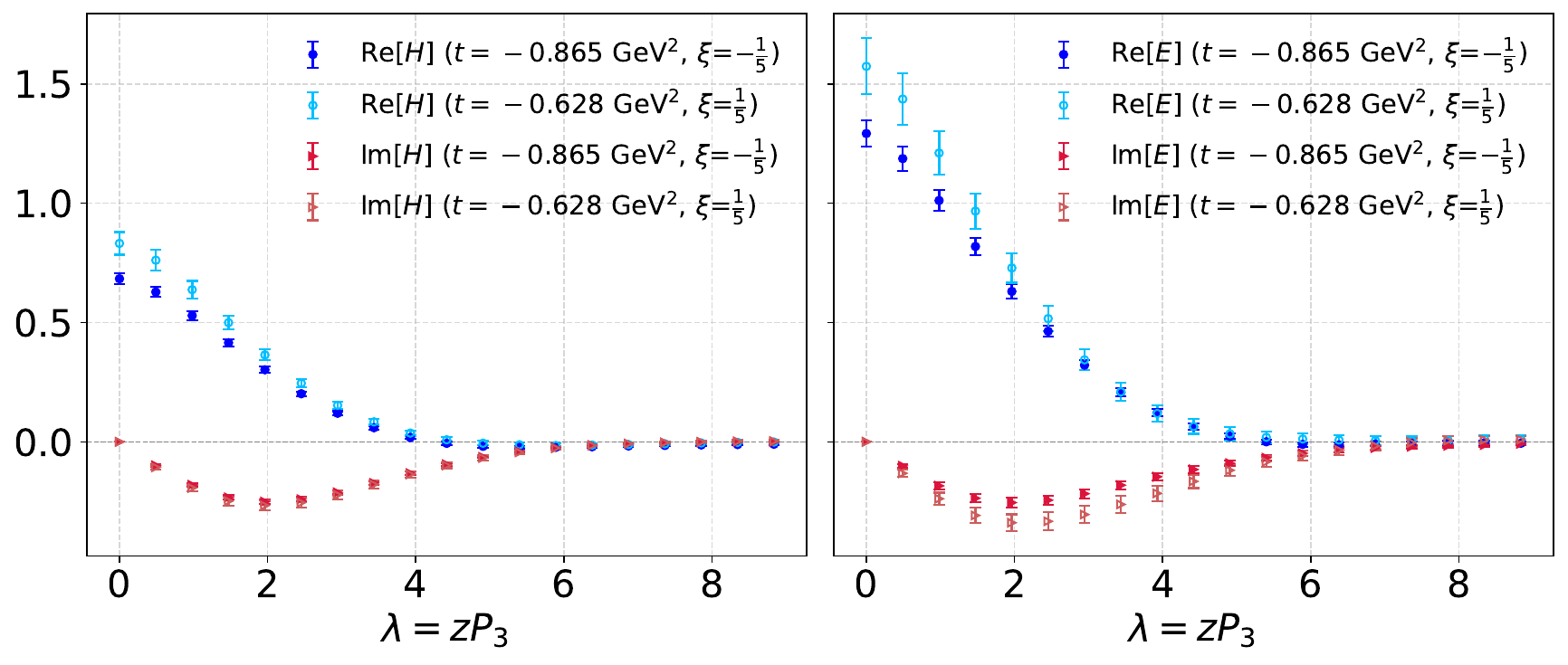}
\vspace*{-3mm}
\caption{Comparison of bare LI GPDs $H$ and $E$ at $\xi=\pm 1/5$ using identical transverse momentum transfers $\Delta_T$, averaged from kinematic setups $\{\Delta_1,\Delta_2\}=\{\pm2,0\}\times\frac{2\pi}{L}$ and $\{0,\pm2\}\times\frac{2\pi}{L}$.}
\label{fig:HE_xipm_Q20}
\end{figure}

Finally, based on the symmetry arguments outlined in Eq.~(\ref{eq:symmetries_A}), the GPDs are expected to be invariant under the transformation $\xi \rightarrow -\xi$. To test this symmetry, we first compare the bare LI GPDs $H$ and $E$ obtained for positive and negative skewness by interchanging $P_{f3}$ and $P_{i3}$ while fixing the magnitude of the momentum transfer vector $\Delta_T$ (averaged over $\{\Delta^1,\Delta^2\} = \{\pm2,0\}$ and $\{0,\pm2\}$). The comparison is presented in Fig.~\ref{fig:HE_xipm_Q20}. Notably, the results for $\xi$ and $-\xi$ do not show agreement. Obviously, this discrepancy arises due to a mismatch in the invariant momentum transfer $t$ associated with these fixed $|\Delta_T|$ choices.

%%%%%%%%%%%%%%%%%%%%%%%%%%%%%%%%%%%%%%%%%%%%%%%%%%%%%%%%%%%%%%%%%%
%%%%%%%%%%%%%%%%%%%%%%%%%%%%%%%%%%%%%%%%%%%%%%%%%%%%%%%%%%%%%%%%%%
%%%%%%%%%%%%%%%%%%%%%%%%%%%%%%%%%%%%%%%%%%%%%%%%%%%%%%%%%%%%%%%%%%
\begin{figure}
\vspace*{-1cm}
\centering
\includegraphics[scale=0.53]{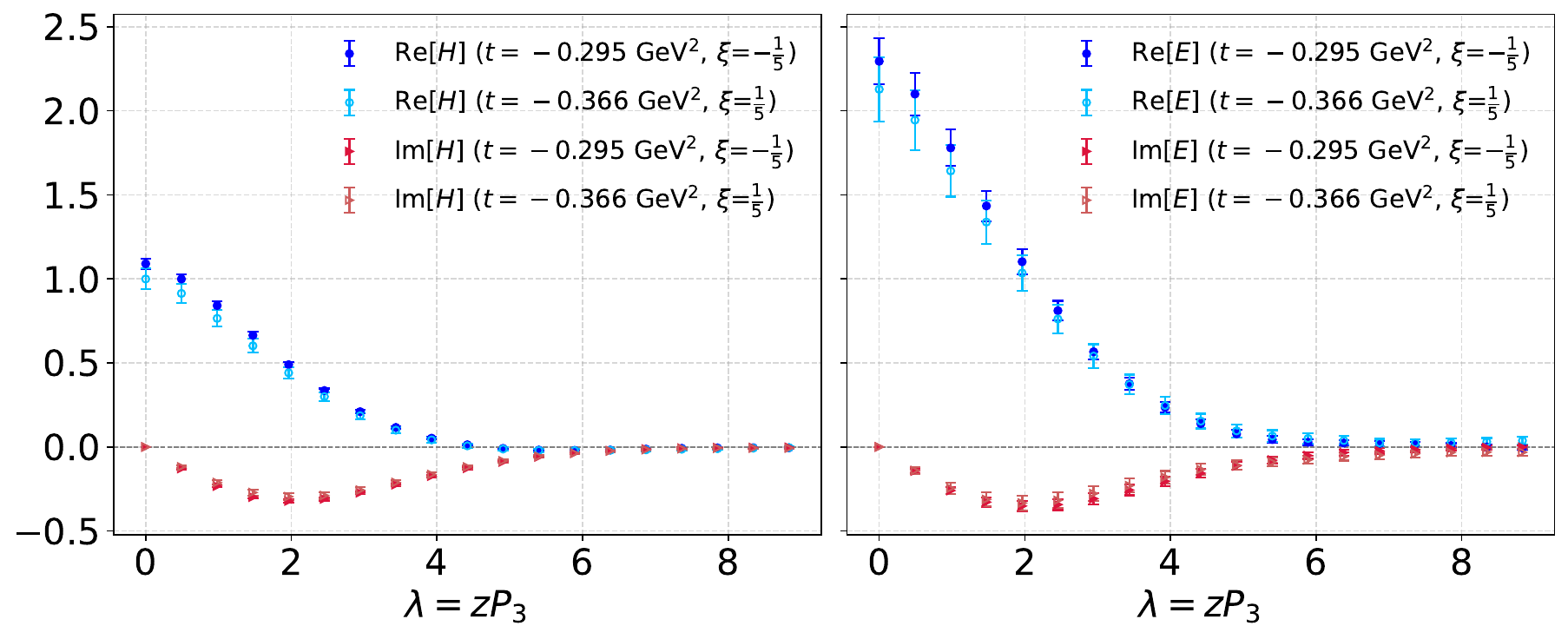}\vspace*{-2mm}
\includegraphics[scale=0.56]{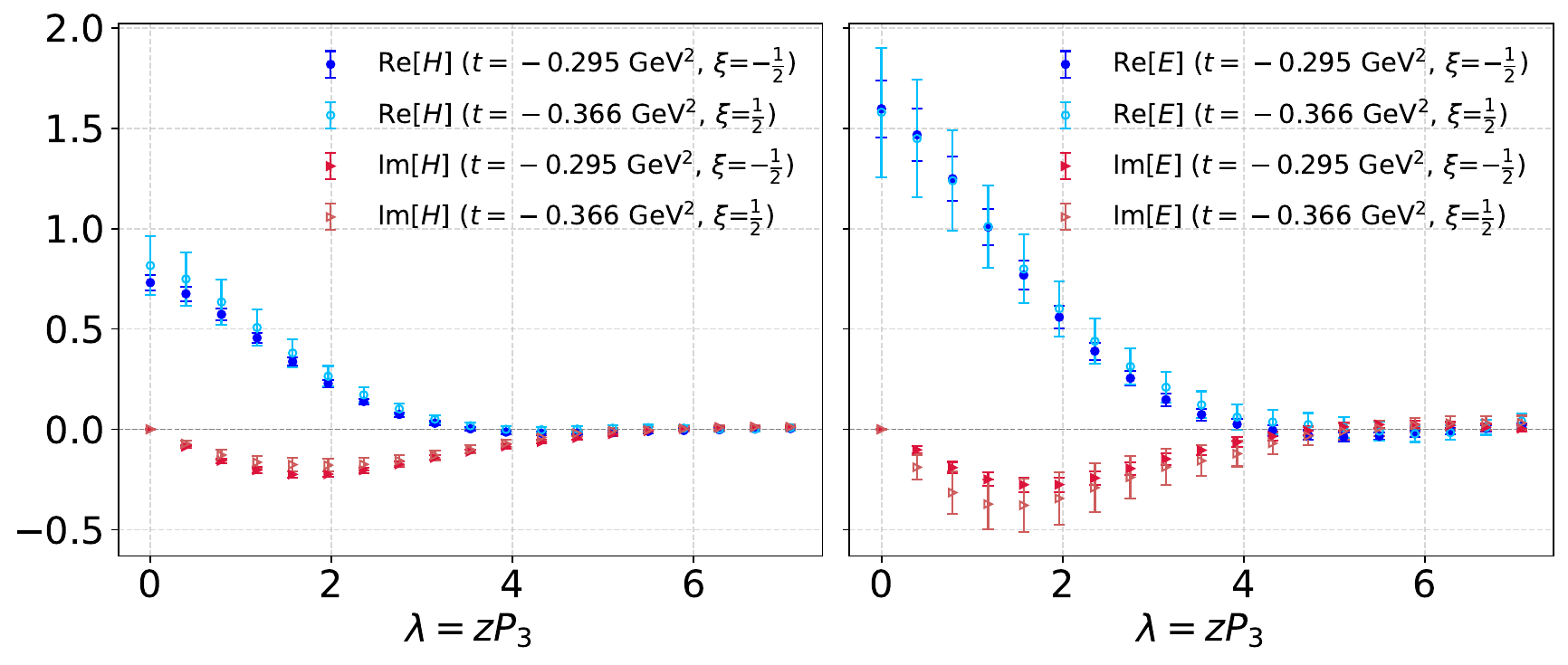}
\vspace*{-3mm}
\caption{Comparison of bare LI GPDs $H$ and $E$ for $\pm\xi$ at similar $t$. The upper panel corresponds to momentum configurations $\{P_{f3},P_{i3}\}=\{3,2\}\times\frac{2\pi}{L}$ and $\{2,3\}\times\frac{2\pi}{L}$, where the momentum transfer $\Delta_T$ is averaged from $(\pm1,\pm1)$ for $\xi=1/5$ and from $(\pm1,0)$ and $(0,\pm1)$ for $\xi=-1/5$. The lower panel corresponds to $\{P_{f3},P_{i3}\}=\{3,1\}$ and $\{1,3\}\times\frac{2\pi}{L}$, with $\Delta_T$ averaged from $(\pm1,\pm1)$ for $\xi=1/2$ and from $(\pm1,0)$ and $(0,\pm1)$ for $\xi=-1/2$.}
\label{fig:HE_xipm}
\end{figure}
%%%%%%%%%%%%%%%%%%%%%%%%%%%%%%%%%%%%%%%%%%%%%%%%%%%%%%%%%%%%%%%%%%

%%%%%%%%%%%%%%%%%%%%%%%%%%%%%%%%%%%%%%%%%%%%%%%%%%%%%%%%%%%%%%%%%%
\begin{figure}
\centering
\includegraphics[scale=0.6]{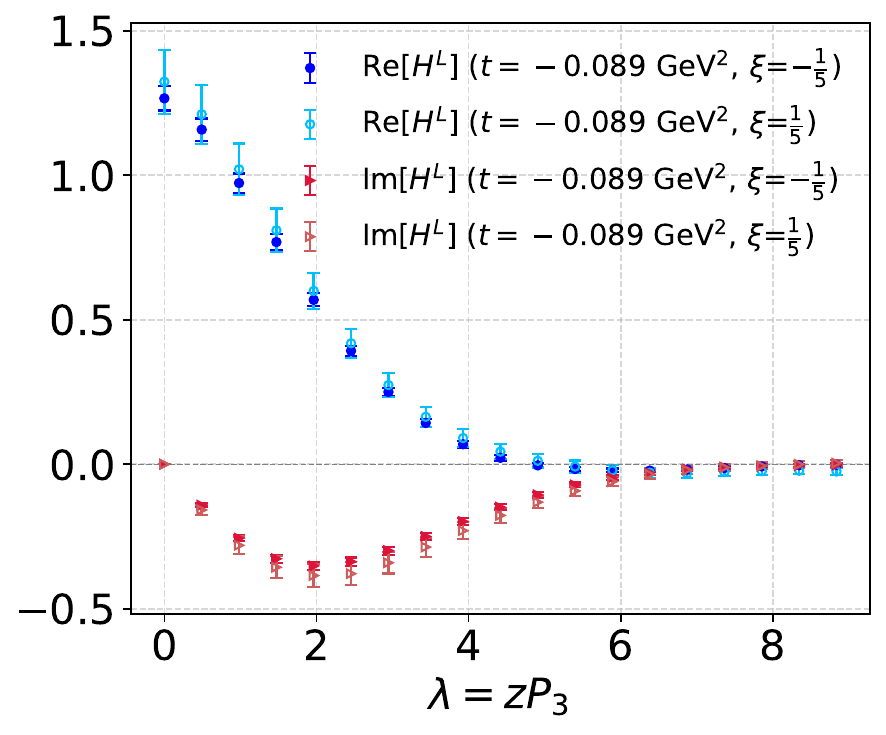}
\caption{Comparison of bare LI GPDs $H^L$ ($\approx H$) from the longitudinal setup $\Delta_T=0$ for $\pm\xi$ at rigorously fixed $t$. $\xi=\pm\frac{1}{5}$ corresponds to momentum configurations $\{P_{f3},P_{i3}\}=\{2,3\}\times\frac{2\pi}{L}$ and $\{3,2\}\times\frac{2\pi}{L}$, respectively.}
\label{fig:HE_xipm_Q00}
\end{figure}
%%%%%%%%%%%%%%%%%%%%%%%%%%%%%%%%%%%%%%%%%%%%%%%%%%%%%%%%%%%%%%%%%%

To rigorously investigate the symmetry properties under the transformation $\xi \rightarrow -\xi$, we select momentum transfers that yield comparable values of $t$ for positive and negative skewness. As illustrated in Fig.~\ref{fig:HE_xipm}, the resulting bare LI GPDs, $H$ and $E$, exhibit excellent numerical agreement between $\xi$ and $-\xi$, thereby confirming their symmetry with respect to $\pm\xi$. Additionally, in Fig.~\ref{fig:HE_xipm_Q00}, we present a comparative analysis of the bare LI GPD $H^L$ (which is close to $H$) at $\xi=\pm1/5$ from the case of the purely longitudinal momentum transfer ($\Delta_T=0$), corresponding rigorously to identical values of $t$. The observed agreement further substantiates the symmetry of the GPDs under the interchange $\xi \leftrightarrow -\xi$.

\subsection{Results of matched GPDs}
\begin{figure}
\centering
\includegraphics[scale=0.55]{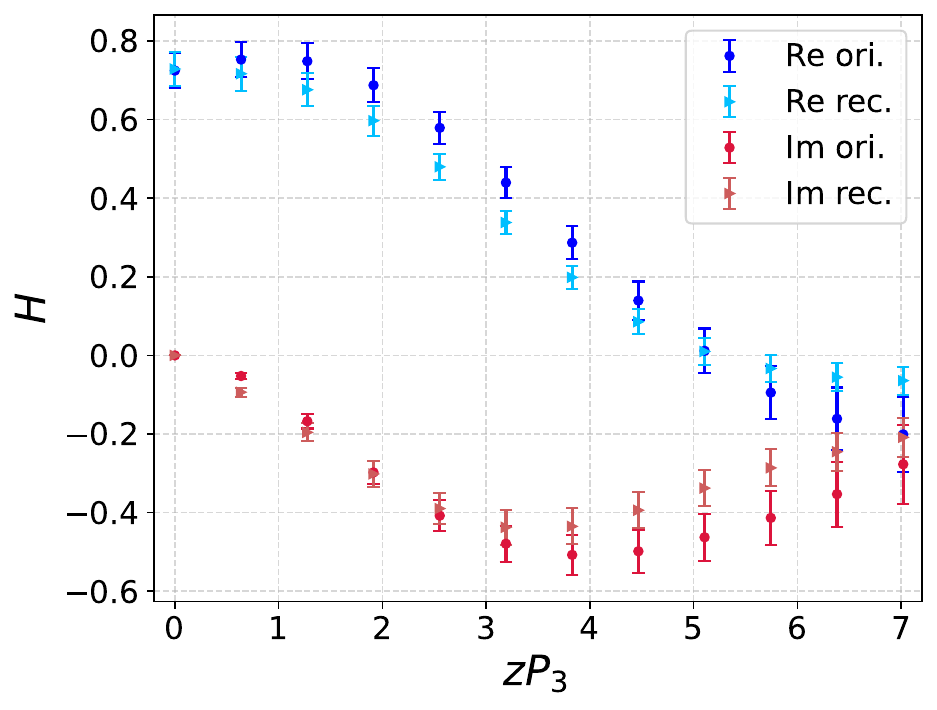}
\includegraphics[scale=0.55]{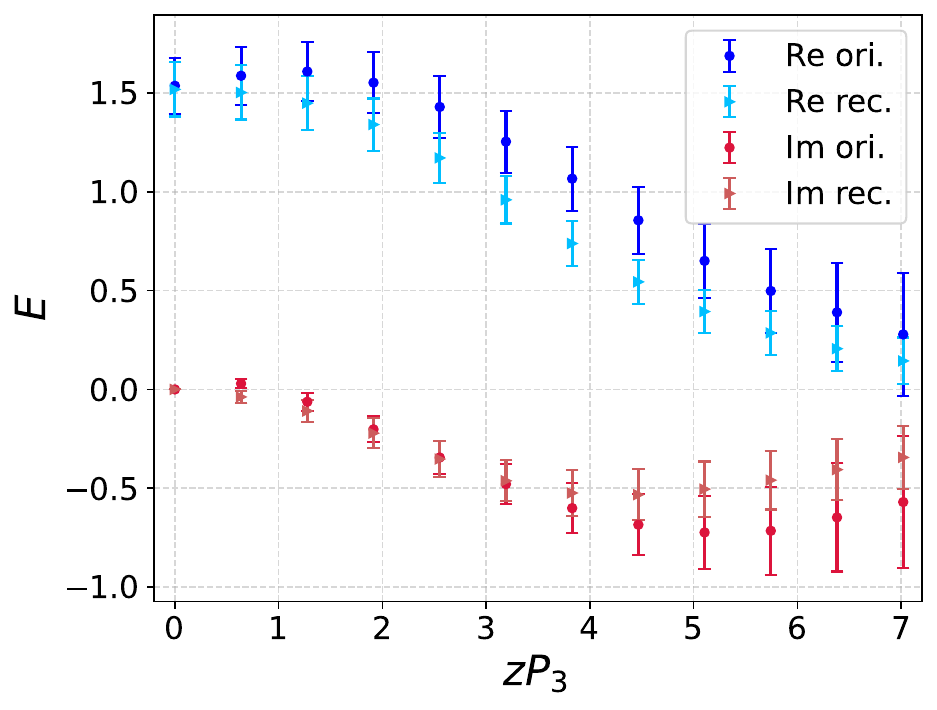}
\caption{Comparison of original renormalized GPDs in coordinate space with ones from Fourier transform of the reconstructed GPDs back to coordinate space. The original GPDs are noted as ``ori.'' in the legends and ones from reconstruction are noted as ``rec.''. The results are shown as an representative example at $t=-0.343\;\rm{GeV}^2$ and $\xi=\frac{1}{7}$.}
\label{fig:BG_check}
\end{figure}

As discussed in Sec.~\ref{sec:lattice}, we first apply the RI/MOM renormalization for bare LI GPDs $H$ and $E$, then adopt the Backus-Gilbert method to extract the quasi-GPDs $H$ and $E$ in momentum space, followed by a one-loop matching procedure to determine the corresponding matched GPDs. Within the BG analysis, we first assessed the reliability of the BG reconstruction by performing a Fourier transform of the reconstructed distributions back to coordinate space and examining their consistency with the original GPDs. The corresponding comparison is shown in Fig.~\ref{fig:BG_check}. The quality of the reconstruction can be quantified by the values of the correlated $\chi^2$ normalized by the number of data points entering the BG procedure, which read 3.7/3.3 for the real/imaginary part of the $H$ GPD and 2.3/1.2 for $E$. This suggests that the BG method is problematic for translating the coordinate-space data to momentum space, as also demonstrated in Ref. \cite{Chu:2025jsi} for PDFs. Nevertheless, the BG approach still provides a reasonable reconstruction, given the simplicity of its model-independent criterion and implementing improvements such as in Ref. \cite{Chu:2025jsi} is outside of the scope of the present work, which concentrates on the asymmetric frame formalism for $\xi\neq0$. Hence, we adopt it in the following and subsequently, we investigate the sensitivity to the truncation parameter $z_{\rm{cut}}$. In this work, we adopt $z_{\rm{cut}} = 12a$ for the following analysis. Further details regarding this choice are provided in Appendix~\ref{sec:app_HE}.

%%%%%%%%%%%%%%%%%%%%%%%%%%%%%%%%%%%%%%%%%%%%%%%%%%%%%%%%%%%%%%%%%%
%%%%%%%%%%%%%%%%%%%%%%%%%%%%%%%%%%%%%%%%%%%%%%%%%%%%%%%%%%%%%%%%%%

\begin{figure}
\centering
\includegraphics[scale=0.7]{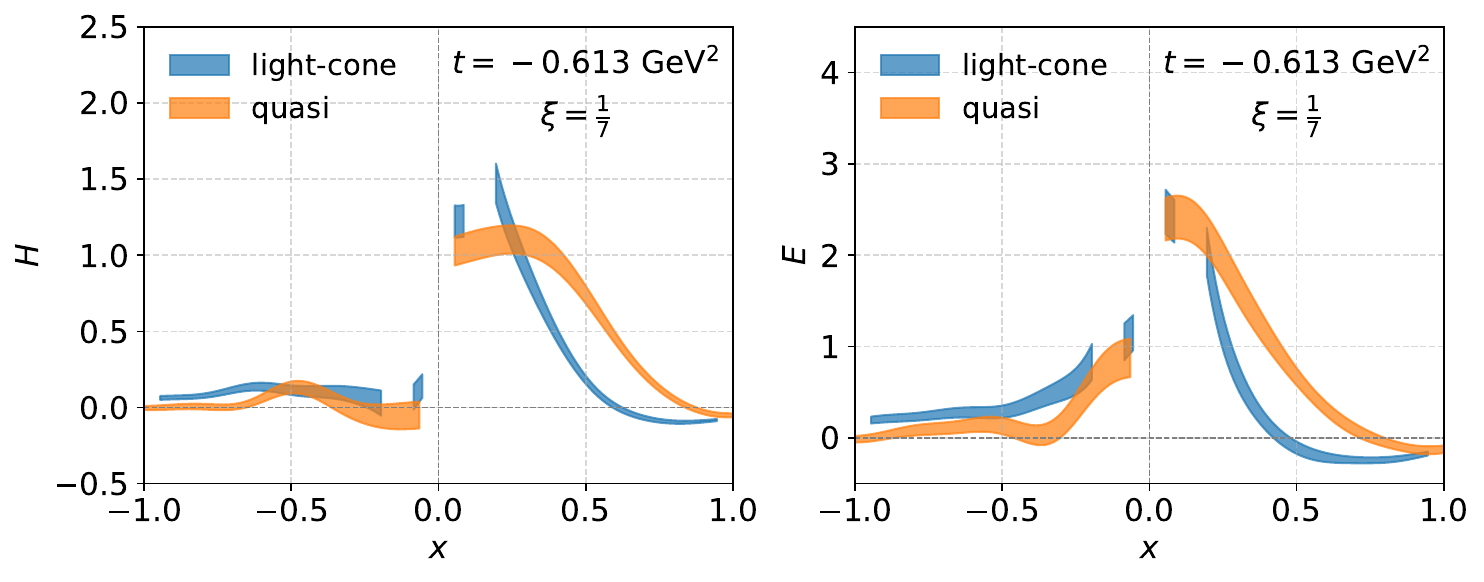}
\includegraphics[scale=0.7]{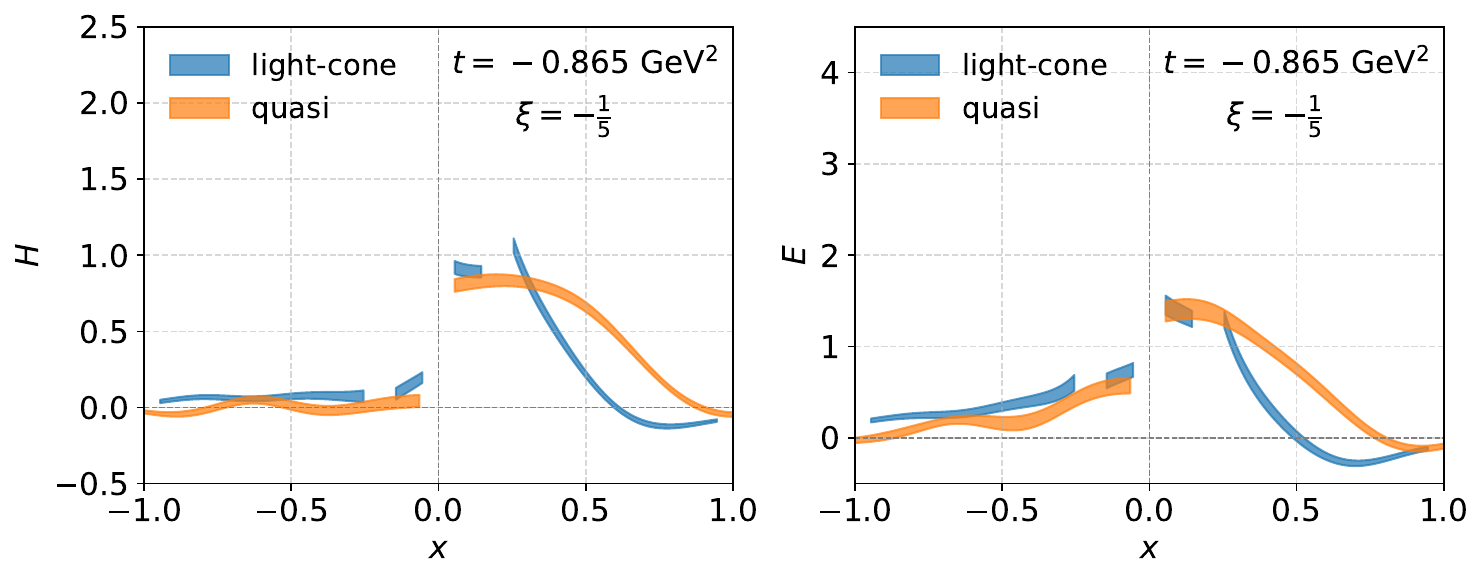}
\includegraphics[scale=0.7]{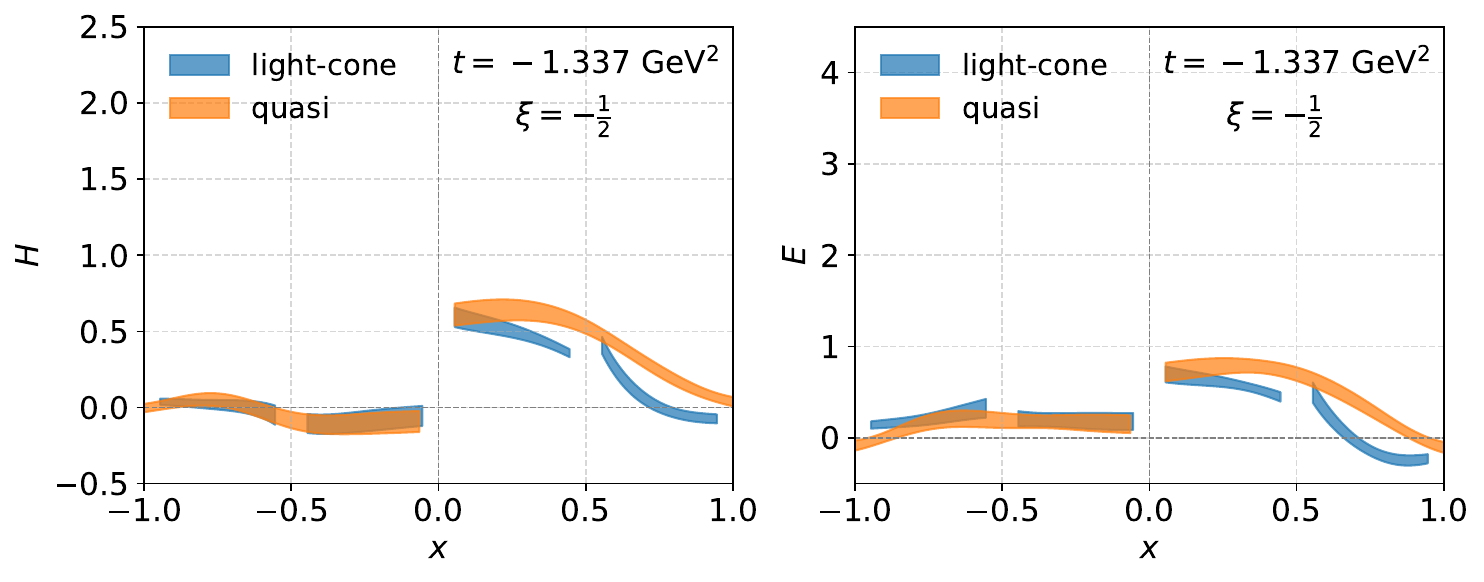}
\caption{Comparisons between quasi-GPDs and the corresponding light-cone GPDs at fixed transverse momentum transfers $\vec{\Delta}=(\pm2,0),(0,\pm2)$ (in units of 2/pi/L), corresponding to different values of $t$ and skewness values $\xi=-1/2,-1/5,1/7$. Values in unreliable regions are excluded, see main text for a discussion.}
\label{fig:HE_lcquasi}
\end{figure}

With the BG method and the matching procedure applied, we first examine the impact of the one-loop matching on the GPDs. Before we show the outcome, an important comment is necessary. In the light-cone results, certain regions are unreliable due to power corrections, namely $1-x_0<|x|<1$, $|\xi|-x_0<|x|<|\xi|+x_0$, where $x_0$ is of order $\Lambda_{\rm QCD}/P_3$ \cite{Holligan:2025baj} and amounts to around 0.2 at our values of the nucleon boost. 
We note that another unreliable region arises in the BG method, which leads to a discontinuous result at $x=0$.
While this is physically sensible for PDFs or zero-skewness GPDs, in the presence of an ERBL region, there should be no discontinuity at zero $x$.
Thus, it can be viewed as an artefact of the BG reconstruction, reflecting the presence of an inverse problem. Physically, one would expect a continuous behavior, but, nevertheless, with a possibly large slope around $x=0$, reflecting the developing discontinuity in the $\xi=0$ limit, see e.g.~Ref.~\cite{Petrov:1998kf} for a study in the chiral soliton model in the large-$N_c$ limit.
Overall, we consider some region of $|x|<x_{\rm BG}$ to be affected by the problem, and it appears in both quasi- and matched GPDs.
To visualize these unreliable regions, we choose $x_0=x_{\rm BG}=0.05$ and exclude them from Fig.~\ref{fig:HE_lcquasi} and all subsequent figures. 
We emphasize that such low value of this cutoff, particularly for $x_0$, underestimates the effects present in the matching (and possibly also in the reconstruction close to $x=0$), but we choose it for illustrative purposes -- to draw attention to the problems close to the discussed values of $x$, but to still present the results obtained with the BG reconstruction and the matching procedure. 

With the above reservations in mind, Fig.~\ref{fig:HE_lcquasi} shows the differences between the quasi-GPDs and the light-cone GPDs at selected skewness values ($\xi=-1/2,\,-1/5$, and $1/7$) for a fixed transverse momentum transfer ($\Delta_T=(2,0)$ with permutations and sign flips).
In all cases, the matching corrections are small for $x<0$. For positive $x$, the corrections become sizable in the DGLAP region, remain small to moderate in the ERBL region, and show a divergence at $x=\xi$.
This leads to the aforementioned discontinuity between the ERBL and the DGLAP regions, reflecting large power corrections.

%%%%%%%%%%%%%%%%%%%%%%%%%%%%%%%%%%%%%%%%%%%%%%%%%%%%%%%%%%%%%%%%%%
%%%%%%%%%%%%%%%%%%%%%%%%%%%%%%%%%%%%%%%%%%%%%%%%%%%%%%%%%%%%%%%%%%

\begin{figure}
\centering
\includegraphics[scale=0.6]{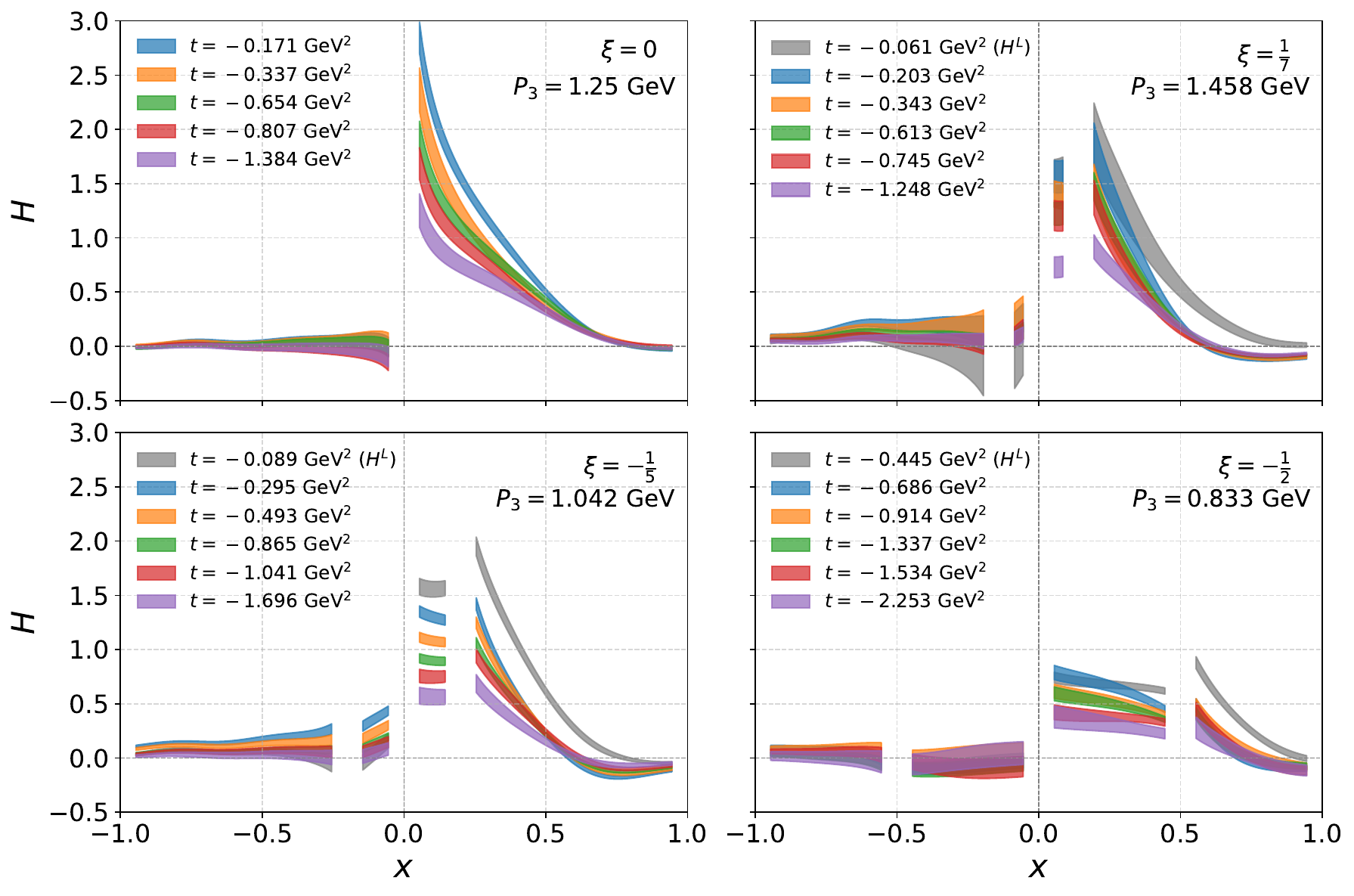}
\caption{Dependence on momentum transfer $t$ for light-cone GPDs $H$ evaluated at different skewness values $\xi$. GPD $H^L$ ($\approx H$) from the purely longitudinal setup ($\Delta_T=0$) is included. $\xi=0$ results are taken from previous studies~\cite{Bhattacharya:2022aob,Cichy:2023dgk}. Values in unreliable regions are excluded, see main text for a discussion.}
\label{fig:H_x_t_dep}
\end{figure}

Next, we investigate the $t$-dependence of the light-cone GPDs at several values of skewness, including the zero-skewness case ($\xi=0$) with data taken from Refs.~\cite{Bhattacharya:2022aob,Cichy:2023dgk}.
Figure~\ref{fig:H_x_t_dep} shows the results for the light-cone GPD $H$ evaluated at $\xi=1/7,\,0,\,-1/5,\,-1/2$, corresponding to $P_3=1.458,\,1.250,\,1.042,\,0.833$~GeV, and also includes $H^L$ ($\approx H$) from the purely longitudinal setup ($\Delta_T=0$). The corresponding results for the GPD $E$ are presented in Appendix~\ref{sec:app_HE}. As expected, the magnitude of $H$ in the ERBL region ($|x|<|\xi|$) decreases monotonically with increasing $-t$, while its $-t$ dependence remains mild in the DGLAP region ($|\xi|\leq x<1$). In addition, the relative discontinuity at $|x|=|\xi|$ becomes less pronounced as $P_3$ increases.
We note that a more realistic estimate of the excluded regions, $x_0\approx0.2$, indicates that the whole ERBL region at $\xi=-1/5,\,1/7$ is unreliable.
Hence, to access the ERBL region in a reliable manner at such low values of skewness, significantly larger nucleon boosts are needed.

%%%%%%%%%%%%%%%%%%%%%%%%%%%%%%%%%%%%%%%%%%%%%%%%%%%%%%%%%%%%%%%%%%
%%%%%%%%%%%%%%%%%%%%%%%%%%%%%%%%%%%%%%%%%%%%%%%%%%%%%%%%%%%%%%%%%%

\begin{figure}
\centering
\includegraphics[scale=0.57]{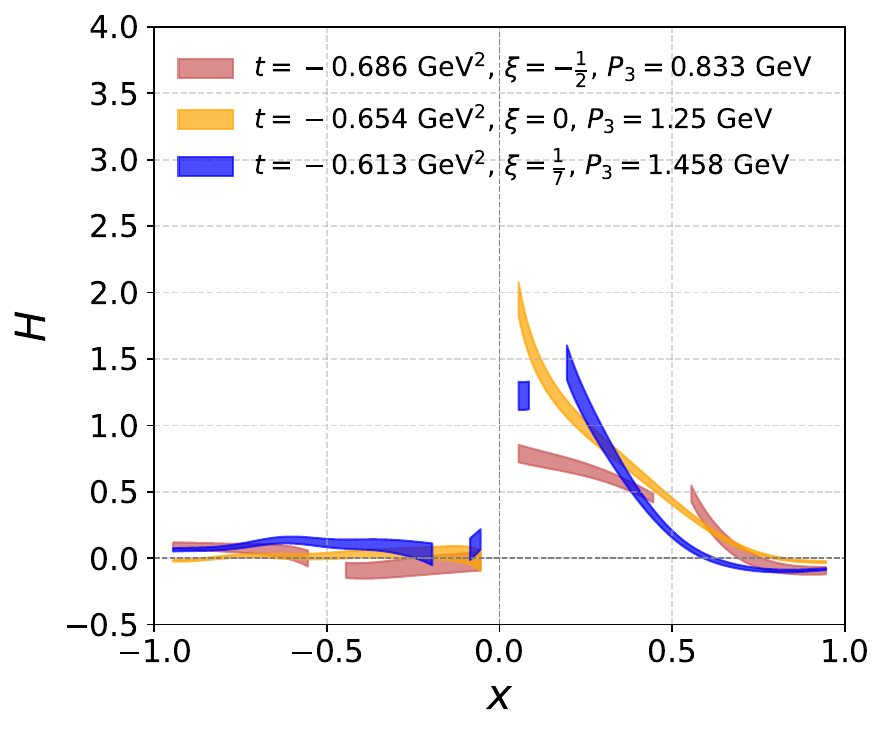}
\includegraphics[scale=0.57]{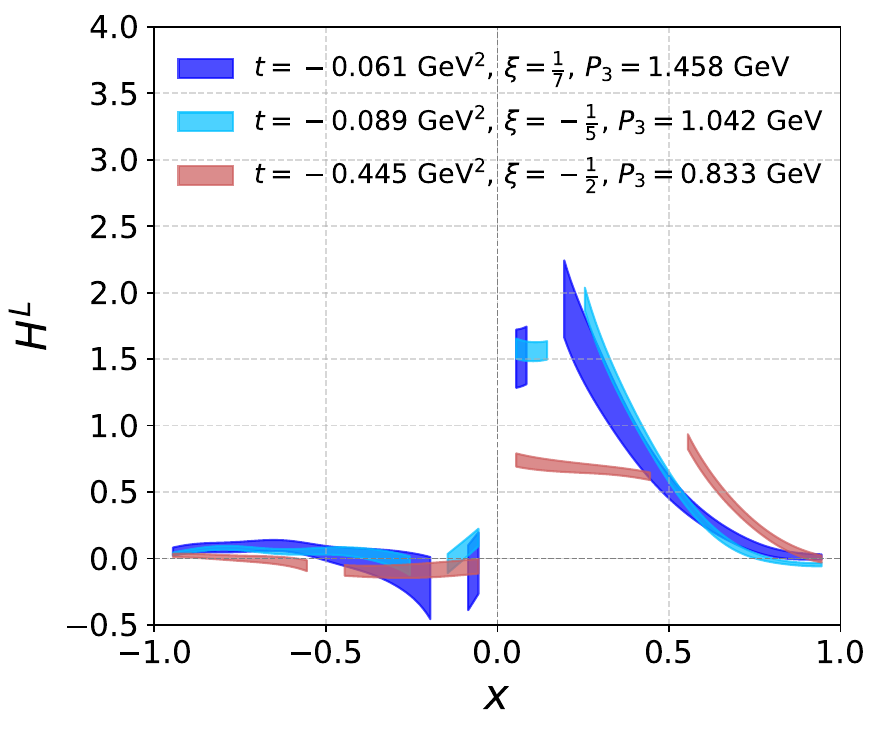}
\caption{Skewness dependence of light-cone GPDs $H$ and $H^L$ ($\approx H$) from the purely longitudinal setup ($\Delta_T=0$). Left panel shows $H$ at comparable momentum-transfer values $t$. Right panel displays $H^L$ at three values of $\xi$ corresponding to different $P_3$. Values in unreliable regions are excluded, see main text for a discussion.}
\label{fig:HE_x_xi_dep}
\end{figure}

Finally, we address the skewness dependence of the light-cone GPDs. As discussed above, calculations performed in the asymmetric frame allow comparison of GPDs at similar $-t$ values across different skewness parameters $\xi$, including $H^L$ ($\sim H$) from the $\Delta_T=0$ case at three values of $\xi$. Fig.~\ref{fig:HE_x_xi_dep} summarizes these results, clearly illustrating a suppression of the light-cone GPDs in the ERBL region as $\xi$ increases, while the DGLAP region shows the opposite tendency. Compared to the $\xi$ dependence of GPDs in coordinate (Fig.~\ref{fig:HE_xi_dep}) and quasi-GPDs in momentum space (Fig.\ref{fig:HE_lcquasi}), this effect and the discontinuity at $|x|=|\xi|$ is mainly caused by the matching. More examples are displayed in Appendix~\ref{sec:app_HE}.

%%%%%%%%%%%%%%%%%%%%%%%%%%%%%%%%%%%%%%%%%%%%%%%%%%%%%%%%%%%%%%%%%%
%%%%%%%%%%%%%%%%%%%%%%%%%%%%%%%%%%%%%%%%%%%%%%%%%%%%%%%%%%%%%%%%%%

Although this work presents skewness-dependent light-cone GPDs determined using the LaMET approach, several important limitations should be acknowledged when interpreting the results. First, we reiterate that power corrections are expected to be significant near the kinematic regions $|x| \to |\xi|$ and $x \to \pm 1$. The former manifests itself as a non-physical discontinuity between the ERBL and the DGLAP regions. As a result, the regions where the extracted GPDs are expected to be reliable are significantly restricted. It indicates that the ERBL region ($|x|<|\xi|$) is effectively excluded from reliable reconstruction at $|\xi|\leq0.2$, while even in the DGLAP region the results may still suffer non-negligible power corrections. Second, a discontinuity is also observed around $x = 0$, which indicates the presence of the inverse problem unaccounted for by the BG method. In $\xi\ne0$ cases of GPDsF, this discontinuity at $x=0$ appears in both the quasi- and light-cone GPDs. It is non-physical and is interpreted as an artifact of the BG reconstruction procedure. Overcoming these limitations requires a much better quality of lattice data, particularly in terms of the attained nucleon boost.

\section{Conclusions and outlook}

In this paper, we extended the asymmetric-frame lattice-QCD program for generalized parton distributions (GPDs) to the case of nonzero skewness ($\xi\neq0$), including the special limit of purely longitudinal momentum transfer. Using a Lorentz-covariant parametrization in terms of frame-independent amplitudes, we extracted the dominant amplitudes from proton matrix elements on an $N_f=2{+}1{+}1$ ensemble of gauge field configurations, determined GPDs $H$ and $E$ in terms of these amplitudes and reconstructed the $x$ dependence with the Backus-Gilbert method. Finally, quasi-GPDs were matched to light-cone GPDs in the $\overline{\text{MS}}$ scheme at $2~\text{GeV}$. We demonstrated that the asymmetric frame enables access to multiple $(t,\xi)$ combinations within a single calculation, making broad kinematic mapping feasible. 

As concerns the extracted amplitudes in coordinate space, $A_1$ and $A_5$ dominate (similarly to $\xi=0$) and, as expected, decrease monotonically with increasing $-t$.
The other amplitudes are suppressed, with several cases compatible with zero within uncertainties.
In the purely longitudinal momentum transfer case, one can only extract a combination of $H$ and $E$ GPDs, denoted by $H^L$. 
However, $H^L\approx H$ in our kinematics, so in practice, this special setup gives access to the $H$ GPD.
We also tested the $\xi\!\to\!-\xi$ symmetry, which cannot be achieved exactly in asymmetric kinematics due to mismatches in values of $t$ when changing the sign of the longitudinal momentum transfer, unless no transverse momentum transfer is present. Nevertheless, the results were found to be consistent with this symmetry when possibly close values of $t$ were used.
Concerning dependence on skewness, increasing its magnitude accelerates the decay of matrix elements, signaling a non-negligible $\xi$ dependence already before matching. For the GPDs themselves, the Lorentz-invariant (LI) definition exhibits better convergence towards the light cone for $E$ and comparable behavior for $H$ relative to the standard definition. 
Thus, the LI variant was chosen to go to momentum space.
After matching, the light-cone $H$ decreases in magnitude with increasing $-t$ in the ERBL region ($|x|<|\xi|$), shows mild $t$ dependence in the DGLAP region, and is suppressed in the ERBL region as $|\xi|$ increases at comparable $t$; the discontinuity at $|x|=|\xi|$ becomes comparatively less pronounced with larger average boost. 
The present light-cone reconstructions are not reliable near $|x|\to|\xi|$ and $x\to\pm1$ due to power corrections; at current boosts, the ERBL region is effectively excluded from robust reconstruction at $|\xi|\leq0.2$, and discontinuities around $x=0$ and $x=\xi$ reflect, respectively, BG properties and limited hadron momentum.

Our results point to a rather clear path forward in lattice extractions of nonzero skewness GPDs. 
A prerequisite for valid reconstruction of the ERBL region in LaMET is to employ much larger nucleon boosts.
This will reduce power corrections, soften matching-induced discontinuities, and enlarge the reliably reconstructible $x$ window. 
However, the momentum space reconstruction can also considerably profit from inclusion of short-distance factorization, utilizing its complementarity with LaMET \cite{Ji:2022ezo}.
In turn, denser and longer coverage in Wilson-line separations $z$ will stabilize BG reconstruction by mitigating the inverse problem. 
Additionally, different reconstruction methods can be adopted, such as ones employing neural networks.
On the renormalization and matching side, adopting improved schemes (e.g., hybrid) together with corresponding matching kernels will further control systematics. A dedicated study of excited-state effects with multiple source-sink separations will help quantify residual contamination. 
Finally, broadening kinematics is also desirable and will lead to a more robust determination of the $t$ dependence.

\vspace*{5mm}
\centerline{\textbf{Acknowledgements}}
M.-H.~Chu, M.~Colaço. and K.~Cichy are supported by the National Science Centre (Poland) grant OPUS No.\ 2021/43/B/ST2/00497.
S.~Bhattacharya gratefully acknowledges the hospitality and support of the Faculty of Physics and Astronomy at Adam Mickiewicz University, where this work was completed.
M.~Constantinou acknowledges financial support by the U.S. Department of Energy, Office of Nuclear Physics,  under Grant No.\ DE-SC0025218.
The work of A.~Metz was supported by the National Science Foundation under the Award No.~2412792. 
F.~Steffens was funded in part by the Deutsche Forschungsgemeinschaft (DFG, German Research Foundation) as part of the CRC 1639 NuMeriQS -- project No.\ 511713970.
The authors also acknowledge partial support by the U.S.~Department of Energy, Office of Science, Office of Nuclear Physics under the umbrella of the Quark-Gluon Tomography (QGT) Topical Collaboration with Award DE-SC0023646.

This research was supported in part by the PLGrid Infrastructure (Prometheus and Helios supercomputers at AGH Cyfronet in Cracow).
Computations were also partially performed at the Poznan Supercomputing and Networking Center (Eagle/Altair supercomputer), the Interdisciplinary Centre for Mathematical and Computational Modelling of the Warsaw University (Okeanos supercomputer), the Academic Computer Centre in Gda\'nsk (Tryton and Tryton Plus supercomputers) and facilities of the USQCD Collaboration, funded by the Office of Science of the U.S. Department of Energy. 
This research used resources of the National Energy Research Scientific Computing Center, a DOE Office of Science User Facility supported by the Office of Science of the U.S. Department of Energy under Contract No. DE-AC02-05CH11231 using NERSC award NP-ERCAP0022961.
The gauge configurations have been generated by the Extended Twisted Mass Collaboration on the KNL (A2) Partition of Marconi at CINECA, through the Prace project Pra13\_3304 ``SIMPHYS".
Inversions were performed using the DD-$\alpha$AMG solver~\cite{Frommer:2013fsa} with twisted mass support~\cite{Alexandrou:2016izb}. 

\bibliography{reference.bib}

\appendix
%%%%%%%%%%%%%%%%%%%%%%%%%%%%%%%%%%%%%%%%%%%%%%%%%%%%%%%%%%%%%%%%%%
\section{Decomposition for longitudinal momentum transfer}
\label{sec:app_longitudinaldecomp}
%%%%%%%%%%%%%%%%%%%%%%%%%%%%%%%%%%%%%%%%%%%%%%%%%%%%%%%%%%%%%%%%%%
In this section, we focus on the case $\vec{\Delta}_T = 0$. 
To obtain a covariant projection of the momentum transfer $\Delta^\mu$ onto the longitudinal plane spanned 
by $P^\mu$ and $z^\mu$, we introduce an auxiliary vector,
\begin{align}
    v^\mu & = P^\mu - \frac{z \cdot P}{z^2} z^\mu,
\end{align}
which is by construction orthogonal to \( z^\mu \), i.e., \( v \cdot z = 0 \). Using \( z^\mu \) and \( v^\mu \), we define the longitudinal projector
\begin{align}
    \Pi^{\mu\nu}_{\parallel} & = \frac{z^\mu z^\nu}{z^2} + \frac{v^\mu v^\nu}{v^2}.
\end{align}
Applying this projector to \( \Delta^\mu \) gives its longitudinal component,
\begin{align}
    \Delta_L^\mu = \Pi^{\mu\nu}_{\parallel} \Delta_\nu 
    = \frac{z^\mu (z \cdot \Delta)}{z^2} + \frac{v^\mu (v \cdot \Delta)}{v^2}.
\end{align}
Using the definition of \( v^\mu \), one can recast this in terms of the available vectors \( P^\mu \) and \( z^\mu \) as
\begin{align}
\Delta_L^\mu & =
\left( \frac{v \cdot \Delta}{v^2} \right) P^\mu + \left( \frac{z \cdot \Delta}{z^2} - \frac{(z\cdot P)(v \cdot \Delta)}{z^2 v^2} \right) z^\mu \nonumber \\
& = \frac{z \cdot \Delta}{z^2 P^2 - (z \cdot P)^2} \big( - ( z \cdot P) P^\mu + P^2 z^\mu \big)
\equiv \alpha P^\mu + \beta z^\mu.
\label{e:alpha_beta_def}
\end{align}
Note that \( \alpha \) is dimensionless and \( \beta \) has mass dimension 2. Since \( z^\mu \sim [\text{mass}]^{-1} \) and \( P^\mu \sim [\text{mass}] \), the full decomposition is dimensionally consistent, and both terms in \( \Delta_L^\mu \) carry the correct mass dimension 1. Furthermore, we observe that both $\alpha$ and $\beta$ are proportional to $\xi$, as expected.

We now analyze how the original Lorentz structures in the matrix element decomposition, Eq.~(\ref{eq:MEparametrization}), reduce under the replacement \( \Delta^\mu \rightarrow \Delta_L^\mu = \alpha P^\mu + \beta z^\mu \). Below, we demonstrate that the terms involving \( A_3, A_5, A_6, A_7, A_8 \) reduce to linear combinations of the existing basis structures \( P^\mu \), \( z^\mu \), and \( i \sigma^{\mu z} \), with their coefficients reorganizing accordingly:

\textbf{(1) Coefficient of \( P^\mu \):}
\begin{align}
\frac{P^\mu}{m} \left[ A_1 + \alpha A_3 - \frac{\alpha^2}{2} A_5 - \frac{\alpha}{2}\, (z \cdot \Delta_L)\, A_6 - \frac{\alpha^2}{2}\, (z \cdot \Delta_L)\, A_8 \right] \quad \equiv \quad \frac{P^\mu}{m} A_1^L.
\end{align}

\textbf{(2) Coefficient of \( z^\mu \):} 
\begin{align}
m z^\mu \left[ A_2 + \frac{\beta}{m^2} A_3 -\frac{\alpha \beta}{2 m^2} A_5 - \frac{\alpha}{2}\, (z \cdot \Delta_L)\, A_7 - \frac{\alpha \beta}{2m^2} (z \cdot \Delta_L) \, A_8 \right] \quad \equiv \quad m z^\mu A_2^L.
\end{align}

\textbf{(3) Coefficient of \( i \sigma^{\mu z} \):}
\begin{align}
i m \sigma^{\mu z} \left[ A_4 + \frac{\beta}{m^2} A_5 \right] \quad \equiv \quad i m \sigma^{\mu z} A_4^L.
\end{align}

\bigskip

\noindent
Therefore, the full matrix element involving only longitudinal momentum transfer reduces to the following covariant decomposition,
\begin{align}
F^{\mu,L} (z,P,\Delta_L) = \bar{u}(P_f,\lambda') \left[ \frac{P^{\mu}}{m} A_1^L + m z^{\mu} A_2^L +  i m \sigma^{\mu z} A_4^L \right] u(P_i, \lambda) ,
\label{e:F_forward}
\end{align}
where we have suppressed the arguments of the amplitudes, $A_i^L \equiv A_i^L(z \cdot P, z \cdot \Delta^L, (\Delta^L)^2, z^2)$. This shows that, under the replacement \( \Delta^\mu \to \Delta_L^\mu \in \text{span}(P^\mu, z^\mu) \), the number of independent structures reduces to three, consistent with the number of independent vectors and tensors that can be constructed from \( P^\mu \), \( z^\mu \), and the Dirac bilinears.

%%%%%%%%%%%%%%%%%%%%%%%%%%%%%%%%%%%%%%%%%%%%%%%%%%%%%%%%%%%%%%%%%%
\section{Extraction of amplitudes from matrix elements}
\label{sec:app_LSM}
%%%%%%%%%%%%%%%%%%%%%%%%%%%%%%%%%%%%%%%%%%%%%%%%%%%%%%%%%%%%%%%%%%
The matrix elements $\Pi_i(\Gamma_j)$ are linearly related to the amplitudes through a coefficient matrix $C$ of size $16 \times 8$.

In the least-squares approach, the 16 decomposition equations in Eqs.~(\ref{eq:Pi0G0_ns}--\ref{eq:Pi3G3_ns}) can be written in matrix form as
\begin{equation}
    \Pi = C \cdot A + \varepsilon,
\end{equation}
where $\Pi$ is a column vector of length 16 containing 16 matrix elements, $A$ is the vector of amplitudes to be determined, and $\varepsilon$ denotes statistical noise. Since the system is overdetermined, the amplitudes are extracted by minimizing the squared residual,
\begin{equation}
    \min_A \|\Pi - C \cdot A\|^2.
\end{equation}
The solution is given by the normal equation,
\begin{equation}
    A = (C^\top C)^{-1} C^\top \Pi.
\end{equation}
This procedure is applied independently for each kinematic setup and bootstrap sample, treating the best-fit values as the extracted amplitudes while neglecting the bias term $\varepsilon$.

In the alternative method, we manually reduce the system to a minimal set of independent equations. First, two matrix elements, $\Pi_0(\Gamma_3)$ and $\Pi_3(\Gamma_3)$, are always discarded as their corresponding coefficient rows are zero. When $\Delta_1 = 0$ and $\Delta_2 \ne 0$, six additional matrix elements—$\Pi_0(\Gamma_2)$, $\Pi_1(\Gamma_0)$, $\Pi_1(\Gamma_1)$, $\Pi_2(\Gamma_2)$, $\Pi_2(\Gamma_3)$, and $\Pi_3(\Gamma_2)$—also have vanishing coefficients. This leaves 8 independent nonzero equations. A similar reduction applies when $\Delta_2 = 0$ and $\Delta_1 \ne 0$.

When $\Delta_1 = \pm \Delta_2$, there exist six pairs of matrix elements with identical or sign-flipped coefficients:
\begin{align}
\Pi_0(\Gamma_1),\mp\Pi_0(\Gamma_2),\nonumber\\
\Pi_1(\Gamma_0),\pm\Pi_2(\Gamma_0),\nonumber\\
\Pi_1(\Gamma_1),\pm\Pi_2(\Gamma_2),\nonumber\\
\Pi_1(\Gamma_2),\pm\Pi_2(\Gamma_1),\nonumber\\
\Pi_1(\Gamma_3),\pm\Pi_2(\Gamma_3),\nonumber\\
\Pi_3(\Gamma_1),\pm\Pi_3(\Gamma_2).
\end{align}
We average each such pair to obtain a single effective equation, thereby reducing the system to 8 independent equations.

In the general case where $|\Delta_1| \ne |\Delta_2|$, a similar pairing can be made between two distinct but related kinematic setups: $\vec{\Delta}^{(1)} = (\Delta_1, \Delta_2, \Delta_3)$ and $\vec{\Delta}^{(2)} = (\Delta_2, \Delta_1, \Delta_3)$. The corresponding matrix elements form 6 symmetry-related pairs:
\begin{align}
\Pi^{(1)}_0(\Gamma_1),\mp\Pi^{(2)}_0(\Gamma_2),\nonumber\\
\Pi^{(1)}_1(\Gamma_0),\pm\Pi^{(2)}_2(\Gamma_0),\nonumber\\
\Pi^{(1)}_1(\Gamma_1),\pm\Pi^{(2)}_2(\Gamma_2),\nonumber\\
\Pi^{(1)}_1(\Gamma_2),\pm\Pi^{(2)}_2(\Gamma_1),\nonumber\\
\Pi^{(1)}_1(\Gamma_3),\pm\Pi^{(2)}_2(\Gamma_3),\nonumber\\
\Pi^{(1)}_3(\Gamma_1),\pm\Pi^{(2)}_3(\Gamma_2).
\end{align}
Apart from them, there are another two pairs across two setups: $\Pi^{(1)}_0(\Gamma_0),\Pi^{(2)}_0(\Gamma_0)$ and $\Pi^{(1)}_3(\Gamma_0),\Pi^{(2)}_3(\Gamma_0)$. By shuffling and averaging over these pairs, the system again reduces to 8 independent equations.

All in all, any case reduces to 8 independent equations, allowing a direct solution by inverting the resulting $8 \times 8$ coefficient matrix.

In practice, we find that the least-squares method and this reduction approach yield numerically consistent results in symmetric kinematics, such as $\Delta_1 \Delta_2 = 0$ or $\Delta_1 = \pm \Delta_2$. This is because, in such cases, there is no mixing between independent kinematic setups, and both methods effectively use the same input. However, in the general case $|\Delta_1| \ne |\Delta_2|$, mixing between the setups prevents such direct reduction within a single setup, whereas the least-squares method still incorporates all available data. However after averaging amplitudes from setups, they give identical results within statistical uncertainties.

%%%%%%%%%%%%%%%%%%%%%%%%%%%%%%%%%%%%%%%%%%%%%%%%%%%%%%%%%%%%%%%%%%
\section{Additional plots of amplitudes}
\label{sec:app_amp}
\begin{figure}
\centering
\includegraphics[scale=0.47]{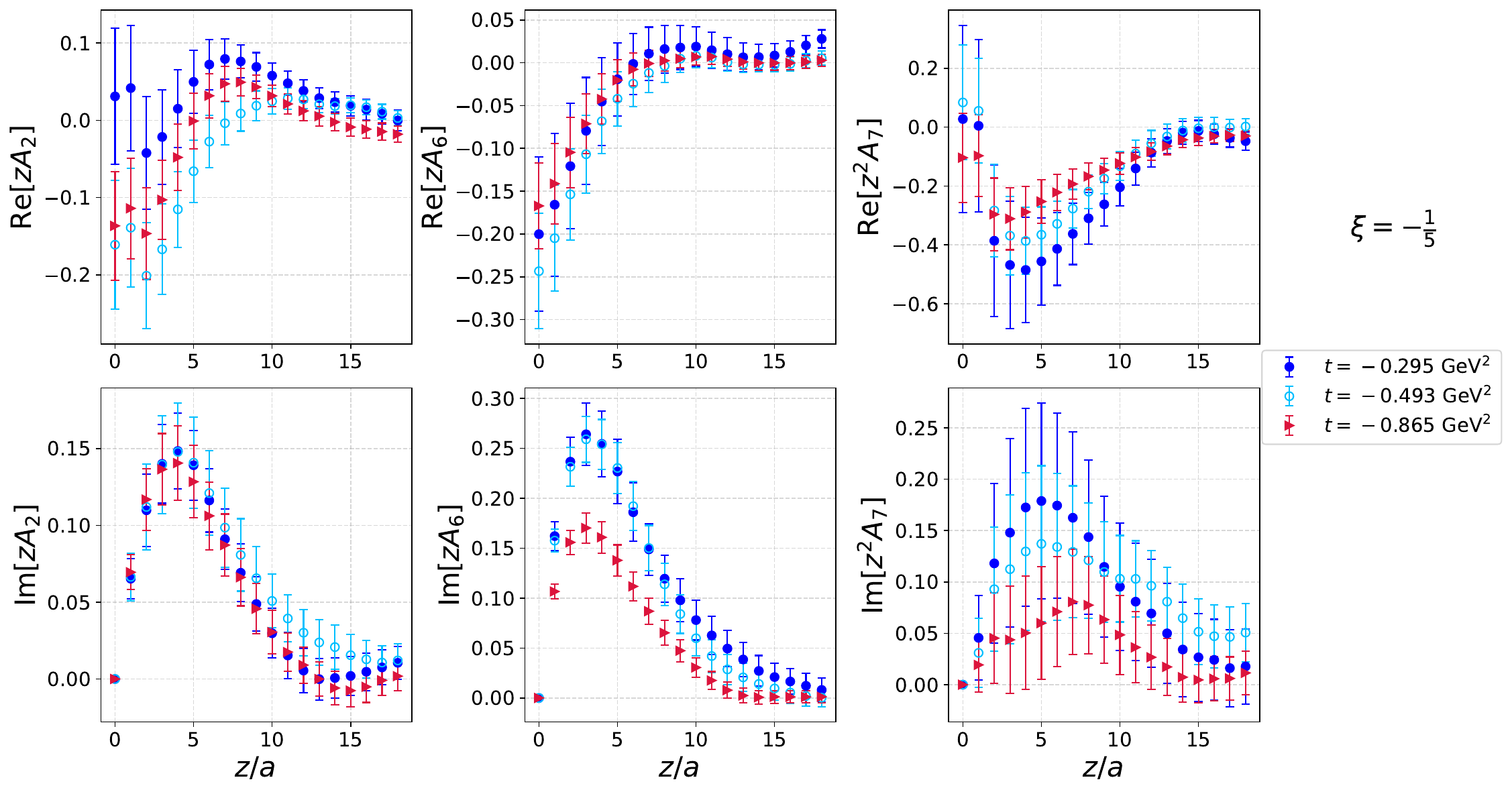}
\caption{Amplitudes $A_2$, $A_6$, and $A_7$ at several values of $t$ for $\xi=-\frac{1}{5}$.}
\label{fig:amp_267_app}
\end{figure}

\begin{figure}
\centering
\includegraphics[scale=0.47]{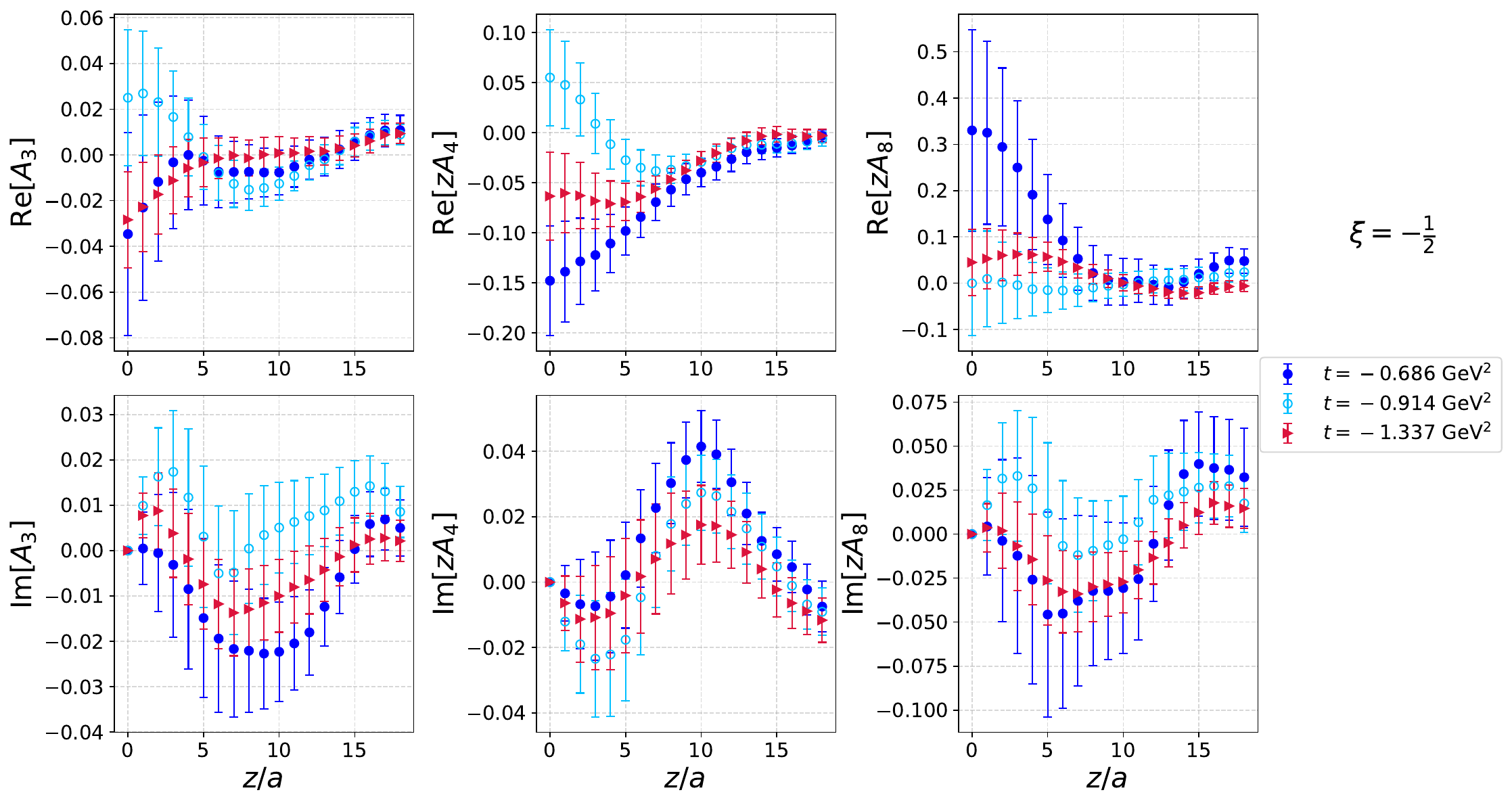}
\includegraphics[scale=0.47]{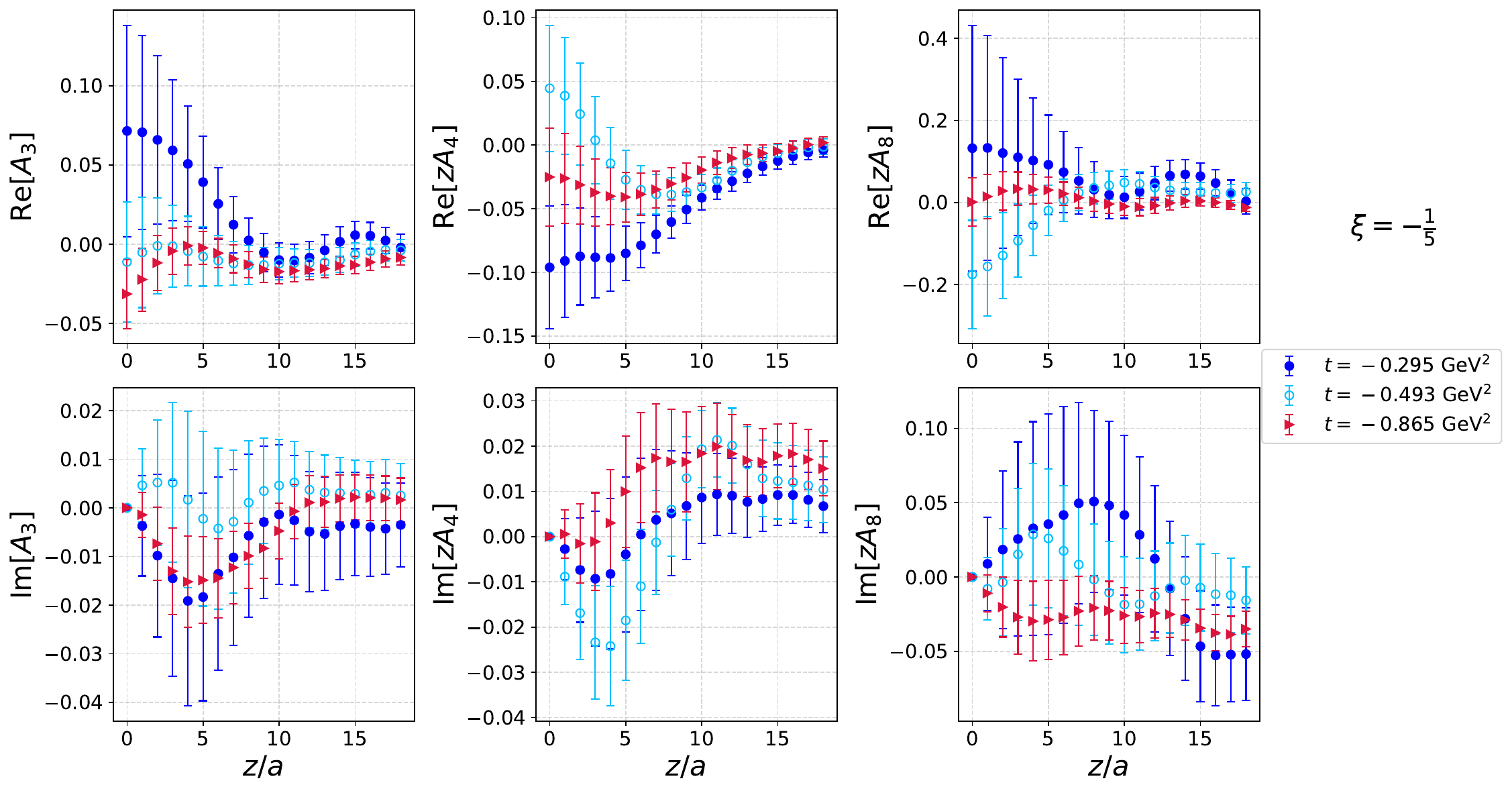}
\caption{Amplitudes $A_3$, $A_4$, and $A_8$ at several values of $t$ and $\xi$. The top six plots correspond to $\xi = -\frac{1}{2}$, and the bottom six to $\xi = -\frac{1}{5}$.}
\label{fig:amp_348_app}
\end{figure}

%%%%%%%%%%%%%%%%%%%%%%%%%%%%%%%%%%%%%%%%%%%%%%%%%%%%%%%%%%%%%%%%%%
To further assess which amplitudes are consistent with zero, we present results for $A_2$, $A_3$, $A_4$, $A_6$, $A_7$, and $A_8$ at several values of $t$ for $\xi = -1/5$ in Fig.~\ref{fig:amp_267_app} and Fig.~\ref{fig:amp_348_app}. Due to large statistical uncertainties, results at $\xi = 1/7$ are inconclusive for these suppressed amplitudes and omitted. Based on the available data, among the real and imaginary parts of these amplitudes, the components that appear significantly nonzero are $\mathrm{Re}[A_6]$, $\mathrm{Im}[A_6]$, $\mathrm{Im}[A_2]$ and $\mathrm{Re}[A_7]$ (the latter found to be consistent with zero in Fig.~\ref{fig:amp_267} at $\xi=-1/2$, but at larger values of $-t$).
The results for $\mathrm{Re}[A_2]$, found to be possibly nonzero in the $\xi=-1/2$ case, seem to point to values consistent with zero here.
Finally, all results for $A_3$, $A_4$ and $A_8$ appear compatible with statistical noise, even though the $\xi\neq0$ setup admits potentially nonzero values of these amplitudes.

%%%%%%%%%%%%%%%%%%%%%%%%%%%%%%%%%%%%%%%%%%%%%%%%%%%%%%%%%%%%%%%%%%
%%%%%%%%%%%%%%%%%%%%%%%%%%%%%%%%%%%%%%%%%%%%%%%%%%%%%%%%%%%%%%%%%%

%%%%%%%%%%%%%%%%%%%%%%%%%%%%%%%%%%%%%%%%%%%%%%%%%%%%%%%%%%%%%%%%%%
%%%%%%%%%%%%%%%%%%%%%%%%%%%%%%%%%%%%%%%%%%%%%%%%%%%%%%%%%%%%%%%%%%

\begin{figure}
\centering
\includegraphics[scale=0.6]{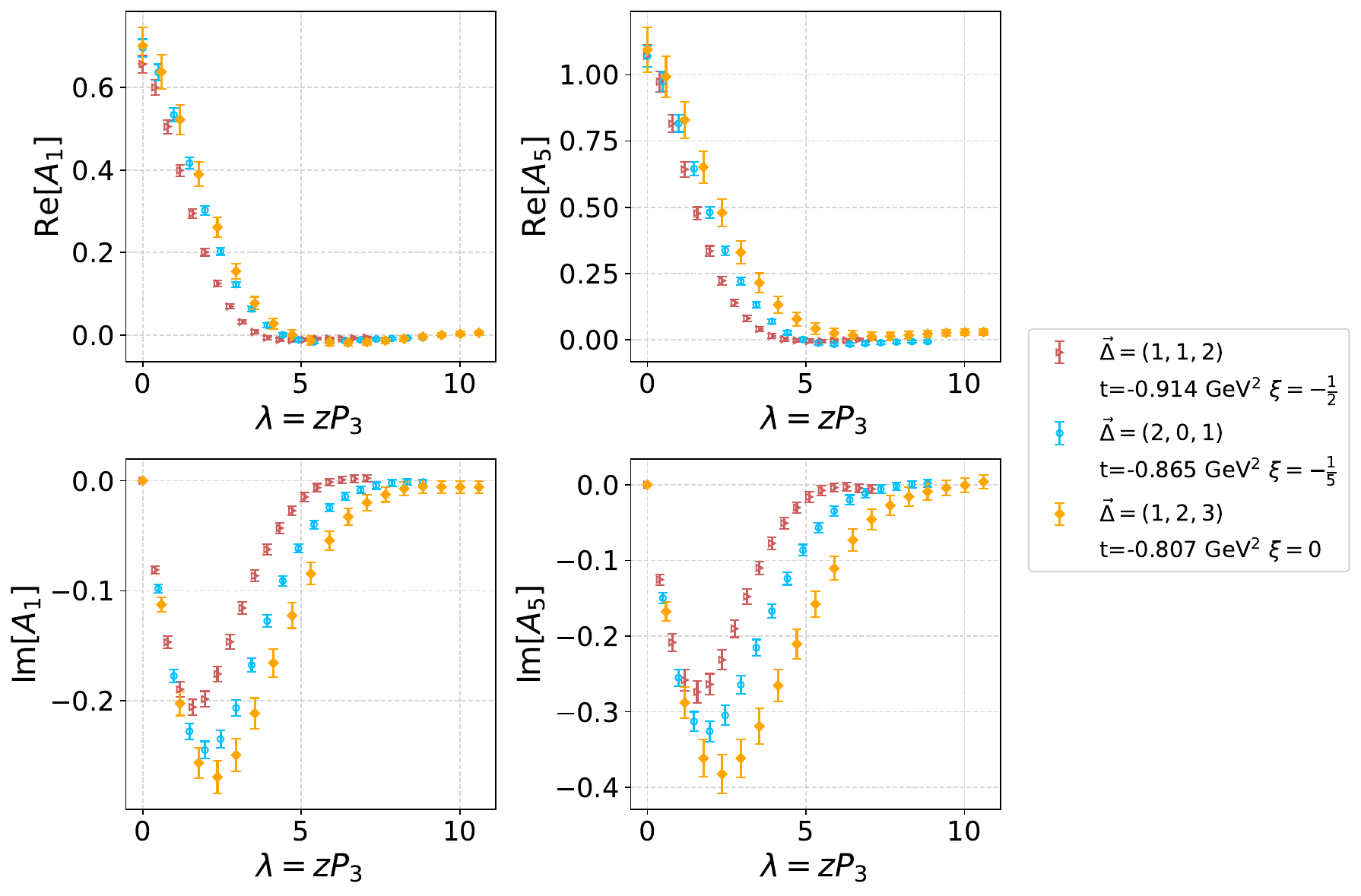}
\includegraphics[scale=0.6]{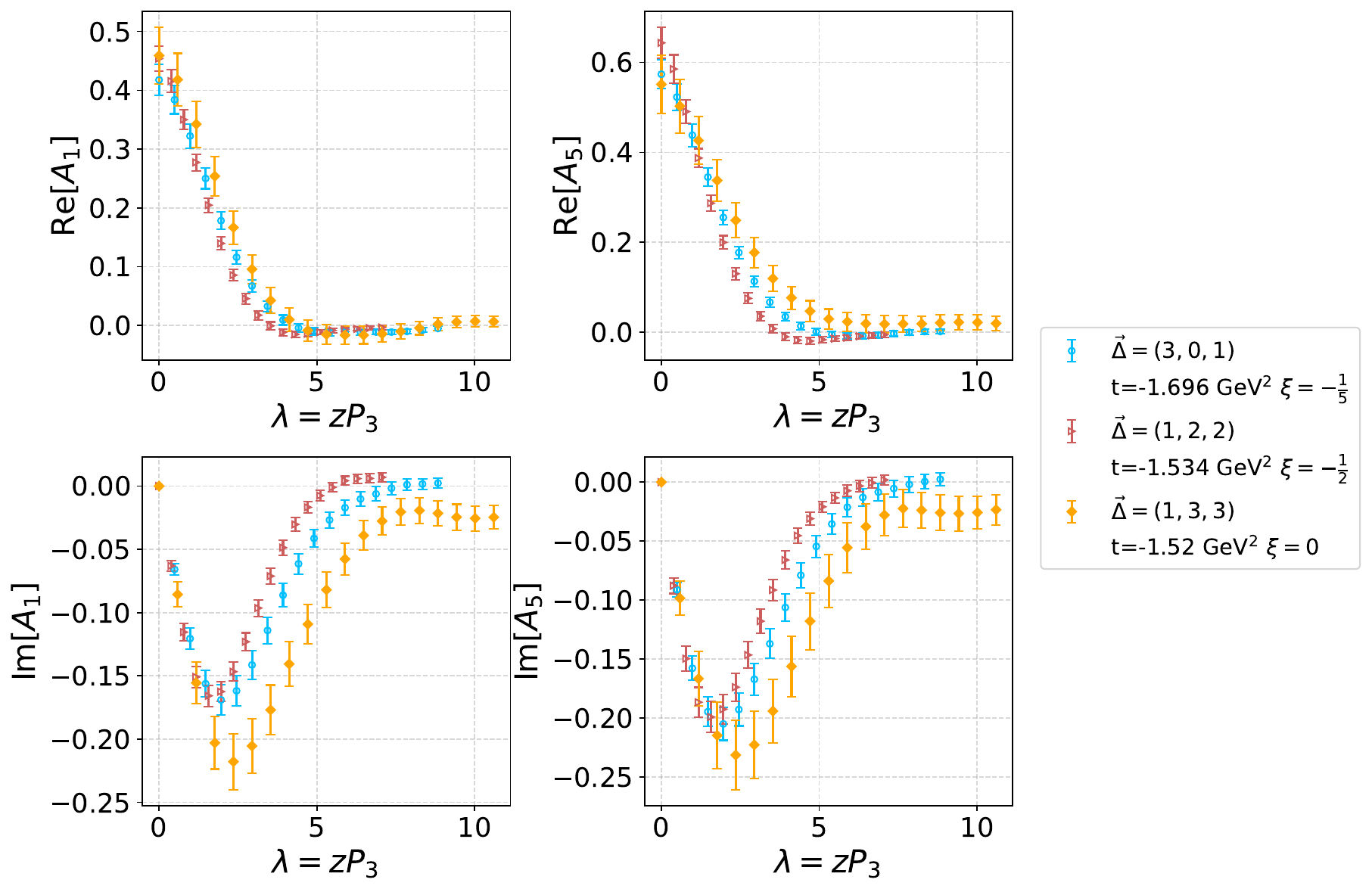}
\caption{$\xi$ dependence of $A_1$ and $A_5$ at similar values of $t$. The upper and lower panel shows the real and the imaginary part, respectively.}
\label{fig:amp_15_xi_dep_1}
\end{figure}

\begin{figure}
\centering
\includegraphics[scale=0.6]{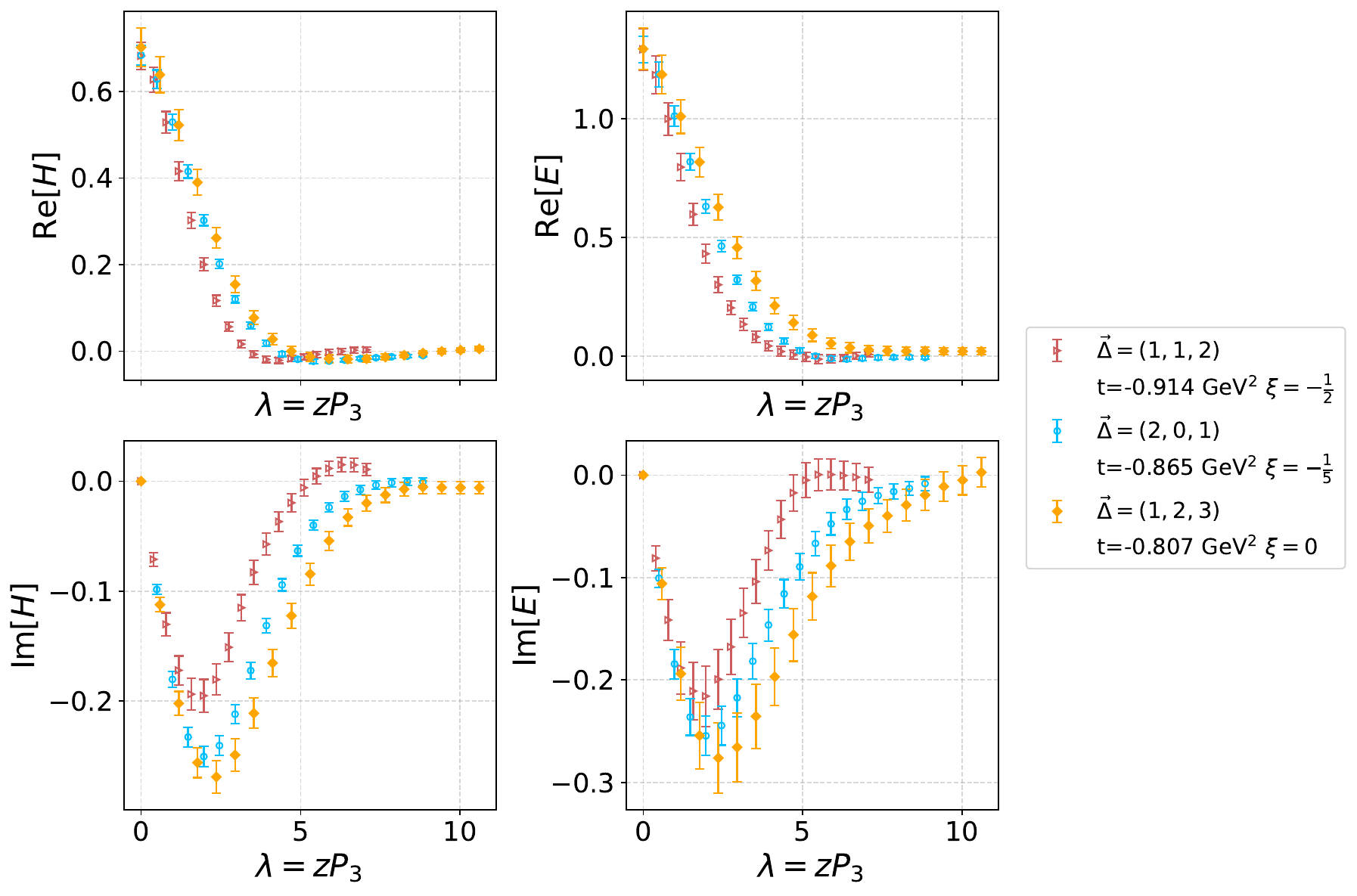}
\includegraphics[scale=0.6]{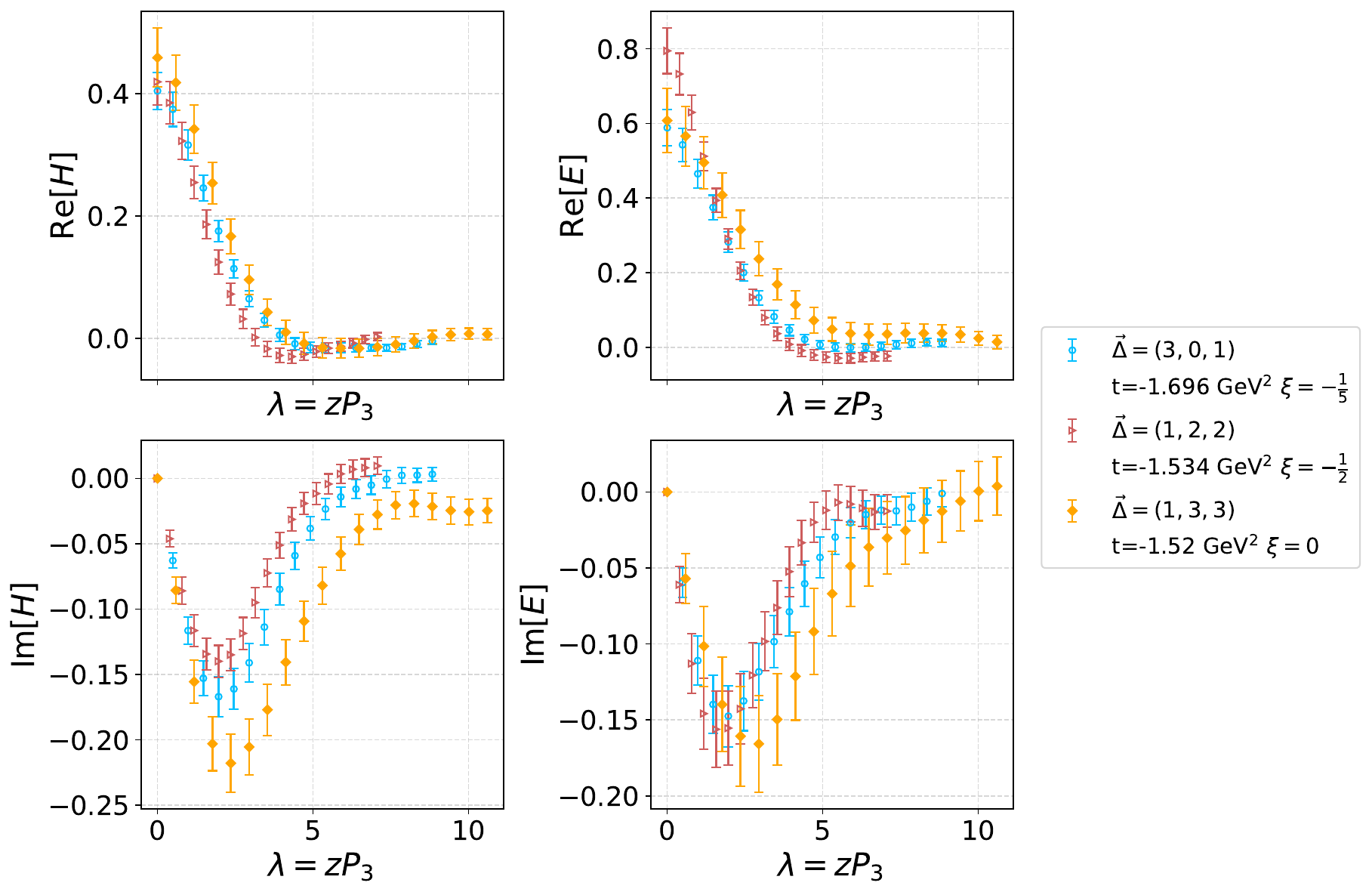}
\caption{Skewness dependence of bare $H$ and $E$ GPDs at similar values of $t$. The upper and lower panel shows the real and the imaginary part, respectively.}
\label{fig:HE_xi_dep_1}
\end{figure}

We further explored the $\xi$ dependence of the amplitudes $A_1$ and $A_5$ at approximately close values of momentum transfer $t$. Two additional examples illustrating this dependence are presented in Fig.~\ref{fig:amp_15_xi_dep_1}. In addition, the corresponding $\xi$ dependence of the generalized parton distributions $H$ and $E$ at similar values of $t$ is demonstrated in Fig.~\ref{fig:HE_xi_dep_1}.

%%%%%%%%%%%%%%%%%%%%%%%%%%%%%%%%%%%%%%%%%%%%%%%%%%%%%%%%%%%%%%%%%%
%%%%%%%%%%%%%%%%%%%%%%%%%%%%%%%%%%%%%%%%%%%%%%%%%%%%%%%%%%%%%%%%%%

%%%%%%%%%%%%%%%%%%%%%%%%%%%%%%%%%%%%%%%%%%%%%%%%%%%%%%%%%%%%%%%%%%
%%%%%%%%%%%%%%%%%%%%%%%%%%%%%%%%%%%%%%%%%%%%%%%%%%%%%%%%%%%%%%%%%%

%%%%%%%%%%%%%%%%%%%%%%%%%%%%%%%%%%%%%%%%%%%%%%%%%%%%%%%%%%%%%%%%%%
\section{Additional plots of GPDs}
\label{sec:app_HE}
\begin{figure}
\vspace*{-1cm}
\centering
\includegraphics[scale=0.55]{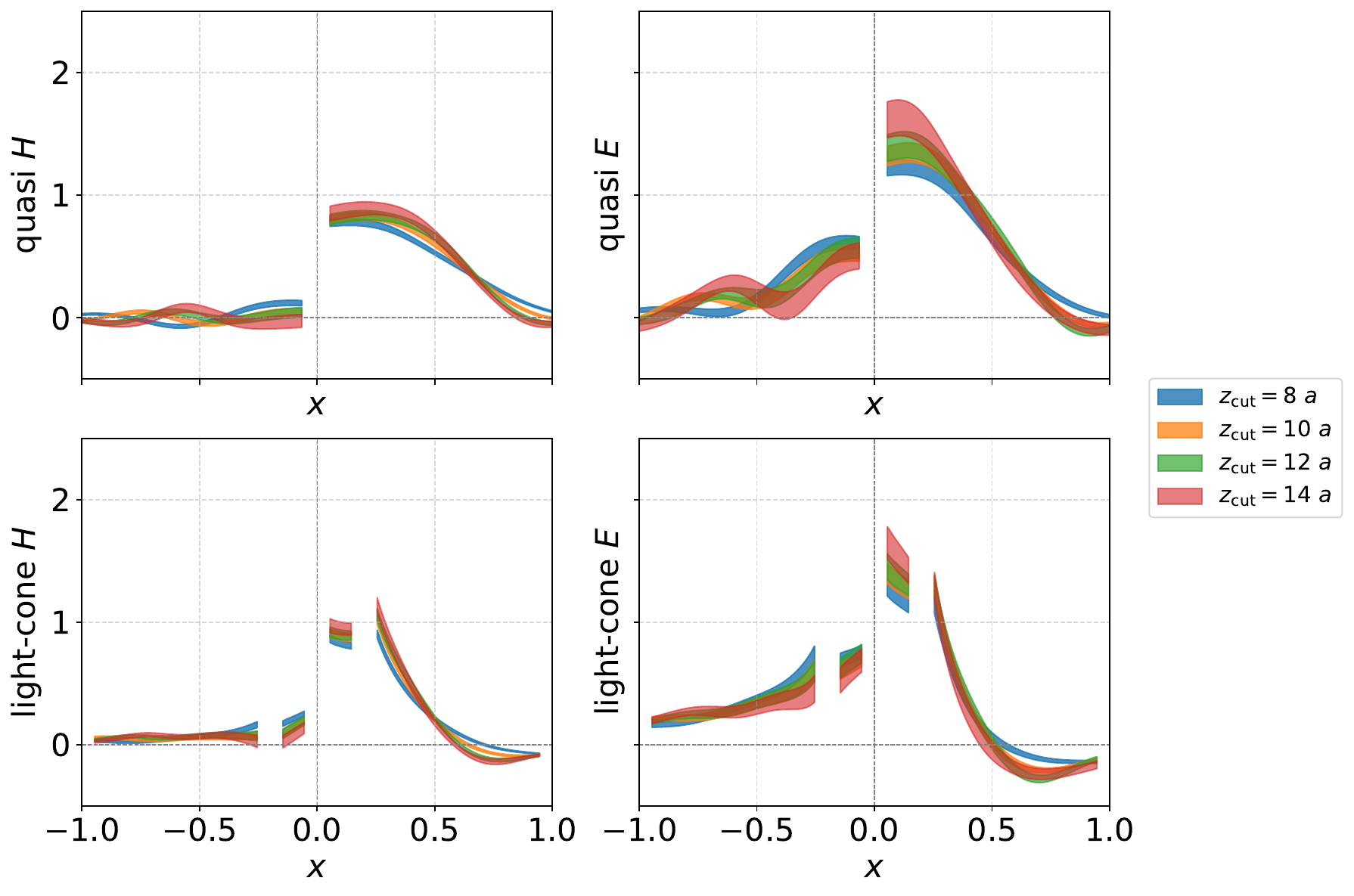}
\vspace*{-3mm}
\caption{Comparison between different truncations on $z$, denoted as $z_{\rm cut}$ in the BG process for quasi-GPDs and light-cone GPDs. They are displayed at averaged momentum transfer $\Delta_T=(\pm2,0),(0,\pm2)$ and skewness $\xi=-1/5$. Values in unreliable regions are excluded.}
\label{fig:HE_zcut}
\end{figure}

%%%%%%%%%%%%%%%%%%%%%%%%%%%%%%%%%%%%%%%%%%%%%%%%%%%%%%%%%%%%%%%%%%
The implementation of the Backus-Gilbert method requires specifying a cut $z_{\rm{cut}}$ in coordinate space. In Fig.~\ref{fig:HE_zcut}, we illustrate the comparison between quasi- and light-cone GPDs, $H$ and $E$ in Lorentz-invariant form, for different values of $z_{\mathrm{cut}}$ in the BG Fourier transform. Here, as a representative example, we choose momentum transfer values averaged over the sets $(0,\pm2)$ and $(\pm2,0)$ at a fixed skewness $\xi = -1/5$. Good convergence is observed for $z_{\mathrm{cut}}$ in the range of $8a$ to $12a$, motivating our choice of $z_{\mathrm{cut}} = 12a$ in subsequent analyses.

%%%%%%%%%%%%%%%%%%%%%%%%%%%%%%%%%%%%%%%%%%%%%%%%%%%%%%%%%%%%%%%%%%
%%%%%%%%%%%%%%%%%%%%%%%%%%%%%%%%%%%%%%%%%%%%%%%%%%%%%%%%%%%%%%%%%%
%%%%%%%%%%%%%%%%%%%%%%%%%%%%%%%%%%%%%%%%%%%%%%%%%%%%%%%%%%%%%%%%%%
\begin{figure}
\vspace*{-0.5cm}
\centering
\includegraphics[scale=0.55]{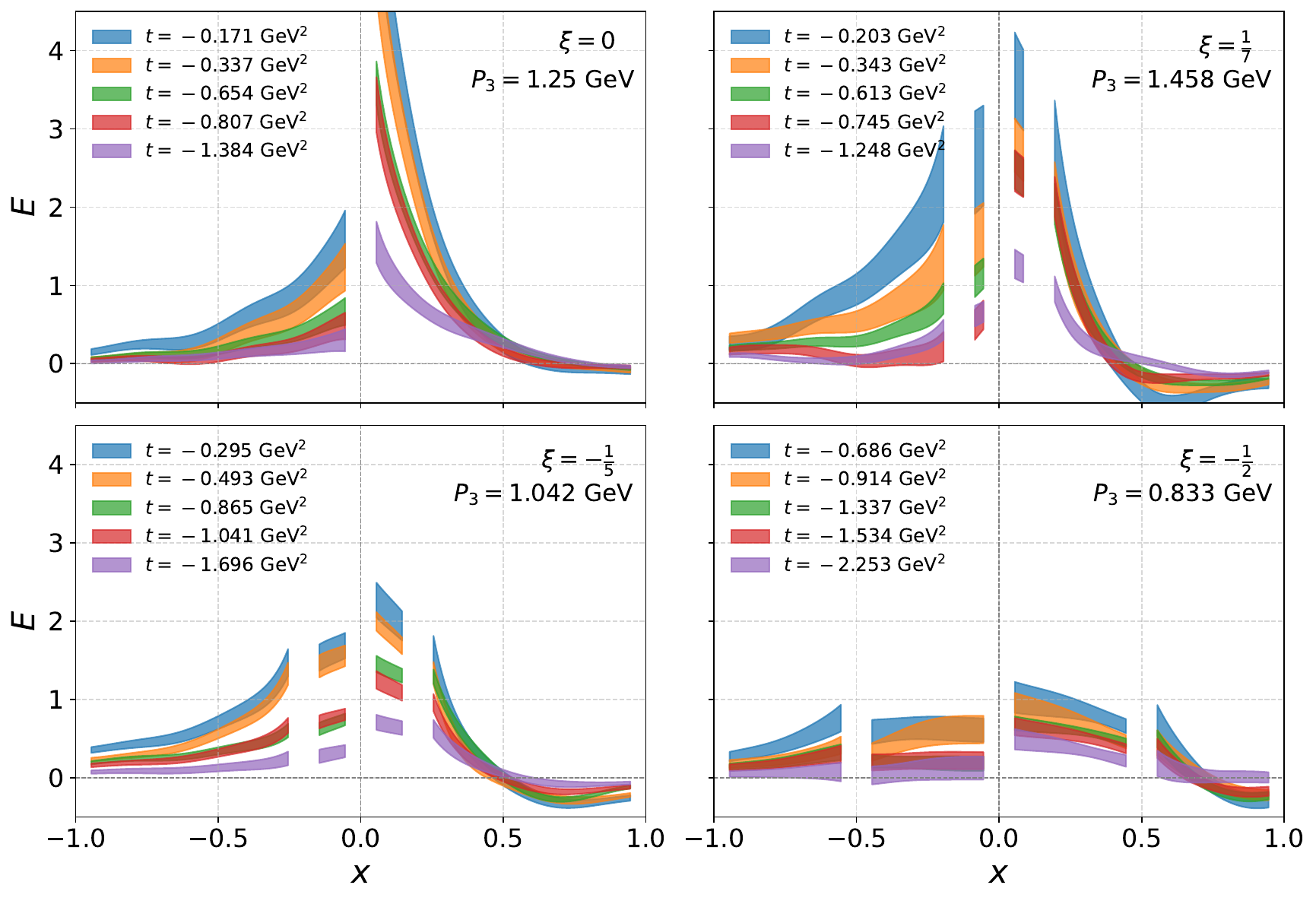}
\vspace*{-3mm}
\caption{Dependence on momentum transfer $t$ for light-cone GPDs $E$ evaluated at different skewness values $\xi$. Values in unreliable regions are excluded.}
\label{fig:E_x_t_dep}
\end{figure}
%%%%%%%%%%%%%%%%%%%%%%%%%%%%%%%%%%%%%%%%%%%%%%%%%%%%%%%%%%%%%%%%%%

The dependence on the values of $t$ for the $E$ GPD is illustrated in Fig.~\ref{fig:E_x_t_dep}. Similar to $H$, it decreases with increasing $-t$ in the ERBL region and has mild dependence on $t$ in the DGLAP region. The relative discontinuity at $|x|=|\xi|$ is also softened as $P^z_{\mathrm{ave}}$ increases.

\begin{figure}
\vspace*{-1cm}
\centering
\includegraphics[scale=0.465]{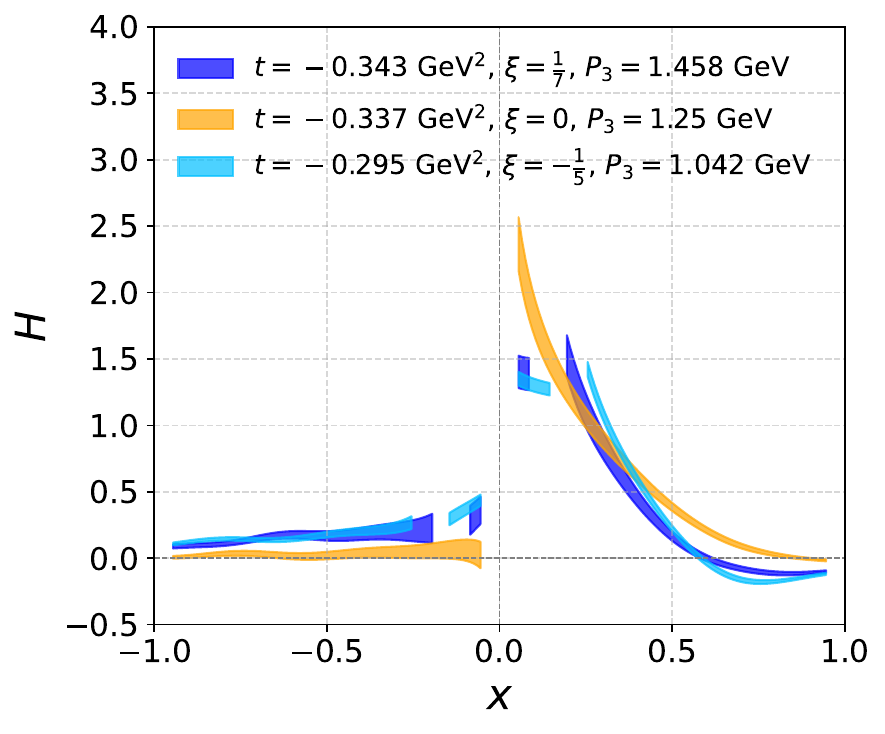}
\includegraphics[scale=0.465]{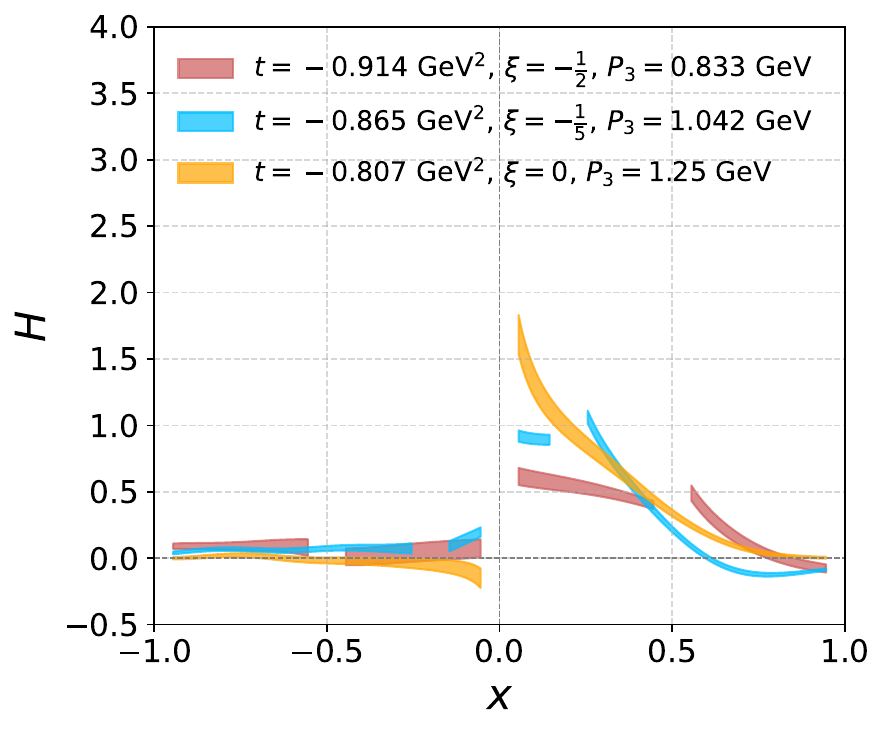}
\vspace*{-0.5cm}
\includegraphics[scale=0.465]{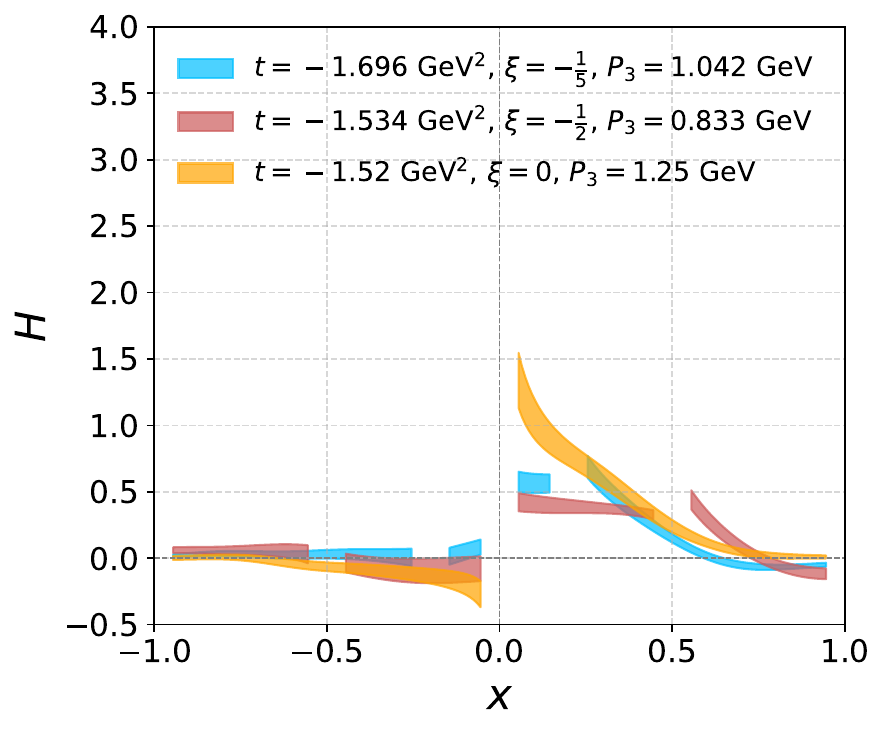}
\caption{Skewness dependence of light-cone GPDs $H$ at comparable values of $t$. Values in unreliable regions are excluded.}
\label{fig:HE_x_xi_dep_H}
\end{figure}

\begin{figure}
\vspace*{-0.5cm}
\centering
\includegraphics[scale=0.465]{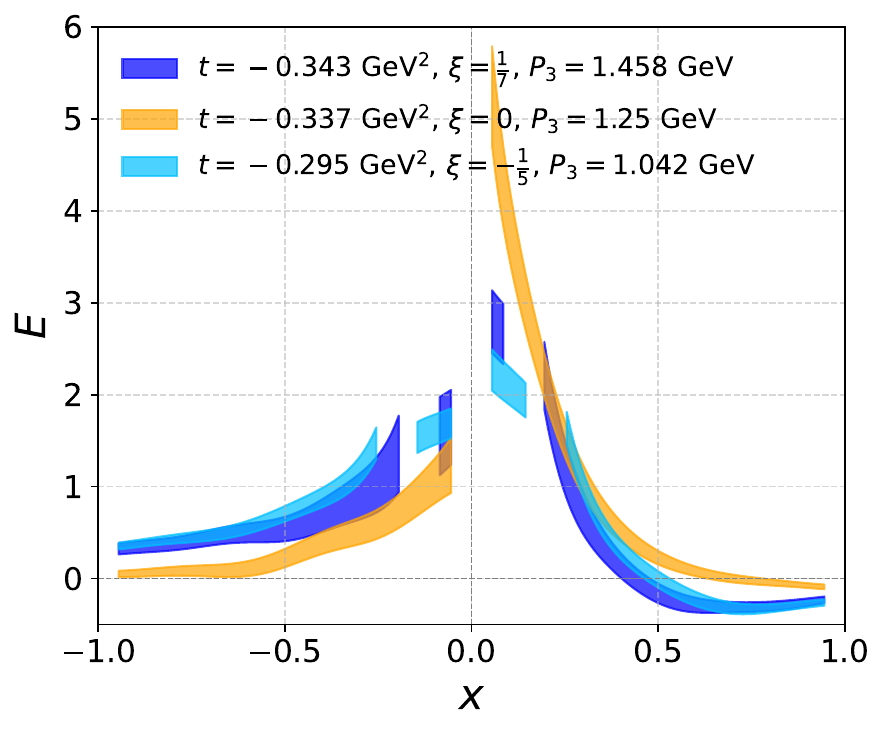}
\includegraphics[scale=0.465]{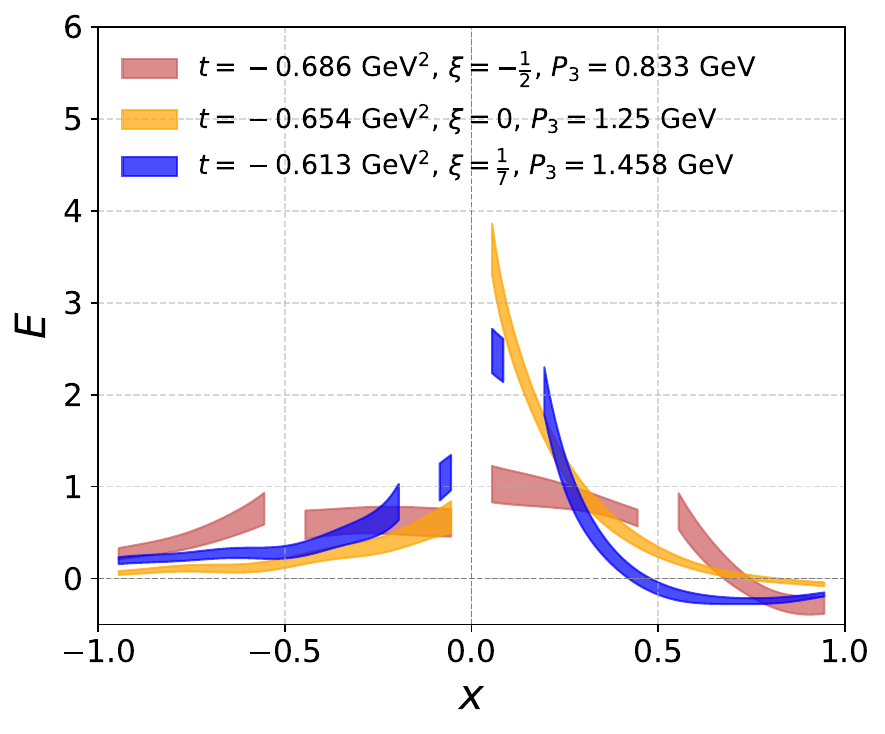}
\vspace*{-0.5cm}
\includegraphics[scale=0.465]{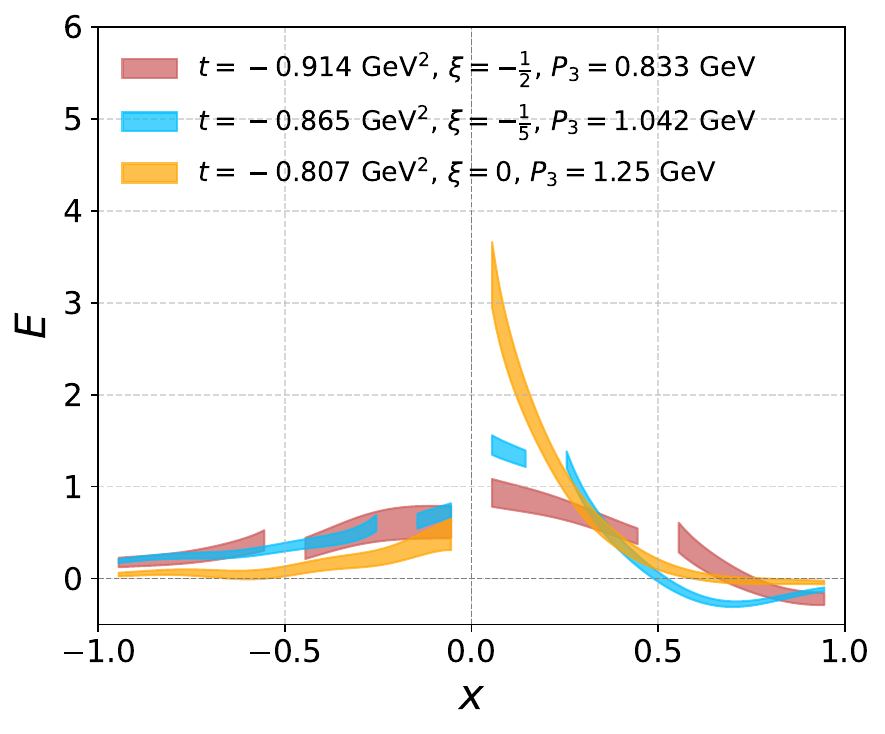}
\includegraphics[scale=0.465]{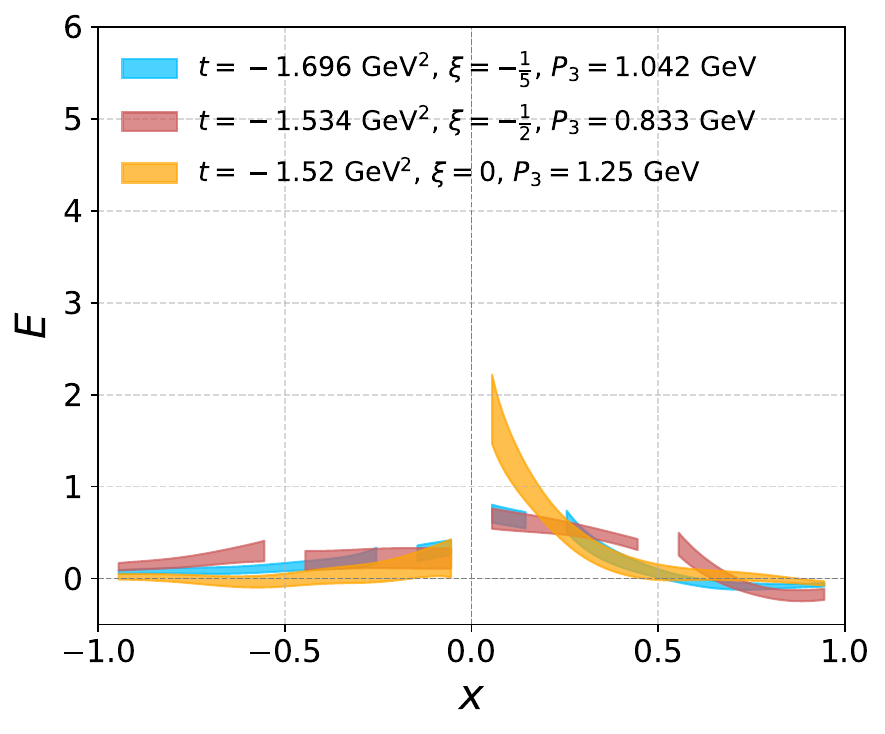}
\caption{Skewness dependence of light-cone GPDs $E$ at comparable values of $t$. Values in unreliable regions are excluded.}
\label{fig:HE_x_xi_dep_E}
\end{figure}

The following plots in Fig.~\ref{fig:HE_x_xi_dep_H} and~\ref{fig:HE_x_xi_dep_E}  illustrate the investigation of the skewness dependence of the light-cone $H$ and $E$ GPDs for additional kinematic cases. 
Conclusions are similar to the ones spelled out in the main text.

%%%%%%%%%%%%%%%%%%%%%%%%%%%%%%%%%%%%%%%%%%%%%%%%%%%%%%%%%%%%%%%%%%
%%%%%%%%%%%%%%%%%%%%%%%%%%%%%%%%%%%%%%%%%%%%%%%%%%%%%%%%%%%%%%%%%%

%%%%%%%%%%%%%%%%%%%%%%%%%%%%%%%%%%%%%%%%%%%%%%%%%%%%%%%%%%%%%%%%%%
%%%%%%%%%%%%%%%%%%%%%%%%%%%%%%%%%%%%%%%%%%%%%%%%%%%%%%%%%%%%%%%%%%

\end{document}